\documentclass[12pt]{article}
\usepackage{amsmath,amsfonts,amssymb}
\usepackage{latexsym}
\usepackage{float}   %For tighter placing of the floats (figures, tables, etc.)
\usepackage{epsf}
\usepackage{graphicx} %For figure environments
\restylefloat{figure} % use it for floats to be put in your specified place [H]
\usepackage[usenames]{color}
\def\be{\begin{equation}}
\def\bea{\begin{eqnarray}}
\def\ee{\end{equation}}
\def\eea{\end{eqnarray}}
\def\non{\nonumber}

\def\eqref#1{(\ref{#1})}
\def\a{\alpha}

\def\ket{\rangle}
\def\e{{\mathrm e}}
\def\im{\mathrm i}

\def\Tr{\hbox{\fontfamily{phv}\selectfont Tr}}
\def\ket#1{|#1\rangle}

\def\dag{\dagger}

\numberwithin{equation}{section}
\begin{document}
%========================================================================

%----------------------------------------------------------------------%
%                     T I T L E
%----------------------------------------------------------------------%

\begin{titlepage}

\begin{centering}

\vspace*{3cm}

\textbf{\Large  Spectrum and thermodynamic properties of
two-dimensional $\mathcal{N}=(1,1)$ super Yang--Mills theory
with fundamental matter and a Chern--Simons term}

\vspace{1.5cm}

\textbf{John R. Hiller,$^a$ Stephen Pinsky,$^b$ Yiannis Proestos,$^b$
Nathan Salwen,$^b$ and Uwe Trittmann$^c$}

\vspace{0.5cm}

\textsl{ ${}^a$Department of Physics \\
University of Minnesota Duluth\\
Duluth, MN 55812}

\vspace{0.5cm}

\textsl{ ${}^b$Department of Physics \\
Ohio State University\\
Columbus, OH 43210}

\vspace{0.5cm}

\textsl{ ${}^c$Department of Physics \& Astronomy\\
Otterbein College\\
Westerville, OH 43081}

\vspace{0.5cm}
%
%{\color{red}Version:}   {\texttt February 09, 2007}
%
%----------------------------------------------------------------------%
%                       A B S T R A C T
%----------------------------------------------------------------------%

\vspace{0.5cm}

\begin{abstract}
We consider  $\mathcal{N}=(1,1)$ super Yang--Mills theory in 1+1 dimensions
with fundamentals at large-$N_c$.  A Chern--Simons term is included to
give mass to the adjoint partons. Using the spectrum of the theory, we calculate
thermodynamic properties of the system as a function of the temperature and
the Yang--Mills coupling. In the large-$N_c$ limit there are two
non-communicating sectors, the glueball sector, which we presented
previously, and the meson-like sector that we present here.  We find that the
meson-like sector dominates the thermodynamics. Like the glueball sector,
the meson sector has a
Hagedorn temperature $T_H$, and we show that the Hagedorn temperature grows
with the coupling. We calculate the
temperature and coupling dependence of the free energy for temperatures
below $T_H$. As expected, the free energy for weak coupling and low temperature  grows quadratically with the
temperature. Also the ratio of the free energies at strong coupling compared to weak
coupling,  $r_{s-w}$, for low temperatures grows quadratically with $T.$ In addition, our data suggest that $r_{s-w}$ tends to zero in the continuum limit at low temperatures.
\end{abstract}

\end{centering}

%\noindent
%PACS number(s):

\vfill

\end{titlepage}
\newpage
%
%----------------------------------------------------------------------%
%                       Introduction
%----------------------------------------------------------------------%
%%%%%%%%%%%%%%%%%%%%%%%%%%%%%%%%%%%%%%%%%%%%%%%%%%%%%%%%%%%%%%%%%%
\section{Introduction}
\label{sectintro}
%%%%%%%%%%%%%%%%%%%%%%%%%%%%%%%%%%%%%%%%%%%%%%%%%%%%%%%%%%%%%%%%%%
The free energy of ${\cal N}=4$ super
Yang--Mills (SYM) theory at a large number of colors $N_c$ is larger at strong
coupling by a factor 4/3 compared to weak
coupling~\cite{Gubser:1996de,Li:1998kd}.
The weak-coupling result is calculable in
perturbation theory, while the strong-coupling result can be derived from black-hole
thermodynamics.
In the light of this finding, one can ask if other SYM theories exhibit
a similar behavior. An analytic calculation of both the strong and the weak
coupling limit of a field theory is
generally not possible, although there have been a number of proposals
for methods to obtain solutions of finite-temperature supersymmetric
quantum field theories~\cite{Das:rx}.  We will use a numerical
approach which is based on Supersymmetric Discrete Light-Cone Quantization
(SDLCQ)~\cite{sakai,Lunin:1999ib}, thereby preserving supersymmetry
exactly.  Currently, this is the only method available for numerically
solving strongly coupled SYM theories. Conventional lattice
methods have difficulty with supersymmetric theories because of the
asymmetric way that fermions and bosons are treated, and
progress~\cite{lattice} in supersymmetric lattice gauge theory has been
relatively slow.

Previously we calculated the thermodynamic properties of pure glue
$\mathcal{N}=(1,1)$ SYM theory in $1+1$ dimensions~\cite{Hiller:2004ft}.
Here we extend the calculations
to include a sector with fundamental partons. In the large-$N_c$ limit the
bound states in this sector of the theory are chains in color space with a
fundamental parton at each end. The links in the chain are adjoint
partons. Bound states of this type
will be called mesonic because they have two fundamental partons, whereas
solutions with only adjoint partons will be called glueballs.
We have also extended the calculation to include a Chern--Simons (CS)
term, which gives mass to the adjoint partons.

The mesons and glueballs constitute
two sectors of the same theory; both contribute to the thermodynamics. In
the  large-$N_c$ limit the sectors decouple.
Due to the cyclic redundancy
of single-trace glueball states, there are many more meson states,
which are in turn
likely to dominate the thermodynamic properties of the system.
In previous work on the glueball sector~\cite{Hiller:2004ft} we found
that the system possesses a Hagedorn temperature.

Recall that SDLCQ makes use of light-cone coordinates, with
$x^+=(x^0+x^3)/\sqrt{2}$ the time variable and $p^-=(p^0-p^3)/\sqrt{2}$
the energy. One must be careful in defining thermodynamic quantities on
the light-cone. It seems natural to define the partition
function~\cite{Brodsky:2001bx} on the light-cone as $e^{-\beta_{_{\rm
LC}}p^-}$. However, as shown by Alves and Das~\cite{Alves:2002tx},
the above prescription leads to singular results for well known
quantities which are finite in the equal-time approach. Their argument is
based on the fact that using $e^{- \beta_{_{\rm LC}} p^-}$ as the
partition function implies that the physical system is in contact with a
heat bath that has been boosted to the light-cone frame. However, this is
\textit{not} equivalent to the physics of a system in contact with a heat
bath at rest.
This can be realized in a more direct way by noting that, since the
light-cone momentum
$$p^+=\frac{(p^0+p^3)}{\sqrt{2}}$$
is conserved, the partition function must include the conserved quantity
and is of the form
$$\mathcal{Z}=\mathrm{e}^{-\beta_{_{\rm LC}}(p^- + \mu p^+)},$$
where $\mu$ is the chemical potential corresponding to the conserved
quantity.  The physical interpretation of the chemical potential is that
of a rotation of the quantization axis. Thus $\mu=1$ corresponds to
quantization in an equal-time frame, where the heat bath is at rest and
the inverse temperature is $\beta=\sqrt{2}\beta_{_{\rm LC}}$, and $\mu \neq
1$ corresponds to quantization in a boosted frame where the heat bath is
not at rest. Thus $\mu$ corresponds to a continuous rotation of the axis
of quantization, and $\mu=0$ would correspond to rotation all the way to
the light-cone frame.

A rotation from an equal-time frame to the
light-cone frame is not a Lorentz transformation. It is known that such a
transformation can give rise to singular results for physical quantities.
This appears to be consistent with the results found
in~\cite{Alves:2002tx}. A number of related issues have been extensively
discussed by Weldon~\cite{Weldon:aq}. The method has also recently been
applied to the Nambu--Jona-Lasinio model~\cite{Beyer:2003qb}.

The difficulties are avoided if we compute the equal-time partition
function $\mathcal{Z}=e^{-\beta p^0}$, as was proposed much earlier by 
Elser and
Kalloniatis~\cite{Elser:1996tq}.  The computation may, of course, still
use light-cone coordinates. Elser and Kalloniatis did this with ordinary
DLCQ~\cite{pb85,bpp98} as a numerical approximation to (1+1)-dimensional
quantum electrodynamics.  

Here we will follow a similar approach using
SDLCQ to calculate the spectrum of ${\cal N}=(1,1)$ super Yang--Mills
theory in 1+1 dimensions~\cite{Antonuccio:1998kz}. Though this calculation
is done in 1+1 dimensions, it is known that SDLCQ can be extended in a
straightforward manner to higher dimensions~\cite{Antonuccio:1999zu,
Haney:2000tk,hpt2001}. As is customary, we will assume that the single-trace
bound states of our large-$N_c$ approximation are single-particle states.

We have discussed the SDLCQ numerical method in a number of other places,
and we will not present a detailed discussion of the method here; for a
review, see~\cite{Lunin:1999ib}. For those familiar with
DLCQ~\cite{pb85,bpp98}, it suffices to say that SDLCQ is similar; both
impose periodic boundary conditions on a light-cone box $x^- \in [-L,L]$
and have discrete momenta and cutoffs in momentum space.  In
1+1 dimensions the discretization is specified by a single integer
$K=(L/\pi) P^+$, the resolution~\cite{pb85}, such that longitudinal momentum fractions
are integer multiples of $1/K$. However, SDLCQ is formulated in such a way
that the theory is also exactly supersymmetric.  Exact supersymmetry
brings a number of very important numerical advantages to the method; in
particular, theories with enough supersymmetry are finite. We have also
seen greatly improved numerical convergence in this approach.

The calculation of thermodynamic quantities requires summing over the
spectrum of available states, which we represent by a density of states (DoS).
We will use a new numerical approach to estimate 
the density of states.  The new approach is more efficient than the method used
in previous work~\cite{Hiller:2004ft},
because it allows us to extract the density of states without 
fully diagonalizing the Hamiltonian, a computationally challenging task.
This innovation will enable us to pursue calculations at higher values
of the resolution $K$.

We find that the two-dimensional SYM theory with fundamentals and 
a Chern--Simons term exhibits a Hagedorn temperature
$T_H$~\cite{Hagedorn_NC_65_68}, and calculate $T_H$ for
several values of the resolution $K$ and Yang--Mills coupling $g$. 
Extrapolating to the continuum limit, we obtain $T_H$ as a
function of the coupling. The Hagedorn temperature is used as an upper 
limit for the temperatures we can use to calculate the 
thermodynamic properties of the system.

In Sec.~\ref{SectDefin} we provide a review of the formulation of super
Yang--Mills theory  with fundamental matter and a Chern--Simons term in 1+1
dimensions.  In Sec.~\ref{SDLCQspect} we discuss some of the properties of the
SDLCQ spectra in some limiting cases and provide comparisons between glueball and
mesonic sectors of the theory. The discussion in Sec.~\ref{Sec:Density}
presents the methods for estimating the density of states and
the Hagedorn temperature.
In Sec.~\ref{sec:Finite temp}, we summarize our formulation of the thermodynamics 
and the formulae we use to calculate
the free energy. We then present the numerical results for the free energy,
which we obtained using the DoS approximation to the spectrum, at various values of
SYM coupling $g$ up to the Hagedorn temperature. Finally,  in
Sec.~\ref{sec:discussion} we conclude by summarizing our
results and the prospects for future work using these methods.

%%%%%%%%%%%%%%%%%%%%%%%%%%%%%%%%%%%%%%%%%%%%%%%%%%%%%%%%%%%%%%%%%%
\section{Super Yang--Mills theory with fundamental matter and Chern--Simons term}
\label{SectDefin}
\subsection{Formulation of the theory and its supercharges}
\label{SubSectSuch}
%%%%%%%%%%%%%%%%%%%%%%%%%%%%%%%%%%%%%%%%%%%%%%%%%%%%%%%%%%%%%%%%%%
We start by considering $\mathcal{N}=1$ supersymmetric gauge theory in
$2+1$ dimensions coupled to fundamental matter and a Chern--Simons
three-form. The action is
\begin{equation}
S_{2+1}=S_{YM}+S_{\mathrm{f.matter}}+S_{CS}, \label{actionA}
\end{equation}
with
\begin{subequations}
\begin{align}
\label{eqn:SYM} S_{YM}&=\int dx^3 \,\Tr
\bigl(-\frac{1}{4}F_{\mu\nu}F^{\mu\nu} +
\frac{\im}{2}{\bar\Lambda} \Gamma^\mu D_\mu \Lambda \bigr),\\
\label{eqn:Sfmatter} S_{\mathrm{f.matter}}&=\int dx^3 \, \bigl(D_\mu \xi^\dagger
D^\mu \xi+ \im\bar{\Psi} D_\mu\Gamma^\mu\Psi -g\left[\bar{\Psi}\Lambda\xi+
\xi^\dagger{\bar\Lambda}\Psi\right]\bigr),\\
\label{eqn:SCS} S_{CS}&=\int dx^3 \,
\frac{\kappa}{2}\biggl(\epsilon^{\mu\nu\lambda}\bigl(
A_{\mu}\partial_{\nu}A_{\lambda}+ \frac{2 \im}{3}g
A_{\mu}A_{\nu}A_{\lambda} \bigr)+ 2 \bar{\Lambda}\Lambda\biggr).
\end{align}
\end{subequations}
The SYM part of the action describes a system of gauge bosons $A_\mu$ and
their superpartners, the Majorana fermions $\Lambda$. Both fields
are $(N_c\times N_c)$ matrices transforming under the adjoint 
representation of $SU(N_c)$; hereafter, unless indicated otherwise, we treat these fields as matrices, 
and thus we suppress the color indices ($i,j,k$). Additionally, we have 
two complex fields, a scalar $\xi$,
and a Dirac fermion $\Psi$, all transforming
according to the fundamental representation of the gauge group. In matrix notation the
covariant derivatives and the gauge field strength are defined as follows:
\begin{align}
\label{eqn:covdev}
D_\mu\Lambda&=\partial_\mu\Lambda+ig[A_\mu,\Lambda], &
D_\mu\xi&=\partial_\mu\xi+ig A_\mu\xi, &
D_\mu\Psi&=\partial_\mu\Psi+ig A_\mu\Psi, \nonumber\\
\\[-1.0mm]
D_\mu \xi^{\dag}&=\partial_\mu\xi^{\dag}-ig\xi^{\dag}A_\mu, &
D_\mu\Psi^{\dag}&=\partial_\mu\Psi^{\dag}-ig\Psi^{\dag}A_\mu, & 
F_{\mu\nu}&=\partial_{[\mu}A_{\nu]}+\im g [A_{\mu}, A_{\mu}]. \nonumber
\end{align}
The action $(\ref{actionA})$ is invariant under supersymmetry
transformations parameterized by a constant two-component
Majorana spinor $\varepsilon\equiv (\varepsilon_{1},\varepsilon_2)^{\rm T};$  $\bar{\varepsilon}\equiv
\varepsilon^{\mathrm{T}}\Gamma^{0}$:
\begin{align}
\label{eqn:susyvars}
\delta A_\mu&=\frac{\im}{2} {\bar\varepsilon}\Gamma_\mu\Lambda,\quad
\delta\Lambda=\frac{1}{4}F_{\mu\nu}\Gamma^{\mu\nu}\varepsilon,\nonumber\\
\delta\xi&=\frac{\im}{2}{\bar\varepsilon}\Psi, \quad
\delta\xi^{\dag}=-\frac{\im}{2}\bar{\Psi}\varepsilon, \\
\delta\Psi&=-\frac{1}{2}\Gamma^\mu\varepsilon D_\mu\xi, \quad
\delta\bar{\Psi}=-\frac{1}{2}D_\mu\xi^{\dag}\bar{\varepsilon}\Gamma^\mu,
\nonumber
\end{align}
where $\Gamma^{\mu\nu},$ the spinor generator of the Lorentz group, is
written as

\[\Gamma^{\mu\nu}=\frac{1}{2}\left[\Gamma^{\mu},\Gamma^{\nu}\right]=\im \epsilon^{\mu\nu\lambda}\Gamma_{\lambda}\,; \quad (\epsilon^{-+2}=1).\]

Using standard Noether
% or N\"{o}ther, but it seems that the family used Noether
techniques, we construct the spinor supercurrent
corresponding to the above supersymmetric field variations
\begin{align}
\label{eqn:sucurrent}
\bar{\varepsilon}q^\mu&\!=\!\frac{\im}{4}\bar{\varepsilon}\Gamma^{\alpha\beta}
\Gamma^\mu\Tr\left(\Lambda F_{\alpha\beta}\right)+
\frac{\im}{2}D^\mu\xi^\dagger
\bar{\varepsilon}\Psi+\frac{\im}{2}\xi^\dagger\bar{\varepsilon}\Gamma^{\mu\nu}
D_\nu\Psi
%\nonumber\\
%&
%\quad 
-\frac{\im}{2}{\bar\Psi}\varepsilon D^\mu\xi+\frac{\im}{2}D_\nu
\bar{\Psi}\Gamma^{\mu\nu}\varepsilon\xi.
\end{align}
For the remainder of the paper we assume that the fields are independent of
the space-like dimension $x^2$, i.e. $\partial_{2}(...)$=0, thereby dimensionally
reducing the theory to two dimensions. Thus the
$\mathcal{N}=1$ supersymmetry in $2+1$~dimensions is naturally expressed
in terms of $\mathcal{N}=(1,1)$ supersymmetry in $1+1$~dimensions.

We will implement light-cone quantization, which means that initial conditions
as well as canonical
(anti-)~commutation relations will be imposed on the light-like surface
$x^+=const$. In particular, we construct the supercharge by integrating the
supercurrent (\ref{eqn:sucurrent}) over the light-like surface
\begin{align}
\label{eqn:sucharge}
\bar{\varepsilon}Q&=\int dx^{-}\biggl(
\frac{\im}{4}{\bar\varepsilon}\Gamma^{\alpha\beta}
\Gamma^+\Tr\left(\Lambda F_{\alpha\beta}\right)+
\frac{\im}{2}D_-\xi^\dagger\
{\bar\varepsilon}\Psi+\frac{\im}{2}\xi^\dagger{\bar\varepsilon}\Gamma^{+\nu}
D_\nu\Psi \nonumber\\
\\[-4mm]
&\quad\qquad\qquad-\frac{\im}{2}\bar{\varepsilon}\Psi^{\dag}
D^+\xi-\frac{\im}{2}
{\bar\varepsilon}\Gamma^{+\nu}D_\nu\Psi^{\dag} \,\xi\biggr)\nonumber.
\end{align}
Note that  because we have taken the fields to be independent of $x^2$,
the integration over this coordinate resulted in a constant factor, which
rescaled our original fields.

By choosing  the following imaginary (Majorana) representation for
the Dirac matrices in three dimensions:
\begin{equation}
\Gamma^0=\sigma_2,\qquad \Gamma^1=\im\sigma_1,\qquad \Gamma^2=\im\sigma_3,
\end{equation}
the Majorana spinor field $\Lambda$ is manifestly real, i.e.
$\Lambda^{\dag}=\Lambda^{T}$. At this point it is convenient to introduce
the component form for the spinors:
\begin{equation}
\Lambda=\bigl(\lambda,{\tilde\lambda}\bigr)^T,\qquad
\Psi=\bigl(\psi,{\tilde\psi}\bigr)^T,\qquad Q=\bigl(Q^+,Q^-\bigr)^T.
\end{equation}
In terms of this decomposition, the superalgebra is realized explicitly in
its $\mathcal{N}=(1,1)$ form, namely
\begin{equation}
\label{eqn:sualg}
\{Q^+,Q^+\}=2\sqrt{2}P^+,\qquad \{Q^-,Q^-\}=2\sqrt{2}P^-,\qquad
\{Q^+,Q^-\}=0,
\end{equation}
where $Q^{+}$ $(Q^{-})$ are left (right) Majorana--Weyl spinors, each
characterizing the smallest spinor representation in $1+1$ dimensions.

To readily eliminate the nondynamical fields, we
impose the light-cone gauge ($A^+=A_{-}=0$). In this case the supercharges 
can be read off $(\ref{eqn:sucharge})$ and  are given by
\begin{align}
\label{eqn:Qplus}
Q^+&=\frac{1}{2}\int dx^-\biggl(\Tr(\lambda\partial_-A^2)+
\frac{\im}{2}\partial_-\xi^\dagger\psi-\frac{\im}{2}
\psi^\dagger\partial_-\xi-
\frac{\im}{2}\xi^\dagger\partial_-\psi+\frac{\im}{2}
\partial_-\psi^\dagger\xi
\biggr),\\
\label{eqn:Qminus} 
Q^-&=\frac{1}{\sqrt{2}}\int dx^-\biggl(\Tr(\lambda\partial_-A^{-}) 
-\im \xi^\dagger D_2\psi +\im D_2\psi^\dagger\xi-\frac{\im}{\sqrt{2}}
\partial_-(\tilde{\psi}^\dagger\xi-\xi^\dagger\tilde{\psi})\biggr).
\end{align}
Notice that the  right-movers ($\tilde\psi$) appear
in the supercharge $Q^{-}$ only in the total derivative term. This is a
consequence of the light-cone formulation, which singles out the
non-dynamical fermion degrees of freedom, leaving in the expression only
the physical spinor fields ($\lambda$ and $\psi$). 
Among the equations of motion that follow from the action
(\ref{actionA}), in the light-cone gauge, three
serve as constraints rather than as dynamical equations. Namely, for
$\tilde{\lambda}$ and  $\tilde{\psi}$, respectively, we have
\begin{align}
\label{eqn:fermionconstraint}
\partial_-\tilde{\lambda}&=-\frac{\im g}{\sqrt{2}}
\bigl([A^2,\lambda]+\im \xi\psi^\dagger -\im \psi\xi^\dagger-\im \kappa
\lambda\bigr),\\
\partial_-\tilde{\psi}&=-\frac{\im g}{\sqrt{2}}A^2\psi+
\frac{g}{\sqrt{2}}\lambda\xi.
\end{align}
While for $A^{-}$ we obtain
\begin{equation}
\label{eqn:gaugeconstraint}
\partial^2_-A^-=gJ,
\end{equation}
with
\begin{equation}
\label{J}
 J=-\im [A^2,\partial_-A^2]+
\frac{1}{\sqrt{2}}\{\lambda,\lambda\}-\im (\partial_-\xi)\xi^\dagger+ \im
\xi(\partial_-\xi^\dagger)+\sqrt{2}\psi\psi^\dagger+ \frac{\kappa}{g}
\partial_{-}A^2.
\end{equation}
Note that the field  $A^-$ has to be eliminated from the
supercharges, in favor of the physical degrees of freedom. This can be
done by inverting $(\ref{eqn:gaugeconstraint}).$
%Apart from the zero mode problem~\cite{zm5}

The only contribution from the
Chern--Simons term enters into the supercharges via equation $(\ref{J})$,
because $\delta \mathcal{L}_{CS}\propto
\partial_{\mu}(\ldots)$ under the supersymmetry transformations
$(\ref{eqn:susyvars});$ the surface term is not shown here. 
The inclusion of a Chern--Simons term in our theory is important, since it
effectively generates mass for the adjoint partons proportional to the
coupling $\kappa.$

\subsection{Bound-state eigenvalue problem}
\label{subSectSpectCon}
%%%%%%%%%%%%%%%%%%%%%%%%%%%%%%%%%%%%%%%%%%%%%%%%%%%%%%%%%%%%%%%%%%

The bound-state spectrum is obtained by solving the
following mass eigenvalue equation:
\begin{align}
\label{eqn:EVP}
2P^+P^-\ket{\varphi}&=\sqrt{2}P^+(Q^-)^2\ket{\varphi}\non\\
&=\sqrt{2}P^{+}\bigl( g( Q^{-}_{\rm SYM}+Q^{-}_{\rm f.matter})+
\im
\kappa Q^{-}_{\rm CS}\bigr)^{2} \ket{\varphi}\equiv M^2\ket{\varphi},
\end{align}
where the various pieces of $Q^-$, after dropping the surface terms and 
eliminating the non-dynamical fields using the 
constraint~$(\ref{eqn:gaugeconstraint})$, may be expressed  as follows: 
\begin{subequations}
\begin{align}
\label{eqn:Qsym}
Q^{-}_{\rm SYM}&=\frac{\im g}{\sqrt{2}} \int dx^- \,\,\left([A^2,
\partial_{-}A^2]+\frac{\im}{\sqrt{2}}
\{\lambda,\lambda\}
\right)\frac{1}{\partial_{-}}\lambda,\\
\label{eqn:Qfmatter} 
Q^{-}_{\rm f.matter}&=\frac{g}{\sqrt{2}}
\int dx^-\biggl(\biggl(\im (\partial_{-}\xi)\xi^\dagger 
                             -\im   \xi(\partial_{-}\xi^\dagger) 
-\sqrt{2}\psi \psi^{\dagger}\biggr)\frac{1}{\partial_{-}}\lambda \non\\ 
&\hspace*{28mm} + \xi^\dagger A^2\psi +\psi^\dagger A^2\xi
%-\frac{\im}{g\sqrt{2}}\partial_{-}({\tilde\psi}^\dagger\xi
%-\xi^\dagger{\tilde\psi})
\biggr),\\
\label{eqn:Qcs}
Q^{-}_{\rm CS}&=\frac{\im \kappa}{\sqrt{2}} 
    \int dx^- (\partial_{-}A^2) \frac{1}{\partial_{-}}\lambda,
\end{align}
\end{subequations}
where a trace over color space is understood.

The strategy for solving equation $(\ref{eqn:EVP})$ is to cast it as a
matrix eigenvalue problem. This is achieved by employing a discrete basis
where the longitudinal light-cone momentum $P^{+}$ is diagonal. The
discrete basis is introduced by first discretizing the 
supercharge\footnote{Note the relative phase between $Q^{-}_{SYM}$ 
and $Q^{-}_{CS}$. $Q^{-}_{SYM}$ is defined as Hermitian and $Q^{-}_{CS}$
is defined to be anti-Hermitian such that $Q^{-}$ remains Hermitian.} $Q^-$
and then constructing $P^-$ from the square of the supercharge: $P^-
=(Q^-)^2/\sqrt{2}$. The two dimensional theory is compactified on a
light-like circle ($-L<x^-<L$), and periodic boundary conditions are
imposed on all dynamical degrees of freedom. This leads to the following
field mode expansions:
\begin{align}
\label{expandA}
A^2_{ij}(0,x^-)&=\frac{1}{\sqrt{4\pi}}\sum_{n=1}^{\infty}\frac{1}{\sqrt{n}}
\left(a_{ij}(n)e^{-\im n\pi x^-/L}+a^\dagger_{ji}(n)e^{\im n\pi x^-/L}\right),\\
\label{expandLambda}
\lambda_{ij}(0,x^-)&=\frac{1}{2^{\frac{1}{4}}\sqrt{2L}}\sum_{n=1}^{\infty}
\left(b_{ij}(n)e^{-\im n\pi x^-/L}+b^\dagger_{ji}(n)e^{\im n\pi x^-/L}\right),\\
\label{expandXi}
\xi_i(0,x^-)&=\frac{1}{\sqrt{4\pi}}\sum_{n=1}^{\infty}\frac{1}{\sqrt{n}}
\left(c_i(n)e^{-\im n\pi x^-/L}+{\tilde c}^\dagger_{i}(n)e^{\im n\pi x^-/L}\right),\\
\label{expandPsi}
\psi_{i}(0,x^-)&=\frac{1}{2^{\frac{1}{4}}\sqrt{2L}}\sum_{n=1}^{\infty}
\left(d_{i}(n)e^{-\im n\pi x^-/L}+{\tilde d}^\dagger_{i}(n)e^{\im n\pi
x^-/L}\right).
\end{align}
In the above expressions\footnote{The inclusion of zero
modes is beyond the scope of the present paper.}
we introduced the discrete longitudinal
momenta $k^+\equiv k$ as fractions $nP^+/K=n\pi/L; \, (n=1,2,3,\ldots)$ of the total longitudinal
momentum $P^+$, where $K$ is the integer that determines the resolution of
the discretization.  The color indices were made explicit as well. Because 
light-cone longitudinal momenta are always
positive, $K$ and each $n$ are positive integers. The number of
constituents is thus bounded by $K$. The
continuum limit is reached by letting $K \rightarrow\infty$.

The time direction in the light-cone formalism  is taken to be the $x^+$
direction.  Thus the (anti-)commutation relations between fields and their
conjugate momenta are assumed on the surface $x^+=0$. Quantization
is achieved by imposing the following relations:
\begin{align}
\label{CanComRelField}
\left[A_{ij}^2(0,x^-),\partial_-A_{kl}^2(0,y^-)\right]&=
\im\left(\delta_{il}\delta_{jk}-\frac{1}{N_c}\delta_{ij}\delta_{kl}\right)
\delta(x^--y^-),\\
\left\{\lambda_{ij}(0,x^-),\lambda_{kl}(0,y^-)\right\}&=\sqrt{2}
\left(\delta_{il}\delta_{jk}-\frac{1}{N_c}\delta_{ij}\delta_{kl}\right)
\delta(x^--y^-),\\
\left[\xi_i(0,x^-),\partial_-\xi_j(0,y^-)\right]&=
\im\delta_{ij}\delta(x^--y^-),\\
\label{CanAntFerm}
\left\{\psi_{i}(0,x^-),\psi_{j}(0,y^-)\right\}&=\sqrt{2}
\delta_{ij}\delta(x^--y^-).
\end{align}
The above (anti-) commutators  can also be expressed, with the help of
equations $(\ref{expandA})-(\ref{expandPsi})$, in terms of
creation-annihilation operators
\begin{equation}
\label{commA}
\left[a_{ij}(k),a^\dagger_{kl}(k')\right]=
\left\{b_{ij}(k),b^\dagger_{kl}(k')\right\}=
\left(\delta_{ik}\delta_{jl}-\frac{1}{N_c}\delta_{ij}\delta_{kl}\right)
\delta(k-k')
\end{equation}
\begin{equation}
\label{commB}
\left[c_{i}(k),c^\dagger_{j}(k')\right]\!=\!\left[{\tilde
c}_{i}(k),{\tilde c}^\dagger_{j}(k')\right]\!=\!
\left\{d_{i}(k),{d}^\dagger_{j}(k')\right\}\!=\! \left\{{\tilde
d}_{i}(k),{\tilde d}^\dagger_{j}(k')\right\}\!=\!\delta_{ij}\delta(k-k').
\end{equation}

The expansion of the supercharge $Q^-$  in terms of creation and annihilation 
operators is a straightforward exercise. 
For instance, the decomposition of  $Q^{-}_{\rm CS}$ in terms of Fourier modes gives 
the following expression:
\begin{equation} 
\label{qcs}
Q^-_{CS}=\biggl(-\frac{\im \kappa\sqrt{L}}{2^{5/4}\sqrt{\pi}}\biggr)
\sum_{n=1}^{\infty}\frac{1}{\sqrt{n}}
\biggl(a_{ij}^{\dagger}(n)b_{ij}(n)+b_{ij}^{\dagger}(n)a_{ij}(n)\biggr)\,.
\end{equation}
Similarly, for the supercharge, $Q^{-}_{\rm f.matter}$, that controls the 
behavior of the fundamental matter fields, we obtain
\begin{align}
\label{Qfmattexp}
Q^{-}_{\rm f.matter}&=\biggl(-\frac{\im g\sqrt{L}}{2^{5/4}\pi}\biggr)
\sum_{n_1,n_2,n_3=1}^{\infty}\Biggl\{
\frac{(n_2+n_3)}{2n_1\sqrt{n_2 n_3}}
\bigg(\tilde{c}^\dagger_i(n_3)\tilde{c}_{j}(n_2)b_{ji}(n_1)\non\\
&-\tilde{c}^\dagger_i(n_2)b_{ij}^\dagger(n_1)\tilde{c}_{j}(n_3)
+b_{ji}^\dagger(n_1) c^\dagger_i(n_2)c_{j}(n_3)
-c^\dagger_i(n_3)b_{ij}(n_1)c_{j}(n_2) \bigg)\non\\
&+\frac{1}{\sqrt{2} n_1}\bigg( \tilde{d}^\dagger_i(n_2)b^\dagger_{ij}(n_1)\tilde{d}_j(n_3)+ \tilde{
d}^\dagger_i(n_3)\tilde{d}_j(n_2)b_{ji}(n_1)
+b^\dagger_{ij}(n_1)d^\dagger_j(n_2) d_i(n_3)
\non\\
&+
d^\dagger_i(n_3) b_{ij}(n_1) d_j(n_2) \bigg)+\frac{\im}{2\sqrt{n_2 n_3}}  
\bigg(c^\dagger_{i}(n_3)a_{ij}(n_2)d_{j}(n_1)\non\\
&+
a^{\dag}_{ij}(n_2)d^{\dag}_{j}(n_1)c_{i}(n_3)
+\tilde{d}^{\dag}_{j}(n_1)a^{\dag}_{ji}(n_2)\tilde{c}_{i}(n_3)
+\tilde{c}^{\dag}_{i}(n_3)\tilde{d}_{j}(n_1)a_{ji}(n_2)\bigg)\non\\
&+\frac{\im}{2\sqrt{n_1 n_2}} \bigg(
a^\dag_{ji}(n_2)c^{\dag}_{i}(n_1)d_j(n_3)+
d^\dagger_{j}(n_3)a_{ji}(n_2)c_{i}(n_1)\non\\
&+\tilde{d}^{\dag}_{j}(n_3)\tilde{c}_{i}(n_1)a_{ij}(n_2)
+\tilde{c}^{\dag}_{i}(n_1)a^{\dag}_{ij}(n_2)\tilde{d}_{j}(n_3)\bigg)
\Biggr\}\,\delta_{n_3,n_1+n_2}.
\end{align}
Our computer code carries out these expansions automatically.

Apart from supersymmetry, the theory we set out to explore possesses
another symmetry,\footnote{Note that when $\kappa=0,$ parity ($\mathcal{P}$) is also conserved.} 
which may be used to reduce the size of the Hamiltonian matrix we need to
produce and diagonalize. Namely, we have a $\mathbf{Z}_{2}$ symmetry $\cal T$
that is associated with the orientation of the large-$N_c$ string of
partons in a state~\cite{Kutasov:1993gq,Gross:1997mx}. It gives a sign when the gauge group
indices are permuted
\begin{equation}
\label{Z2-S}
a_{ij}(k)\overset{\cal T}{\rightarrow} -a_{ji}(k) , \qquad
      b_{ij}(k)\overset{\cal T}{\rightarrow} -b_{ji}(k) .
\end{equation}

In this paper we will discuss numerical results obtained in the
large-$N_c$ limit, i.e. terms of order $1/N_c$ in the above
expressions are dropped. Note that corrections on the order of $1/N_c$
are expected to lead to interesting effects~\cite{Antonuccio:1998uz};
however, they are beyond the scope of this work.
%
%%%%%%%%%%%%%%%%%%%%%%%%%%%%%%%%%%%%%%%%%%%%%%%%%%%%%%%%%%%%%%%%%%
\section{Meson and glueball spectra}
\label{SDLCQspect}
%%%%%%%%%%%%%%%%%%%%%%%%%%%%%%%%%%%%%%%%%%%%%%%%%%%%%%%%%%%%%%%%%%
\subsection{Limiting cases}
\label{limitingcas}
%%%%%%%%%%%%%%%%%%%%%%%%%%%%%%%%%%%%%%%%%%%%%%%%%%%%%%%%%%%%%%%%%%

We first investigate the strong-coupling ($g\gg\kappa$) and weak-coupling ($g\rightarrow0$)
limits of the theory.
In the strong-coupling limit, we previously found~\cite{Hiller:2002cu}
that there are approximate BPS glueball bound states with masses (squared) 
\[M^2_{aBPS}=n^2 \kappa^2, \quad n=2,3,\ldots, (K-1).\]
In the meson sector under investigation in the present paper, nearly all 
the masses grow with $g$. However, we see evidence for a 
state that remains near zero mass as $g \to \infty.$

%%%%%%%%%%%%%%%%%%%%%%%%%%%%%%%%%%%%%%%%%%%%%%%%%%%%%%%%%%%%%%%%%%%%%%%%%%%%
%\subsubsection*{Free mesonic spectrum}
%%%%%%%%%%%%%%%%%%%%%%%%%%%%%%%%%%%%%%%%%%%%%%%%%%%%%%%%%%%%%%%%%%%%%%%%%%%%

\begin{figure}
  \begin{center}
     \hspace*{-0.5cm}
     \begin{tabular}{cc}
   % \hline\hline\\[3mm]
    \resizebox{75mm}{!}{\includegraphics{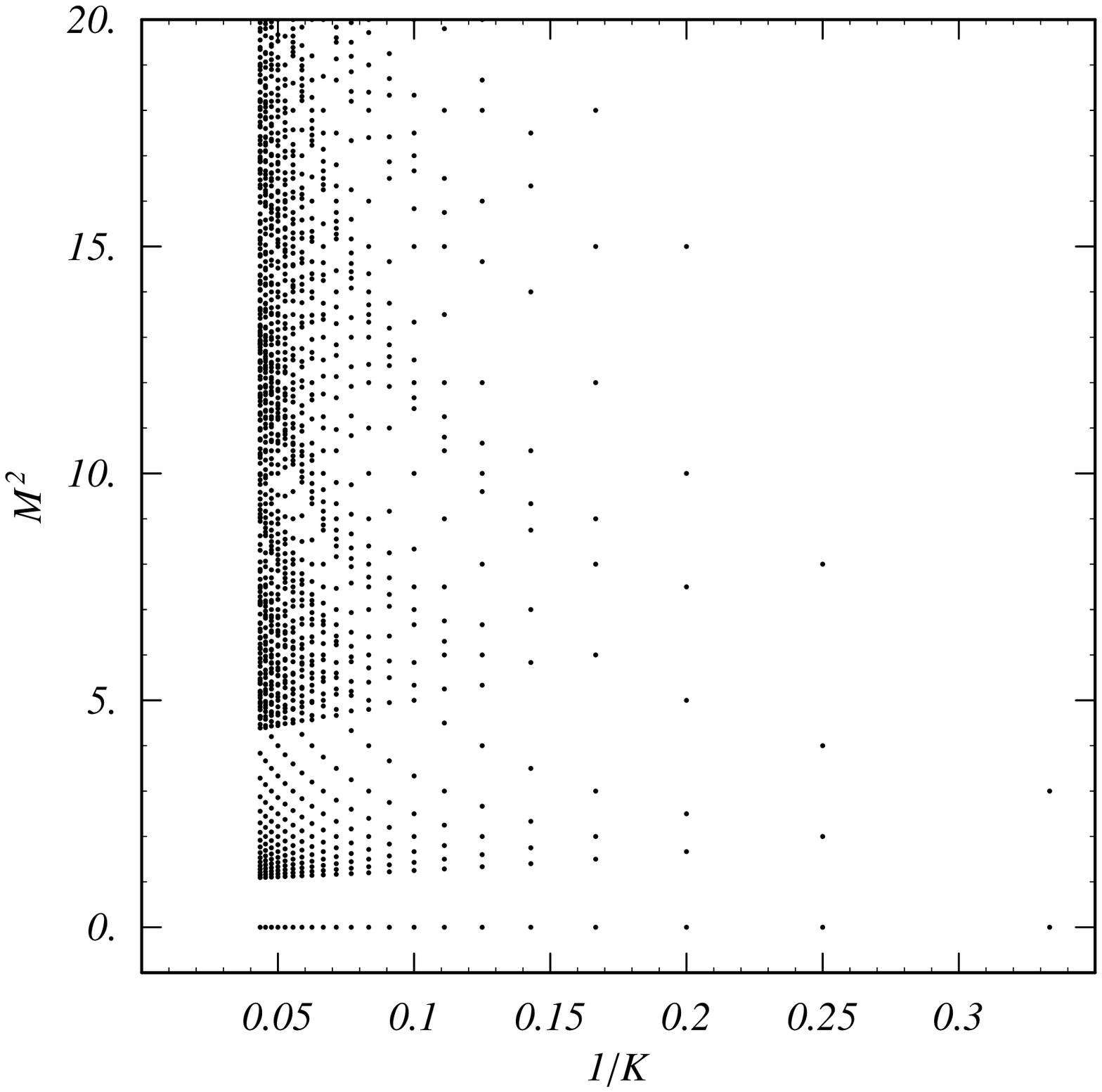}} &
\resizebox{75mm}{!}{\includegraphics{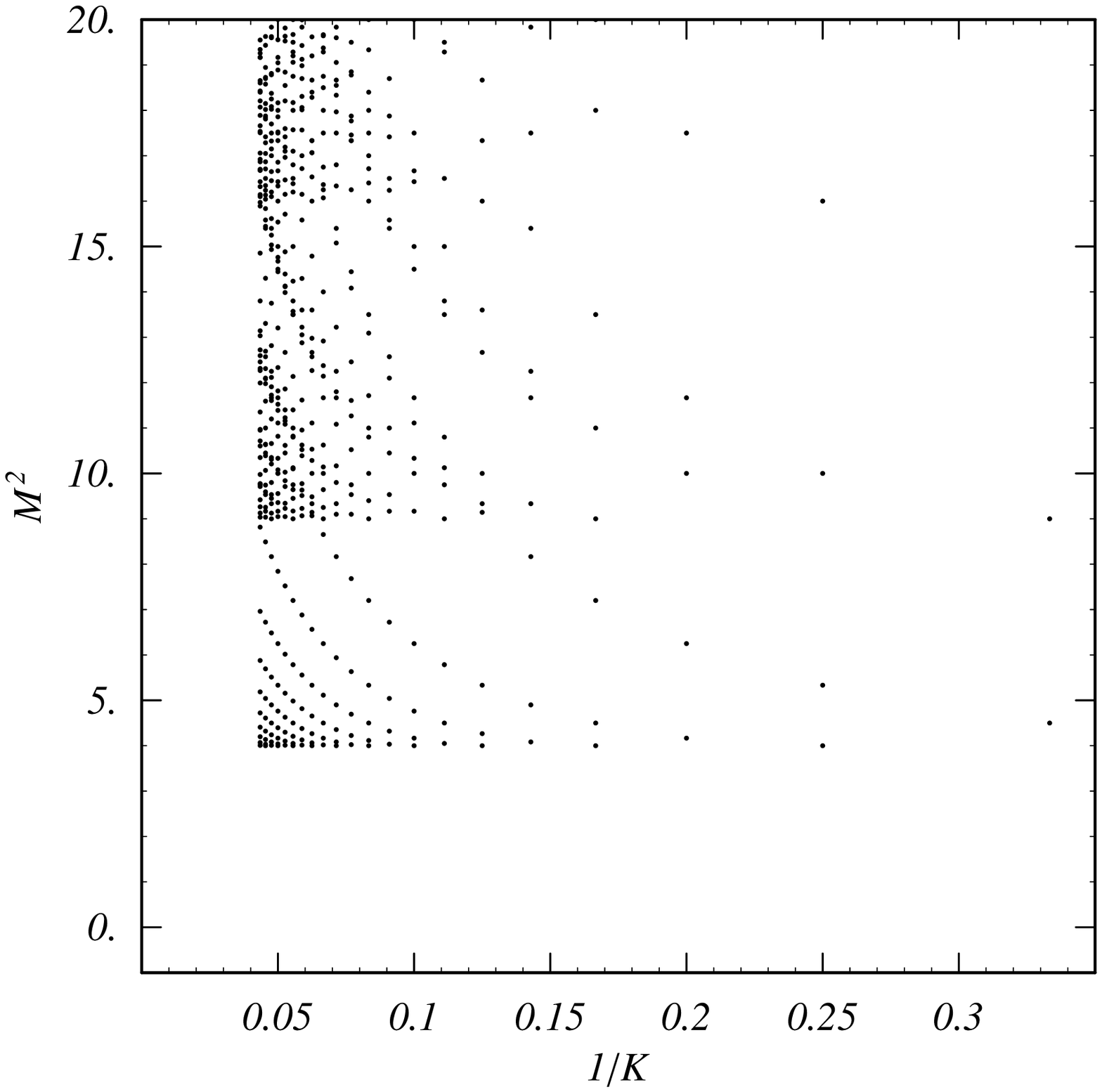}} \\
     \qquad (a)& \qquad (b)\\[3mm]
    %  \hline\hline
    \end{tabular}
\vspace{-6mm}
    \caption{Mass spectra for (a) mesonic bound states and (b) glueball bound states
     as a function of the inverse resolution for  $3\leq K\leq 23$ when
$\kappa=1$ and $g=0$  (free theory). \label{CS1g0}}
\end{center}
\end{figure}

The free theory can be  solved analytically, and the results are shown
in Fig.~\ref{CS1g0}.
Its free-meson spectrum, Fig.~\ref{CS1g0}(a), has $K-1$
massless states in each symmetry sector. Each such state is made out
of two fundamental partons. All states in the corresponding
glueball sector, Fig.~\ref{CS1g0}(b), are massive.
The mass scale for all states is set by the CS coupling $\kappa$.
To obtain the spectrum,
we consider sets of free partons that form mesonic color-singlet
multi-parton combinations. There are many other combinations of such free
partons that belong to
the non-singlet sector of the Hilbert space, which we can omit in the
large-$N_c$ limit; for $g\neq 0$ the
only viable states are color-singlets. In other words, the free
color-singlet combinations are expected to become bound states as soon
as the coupling is turned on.
Mesonic multi-parton color-singlet states are of the form
\[
f^{\dag}_{i}(m_1)\tilde{f}^{\dag}_{i}(m_2)|0\rangle,\,
f^{\dag}_{i}(m_1)a^{\dag}_{ij}(n_1)\tilde{f}^{\dag}_{j}(m_2)|0\rangle, \,
f^{\dag}_{i}(m_1)a^{\dag}_{ik}(n_1)a^{\dag}_{kl}(n_2)\tilde{f}^{\dag}_{l}(m_2)|0\rangle, \ldots,
\]
where the operators $f^{\dag}, \tilde{f}^{\dag}$ create 
fundamental partons,
while $a^{\dag}$ create adjoint partons. In our theory
we have four types of fundamental partons and
two types of adjoint partons. In the large-$N_c$ limit
we can have the following four types of mesonic multi-parton color-singlet
states:
\[c^{\dag}_i(\mathrm{adj.})_{ij}\tilde{c}_{j}^{\dag}|0\rangle,
c^{\dag}_i(\mathrm{adj.})_{ij}\tilde{d}_{j}^{\dag}|0\rangle,
d^{\dag}_i(\mathrm{adj.})_{ij}\tilde{d}_{j}^{\dag}|0\rangle,
d^{\dag}_i(\mathrm{adj.})_{ij}\tilde{c}_{j}^{\dag}|0\rangle,\]
where $(\mathrm{adj})_{ij}$ can be any string of adjoint partons.
Only the adjoint partons contribute to the mass of a state 
and do so proportional
to the CS coupling $\kappa$. Note also that the first
pair of fundamental partons above forms a massless combination; there are
$4(K-1)$ of these at each value of the resolution.

To construct the spectrum and find the degeneracies of each mass state,
we utilize combinatorics and the DLCQ multi-particle formula
\begin{align}
\label{eqn:freepartons}
M^{2}_{j_{max}}(K)=\kappa^{2}K \sum_{l=1}^{j_{max}} \frac{1}{n_{l}}.
\end{align}
Here $1\leq j_{max}\leq (K-2)$ is the number of adjoint partons.
Specifically, we calculate
the  compositions  $C^{K}_{j}=\binom{K-1}{j-1}$ (i.e. ordered partitions)
of the integer $K$ into $j=(2+j_{max})$ parts, where the factor two counts
the two massless fundamental partons in a particular state. For
example, at $K=5$ the compositions that give massive multi-particle
states are:
$\big\{ C^5_3, C^5_4, C^5_5  \big\}.$ The first
composition in the list, $ C^5_3,$  yields (modulo parton type) three  
states with $M^2_{1}=5\kappa^2$, two  states with
$M^2_{1}=\frac{5}{2}\kappa^2$, and one  state with
$M^2_{1}=\frac{5}{3}\kappa^2$. Thus, taking into account 
the several types of partons that can form $C^5_3$, the total 
number of states we get for this
case is $2^3(3+2+1)=48.$  The
total number of states -- including the massless ones -- as a function of $K$
is thus
\begin{align}
\label{eqn:totfreemeson}
N(K)&=\sum_{k=2}^{K} 2^k C^K_k.
\end{align}
The dimension $N(K)$ of the Hilbert space of states grows exponentially with
$K$, e.g. $N(16)=28,697,812$.

As an example, consider  a four parton state, consisting of two
fundamental and two adjoint partons, with mass
\[
M^2_{4}=\frac{n+m}{nm}K, \qquad n,m = 1,2,3,\ldots.
\]
The two adjoint partons have mass unity
$(\kappa^2=1)$, while the two fundamentals are massless.
Thus at $K=7$, we  have four-parton states starting at $M^2=\frac{35}{6}$.

%%%%%%%%%%%%%%%%%%%%%%%%%%%%%%%%%%%%%%%%%%%%%%%%%%%%%%%%%%%%%%%%%%%%%%%%
%\subsubsection*{Free glueball spectrum}
%%%%%%%%%%%%%%%%%%%%%%%%%%%%%%%%%%%%%%%%%%%%%%%%%%%%%%%%%%%%%%%%%%%%%%%%
The glueball spectrum is evaluated in a similar fashion.
However, the free glueball multiparticle color-singlet states form
closed loops made out of adjoint fermion $(b^{\dag}_{ij})$ and
boson $(a^{\dag}_{ij})$ partons, with a mass
easily obtained by $(\ref{eqn:freepartons}).$  Obviously,
there are no massless states in the glueball sector. The cyclic symmetry
of the color trace reduces the total number of states that are available,
so the free mesons will dominate the free energy.
Due to supersymmetry it suffices to count only the fermions.
Using the combinatorics above, we arrive at the number of fermionic states
with $j$ partons~\cite{Hiller:2004rb},
\begin{align}
\label{eqn:Kintoj}
N_{f}(K; j)=\sum_{q=0}^{\infty} \frac{(2q+1)}{j} \,\tilde{C}_{f}\bigg(\frac{K}{2q+1};\, 
               \frac{j}{2q+1}\bigg).
\end{align}
The function $\tilde{C}_{f}$ is defined recursively as
\[\tilde{C}_{f}(K;\, j)=
2^{j-1} C^{K}_{j} - \sum_{q=1}^{\infty} \tilde{C}_{f}\bigg(\frac{K}{2q+1};\, 
              \frac{j}{2q+1}\bigg).\]
Note that $\tilde{C}_{f}\big(\frac{K}{2q+1};\, \frac{j}{2q+1}\big)$ is zero if none of
its arguments is an integer. The total number of states at a specific $K$ is
found by summing over the number of partons $j.$ For example, at
$K=5$ and $j=3,$ we have a total of 16 states. Eight of these states
have mass squared $M^2=10 \kappa^2$ and the other eight
have $M^2=\frac{35}{3}\kappa^2.$
\begin{figure}
 \begin{center}
 \hspace*{-0.5cm}
    \begin{tabular}{cc}
 % \dblrule\\[5mm]
  \resizebox{75mm}{!}{\includegraphics{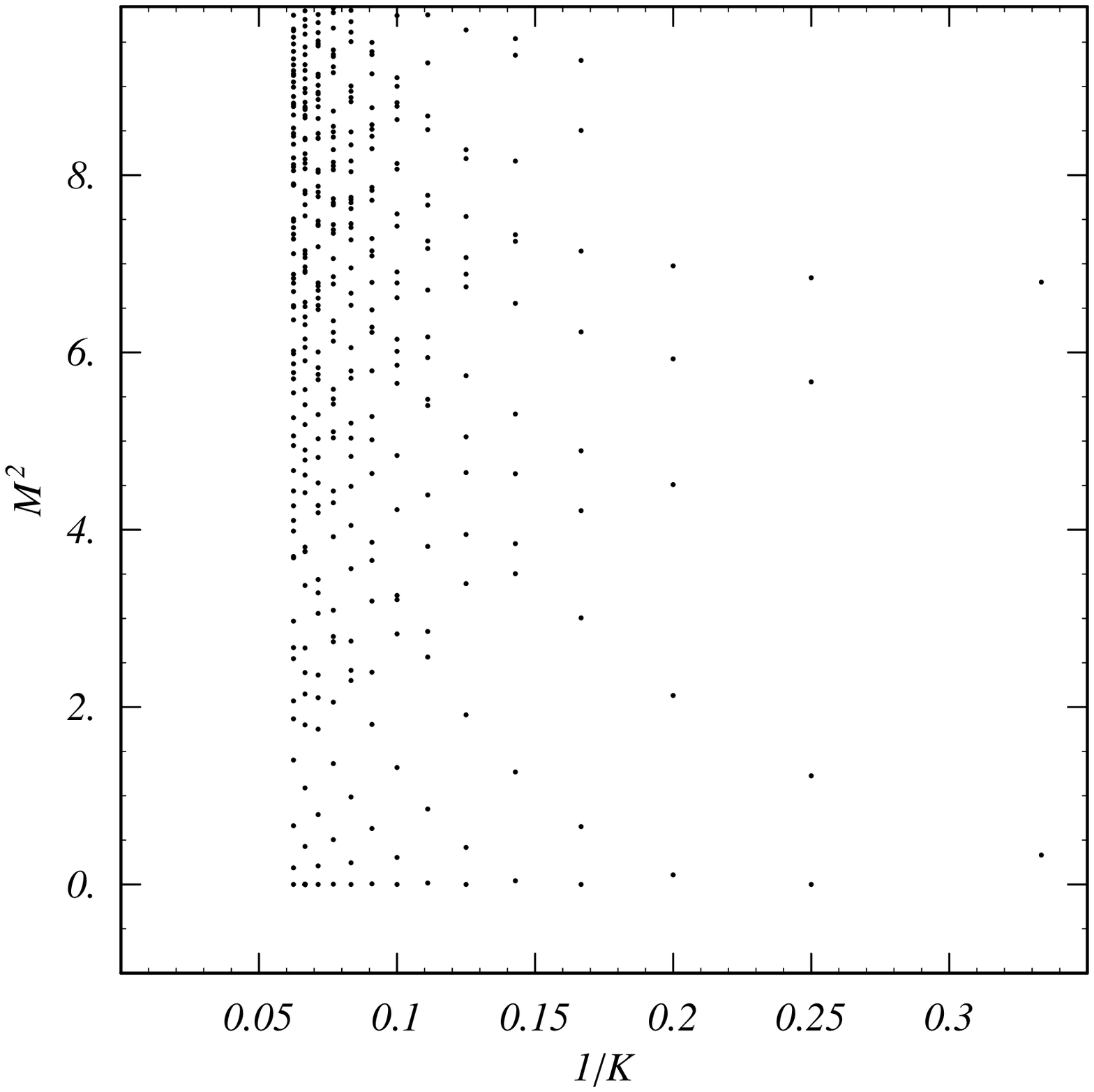}} &
     \resizebox{76.5mm}{!}{\includegraphics{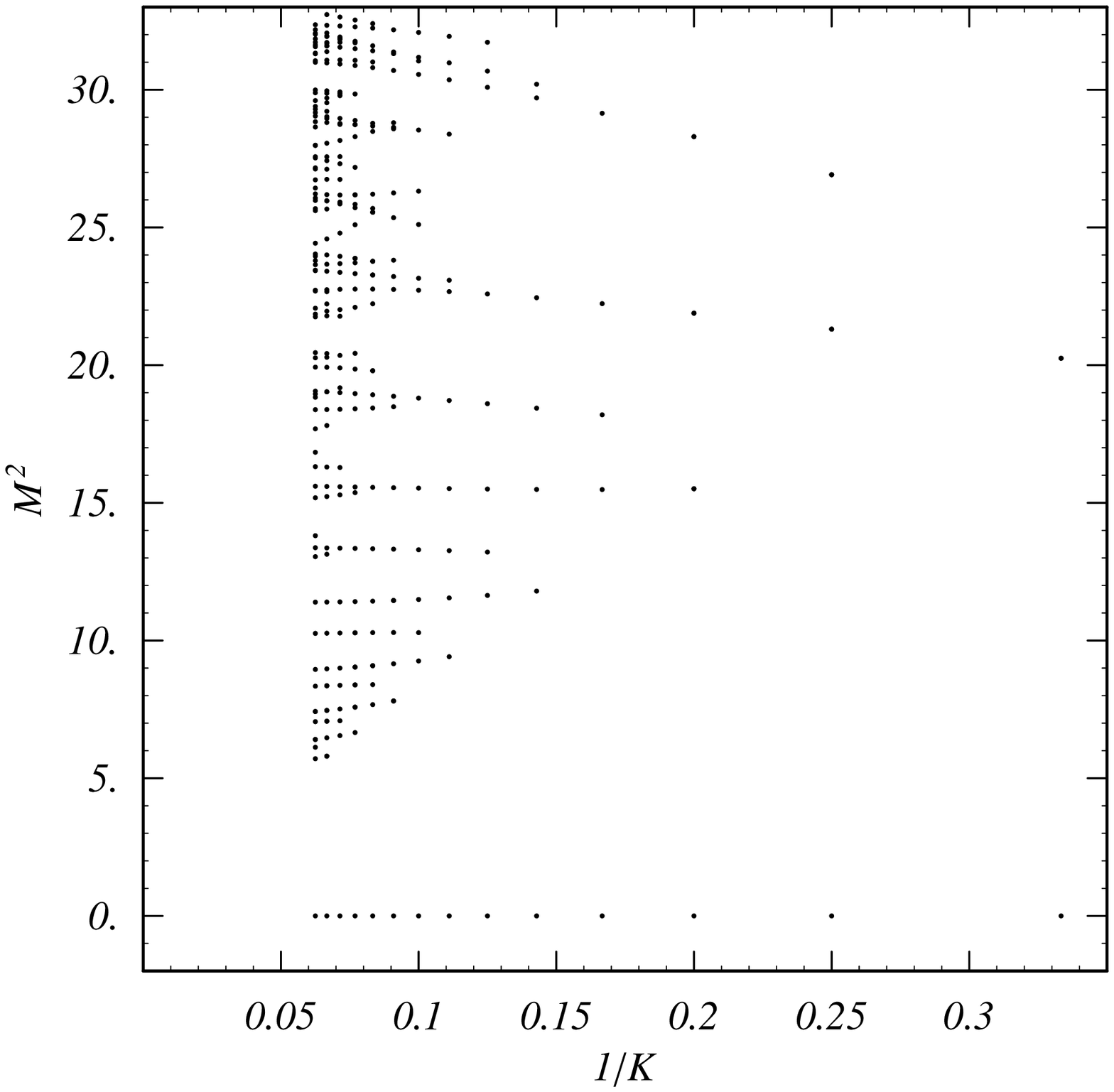}}
     \\
     \qquad (a)& \qquad (b)\\[3mm]
    %  \hline\hline
    \end{tabular}
\vspace{-6mm}
    \caption{Mass spectra for (a) the meson ($\mathcal{P}$-even, $\mathcal{T}$-even) sector and (b) the glueball ($\mathcal{P}$-even, $\mathcal{T}$-even) sector
    as a function of the inverse resolution for $3\leq K\leq 16$ when $\kappa=0$ and $g=1$.
\label{CS0g1}}
  \end{center}
\end{figure}
\begin{figure}
 \begin{center}
    \hspace*{-0.5cm}
     \begin{tabular}{cc}
 % \dblrule\\[5mm]
  \resizebox{75mm}{!}{\includegraphics{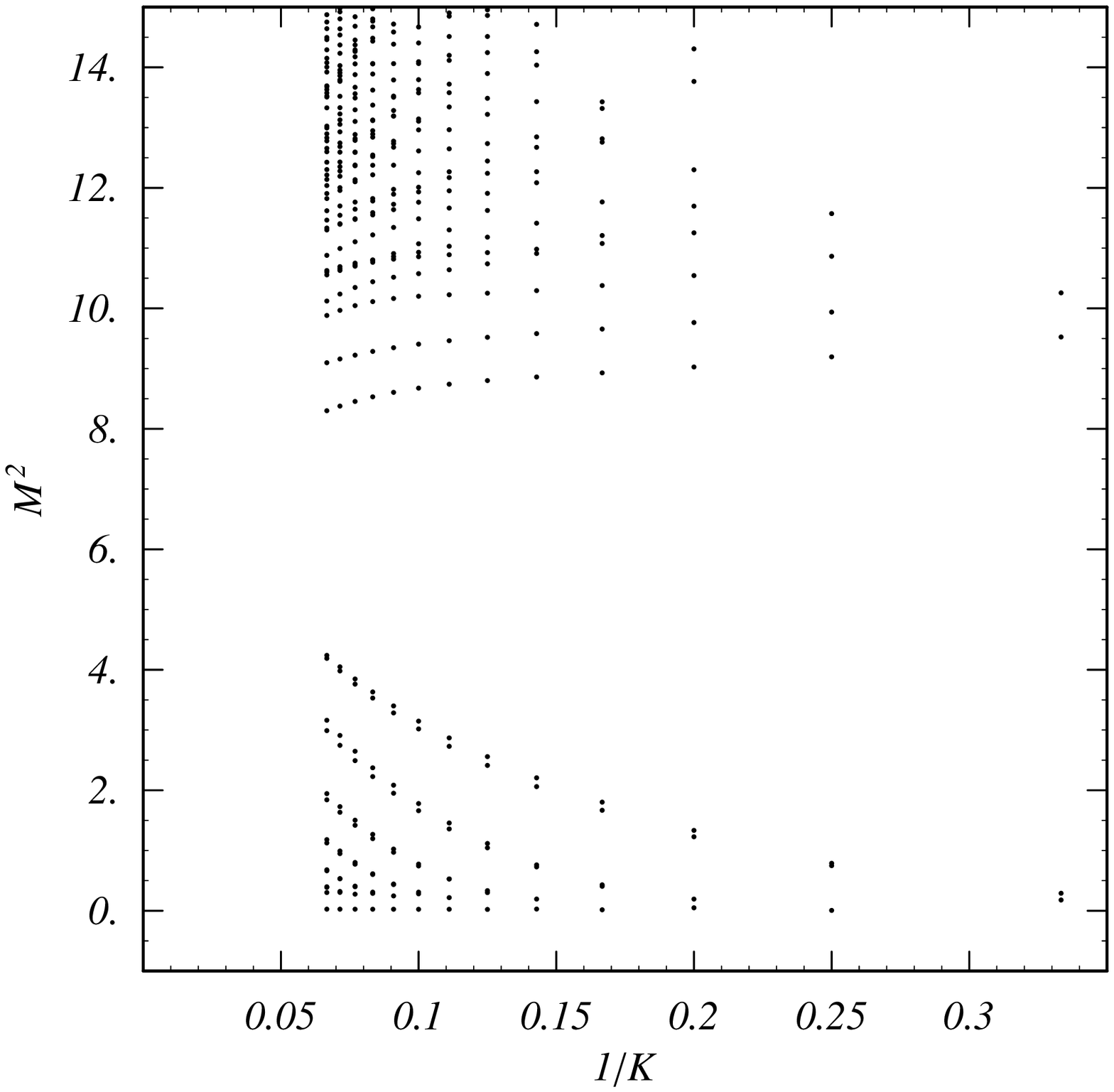}}&
     \resizebox{75mm}{!}{\includegraphics{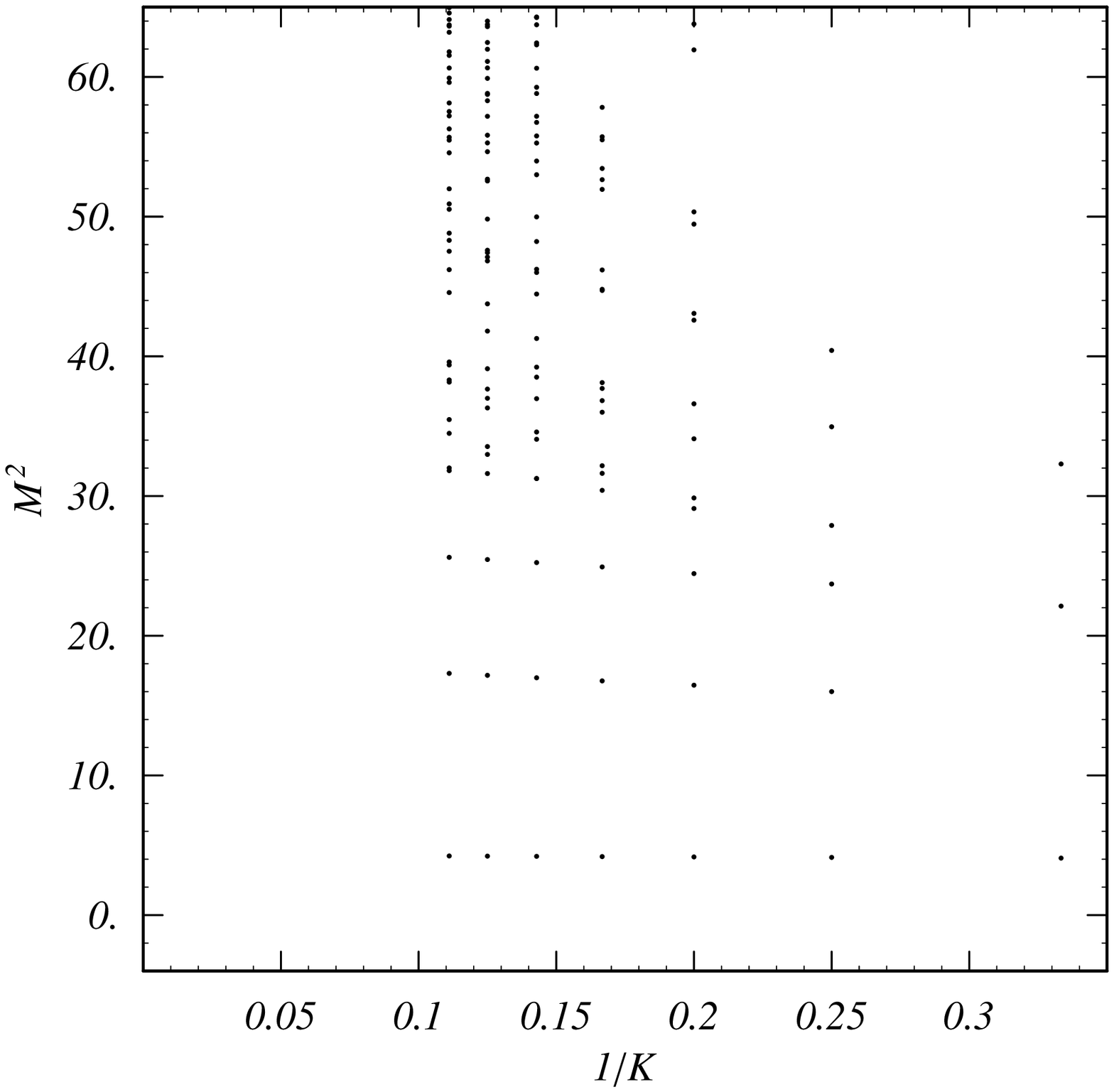}}
      \\
     \qquad (a)& \qquad (b)\\[3mm]
    %  \hline\hline
    \end{tabular}
\vspace{-6mm}
    \caption{Mass spectra for (a) the meson($\mathcal{T}$-even) sector and (b) the glueball ($\mathcal{T}$-even) sector
    as a function of the inverse resolution for $3\leq K\leq 16$ when
$\kappa=1$ and $g=1$. We note also a small splitting in the masses due to the presence of CS term which breaks explicitly the $\mathcal{P}$ symmetry.}
    \label{CS1g1}
  \end{center}
\end{figure}
%
%%%%%%%%%%%%%%%%%%%%%%%%%%%%%%%%%%%%%%%%%%%%%%%%%%%%%%%%%%%%%%%%%%%%%
\subsection{Comparison of meson and glueball spectra}
\label{gluesect}
%%%%%%%%%%%%%%%%%%%%%%%%%%%%%%%%%%%%%%%%%%%%%%%%%%%%%%%%%%%%%%%%%%%%%

The generic meson and glueball spectra for nonzero $g$ are shown in Figs.~\ref{CS0g1}
and \ref{CS1g1}.
In the glueball sector with finite coupling but vanishing CS coupling, there
are
$2(K-1)$ massless BPS states. The number of partons in these states grows
with the
 resolution $K,$ and there is a mass gap between these massless
states
and the lowest massive states that decreases with increasing resolution.
When the
CS coupling is not vanishing $(g\neq 0, \kappa\neq 0)$, the BPS massless
glueball
bound states become approximate BPS states~\cite{Hiller:2002cu}, with
bound-state masses nearly
independent of the gauge coupling. The masses of the remaining states in
this sector
grow rapidly with the coupling.  

In earlier work~\cite{Hiller:2004ft} we studied
the thermodynamics of this sector with vanishing CS coupling.
Here, we are considering the mesonic sector of this SYM theory. From a previous
work~\cite{Hiller:2003jd,Hiller:2003qe} we know that for non-zero coupling
there
is a mass gap in the low-mass sector. The low-mass sector consists of
the
states that become massless bound states of two fundamental partons in the
limit
that the coupling goes to zero. This mass gap decreases as the resolution
increases.
 Of the $K-1$ massless states in each symmetry sector at vanishing coupling
only
one remains at finite coupling.

In the large-$N_c$ limit, the mesonic and glueball sectors decouple.
The
thermodynamics of the theory is generated by the partition function which
is the
product of the partition functions of the two sectors, and the free energy
is the
sum of the two free energies. The glueball bound states are closed
loops in
color space; their cyclic symmetry greatly reduces the number
of basis
states. Therefore, the number of glueball states relative to the meson
states at a particular $K$ is very small. One would thus
expect the mesonic
bound states to dominate the thermodynamics.

There are several ways that the glueball bound states may
affect the thermodynamics of the
full theory. At very low temperature, the thermodynamics will be dominated by
the very
low mass states.  At small coupling there are many more light mesonic states
than glueballs. At strong coupling there are many
approximate BPS glueball bound states, while only one of the mesonic states
remains
massless. Thus at strong coupling and at temperature high enough to be
influenced
by the approximate BPS states, the thermodynamics will eventually be dominated by
the glueball sector.

\section{Density of states and Hagedorn temperatures}
\label{Sec:Density}
%\subsection{Brief explanation of the approach}
\label{subSec:explainDensity}
We calculate thermodynamic quantities from the partition function, which we
express as a sum of Boltzmann factors weighted by the density of states (DoS),
$\rho(M^2,K)$. The
discrete spectrum is estimated numerically, and a fit to the data is used to
calculate the DoS. The discrete spectrum is used to calculate the cumulative
distribution function (CDF), $N(M^2,K)$, which is the number of states
with mass squared below $M^2$ at resolution $K.$
The DoS is related to the CDF by
\begin{equation}
\label{eqn:rho}
\rho(M^2,K)\equiv \rho_{K}(M^2) = \frac{dN\left(M^2,K\right)}{dM^2},
\end{equation}
with dimensions of $L^{2}$.

A comment regarding the units of the invariant mass squared eigenvalues $M^2$
is in order. From
$(\ref{eqn:EVP})$ it is inferred that the Hamiltonian is of the form
\begin{align}
\label{masssqrti}
P^-&=g^{2} A_{n} + \kappa g B_{n}+\kappa^{2} C_{n}=
\kappa^{2}\biggl(\frac{g^2}{\kappa^2} A_{n} +  \frac{g}{\kappa} B_{n}+ C_{n}\biggr),
\end{align}
where $g$ stands for  $g\sqrt{N_{c}/\pi}$. Thus it is a function of a
dimensionless ratio, $g/ \kappa$, and the
dimensionful\footnote{Choosing the CS coupling to
set the mass scale is quite natural for the problem at hand, since among others we investigate the
case where $g=0$ and $\kappa=1$,  where the mass
squared eigenvalues are proportional to $\kappa^2,$ see $(\ref{eqn:freepartons}).$ In general we have the
freedom of choosing the parameters such that they suit the
problem.} parameter $\kappa.$ The latter sets the mass
scale. Here we simply fix the value of $\kappa$ to unity, while we numerically
investigate the spectra for several values of $g$. So the quantities we
calculate are expressed in units where $\kappa=1.$

It is interesting that for $g$ large and $\kappa =0$ we have
$M^2_i=g^2 A_i,$ so the
eigenvalues scale with $g^2.$ Therefore, we may determine the strong coupling
properties of the theory from the solution of the $g=1$, $\kappa=0$ theory,
a numerically much simpler problem. For some associated 
numerical results, see Sec.~\ref{subSec:NumResultsFE}.
%
%Lanczos new numerical technique
%

\subsection{Estimating the density of states} \label{EstimatingDoS}

The DoS can be estimated by diagonalizing $P^-$, computing the CDF from
the spectrum, and differentiating a smooth fit to the CDF.  This is
what was done in previous work~\cite{Hiller:2004ft}.  The size of
the matrix representation of $P^-$ increases with $K$ and with
the number of fields. Eventually,
the computational cost becomes too high.  To ameliorate the situation, we
adapted a Lanczos-based algorithm to estimate the CDF directly.

We start by writing the density of states as
\be \rho(M^2)=\sum_n d_n \delta(M^2-M_n^2), \ee
where $d_n$ is the degeneracy of the mass eigenvalue $M_n$. The CDF is just
\be N({\rm mass}^2 \leq M^2)=\int^{M^2}d\bar{M}^2 \rho(\bar{M}^2). \ee
The density can be written in the form of a trace over $e^{-\im P^-x^+}$ as
follows:
\begin{align}
\rho(M^2)&=\frac{1}{2P^+}\sum_n d_n\delta(M^2/2P^+-P_n^-)
  =\frac{1}{4\pi P^+}\int_{-\infty}^\infty e^{\im M^2x^+/2P^+}
        \sum_n d_n e^{-\im P_n^- x^+} dx^+ \non\\
  &=\frac{1}{4\pi P^+}\int_{-\infty}^\infty e^{\im M^2x^+/2P^+}
           \Tr \, e^{-\im P^-x^+} dx^+.
\end{align}
To approximate the trace, we use an average over a random sample of
vectors~\cite{AlbenHams}.  Define a local density for a single vector
$|s\rangle$ as
\begin{align}
\rho_s(M^2)&=\frac{1}{4\pi P^+}\int_{-\infty}^\infty e^{\im M^2x^+/2P^+}
\langle s|e^{-\im P^-x^+}|s\rangle dx^+,
\end{align}
so that the average can be written
\be \rho(M^2)\simeq\frac{1}{S}\sum_{s=1}^S \rho_s(M^2). \ee
The sample eigenstates $|s\rangle$ can be chosen as random phase
vectors~\cite{Iitaka}, meaning that the coefficient of each Fock state in
the basis is a random number of modulus one.

\begin{figure}
 \begin{center}
   % \begin{tabular}{lr}
  %\dblrule\\[5mm]
  \resizebox{100mm}{!}{\includegraphics{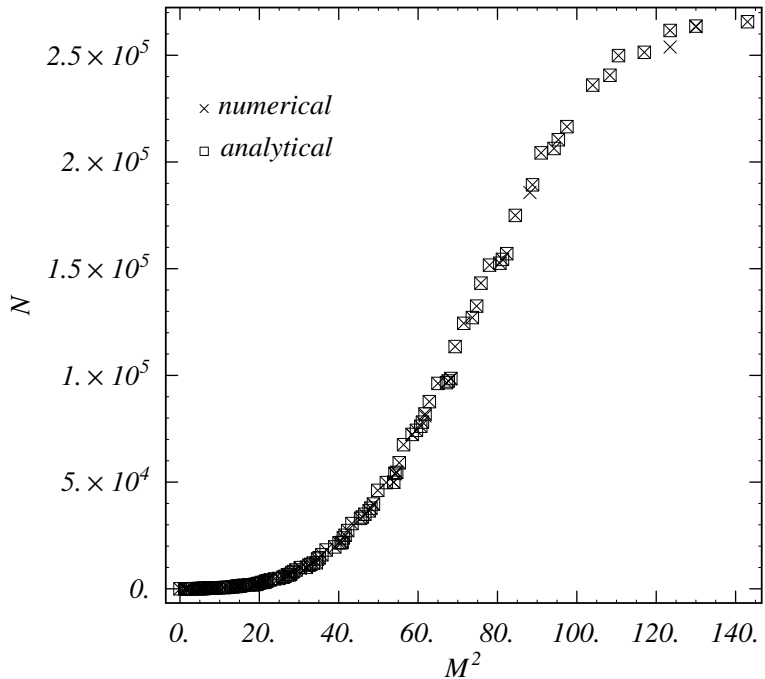}}
%  \resizebox{100mm}{!}{\includegraphics{compK13_analytic_dataBOUT.eps}}
      \\[-2mm]
     % \dblrule\\[-6mm]
    %\end{tabular}
    \caption{CDF of the free $(g=0)$ mesonic $\mathcal{T}$-odd $(\mathcal{T}^{-})$
sector at $K=13$. Crosses (boxes) refer to numerical (analytical) calculation
of the bound-state masses.}
\label{fig:compAnNum}
  \end{center}
\end{figure}

The matrix element $\langle
s|e^{-iP^-x^+}|s\rangle$ can be approximated by Lanczos
iteration~\cite{JaklicAichhorn}.
Let $D$ be the square of the norm of $|s\rangle$, and define
$|u_1\rangle=\frac{1}{\sqrt{D}}|s\rangle$ as the initial Lanczos vector.
Then we have
\begin{equation}
\rho_s(M^2)=\frac{D}{4\pi P^+}\int e^{\im M^2x^+/2P^+}
           \langle u_1|e^{-\im P^-x^+}|u_1\rangle dx^+,
\end{equation}
and $\langle u_1|e^{-\im P^-x^+}|u_1\rangle$ can be approximated by the
$(1,1)$ element of the exponentiation of the Lanczos tridiagonalization of
$P^-$.  Let $P_s^-$ be this tridiagonal matrix, and solve the eigenvalue
problem
\be P_s^-\vec{c}_n^{\,s}=\frac{M_{sn}^2}{2P^+}\vec{c}_n^{\,s}. \ee
A diagonal matrix $\Lambda$ is related to $P_s^-$ by the usual similarity
transformation $P_s^-=U\Lambda U^{-1}$, where $U_{ij}=(c_j^s)_i$ and
$\Lambda_{ij}=\delta_{ij}\frac{M_{sn}^2}{2P^+}$. This means that the
$(1,1)$ element is given by
\begin{equation}
\left(e^{-\im P_s^-x^+}\right)_{11}=\sum_n|(c_n^s)_1|^2e^{-\im M_{sn}^2
x^+/2P^+}.
\end{equation}
The local density is
\begin{align}
\rho_s(M^2)&\simeq \frac{D}{4\pi P^+}\int e^{\im M^2x^+/2P^+}
\sum_n|(c_n^s)_1|^2e^{-\im M_{sn}^2 x^+/2P^+} dx^+ \non\\
&\simeq \frac{D}{4\pi P^+}\sum_n|(c_n^s)_1|^2 2\pi\delta(M^2/2P^+-M_{sn}^2/2P^+)  \\
&\simeq \sum_n w_{sn} \delta(M^2-M_{sn}^2)\non,
\end{align}
where $w_{sn}\equiv D|(c_n^s)_1|^2$ is the weight of each Lanczos
eigenvalue. Note that only the extreme Lanczos eigenvalues are good
approximations to eigenvalues of the original $P^-$; however, the other
Lanczos eigenvalues and eigenvectors provide a smeared representation of
the full spectrum.

The contribution to the cumulative distribution function is
\begin{equation}
N_s(M^2)\equiv \int^{\bar{M}^2} d\bar{M}^2\rho(\bar{M}^2)\simeq\sum_n
w_{sn}\theta(M^2-M_{sn}^2).
\end{equation}
The full CDF is then approximated by the average
\be N(M^2)\simeq\frac1S\sum_s N_s(M^2). \ee

In forming the full CDF, one has to decide how to combine theta functions.
This is done by using the first sample run as a template for values
$M_{1n}^2$ at which to evaluate $N$.  The contributions of the other
samples to $N$ at these values are estimated by linear interpolation in
cases where the Lanczos eigenvalues $M_{sn}^2$ are not the same as those
in the first set.  Also, in cases where duplicate eigenvalues are
generated by the Lanczos iterations, only one is included in the template
and the associated weights are added together.

The convergence of the approximation is dependent on the number of Lanczos
iterations per sample, as well as the number $S$ of samples. Test runs
indicate that the recommended value~\cite{AlbenHams} of 20 samples is 
sufficient.  The number of Lanczos iterations is kept at 1000 per sample;
using only 100 leaves errors on the order of 1-2\%.

A check for the validity of this approach is the
comparison between the CDFs for the free theory,
where the analytic solution is available.
Figure~\ref{fig:compAnNum} clearly shows that the numerical technique
introduced here gives a CDF almost identical to the one obtained by
the analytical calculation. The very few points that appear to be
extraneous have no impact on the fitting algorithm we use to calculate the
fits to the CDFs.

\subsection{Fits to the spectrum}

In an earlier work related to thermodynamics~\cite{Hiller:2004ft}, we  split the spectrum into low
and high-mass regions, separated by the mass gap. The bound-state
spectrum for this problem has similar characteristics.  For instance, 
for small values of the coupling, namely $g \lesssim 1,$ the mass gap separates
the $K-1$ nearly massless color-singlet
states evolving from the massless states of the free theory from the
rest of the spectrum. For those values of the coupling these states have $M^2
\lesssim 1.$
Thus our density of states, $\rho_{K}(M^2)$, is zero in the mass gap 
$(M_1^2,M_2^2)$, and the CDF has the following generic form:
\begin{equation}
N(M^2,K)=\,\begin{cases}
\, N_{1}(M^2,K), & M^2_{\rm min}\leq M^2\leq  M^2_{1}\\
\, const., &   M^2_{1}< M^2<  M^2_{2}  \\
\, N_{2}(M^2,K), &    M^2_{1} \leq M^2 \leq M^2_{\rm max} .\\
\end{cases}
\end{equation}
%
%cdf&DoS;g=0.1
%
\begin{figure}[H]
  \begin{center}
     \hspace*{-0.5cm}
     \begin{tabular}{ccc}
   % \hline\hline\\[3mm]
%\psfrag{+}{}
%\psfrag{Z}{}
%\psfrag{N}[c][Bl][1.95][180]{$N_{T^{+}}$}
      \resizebox{50.0mm}{!}{\includegraphics{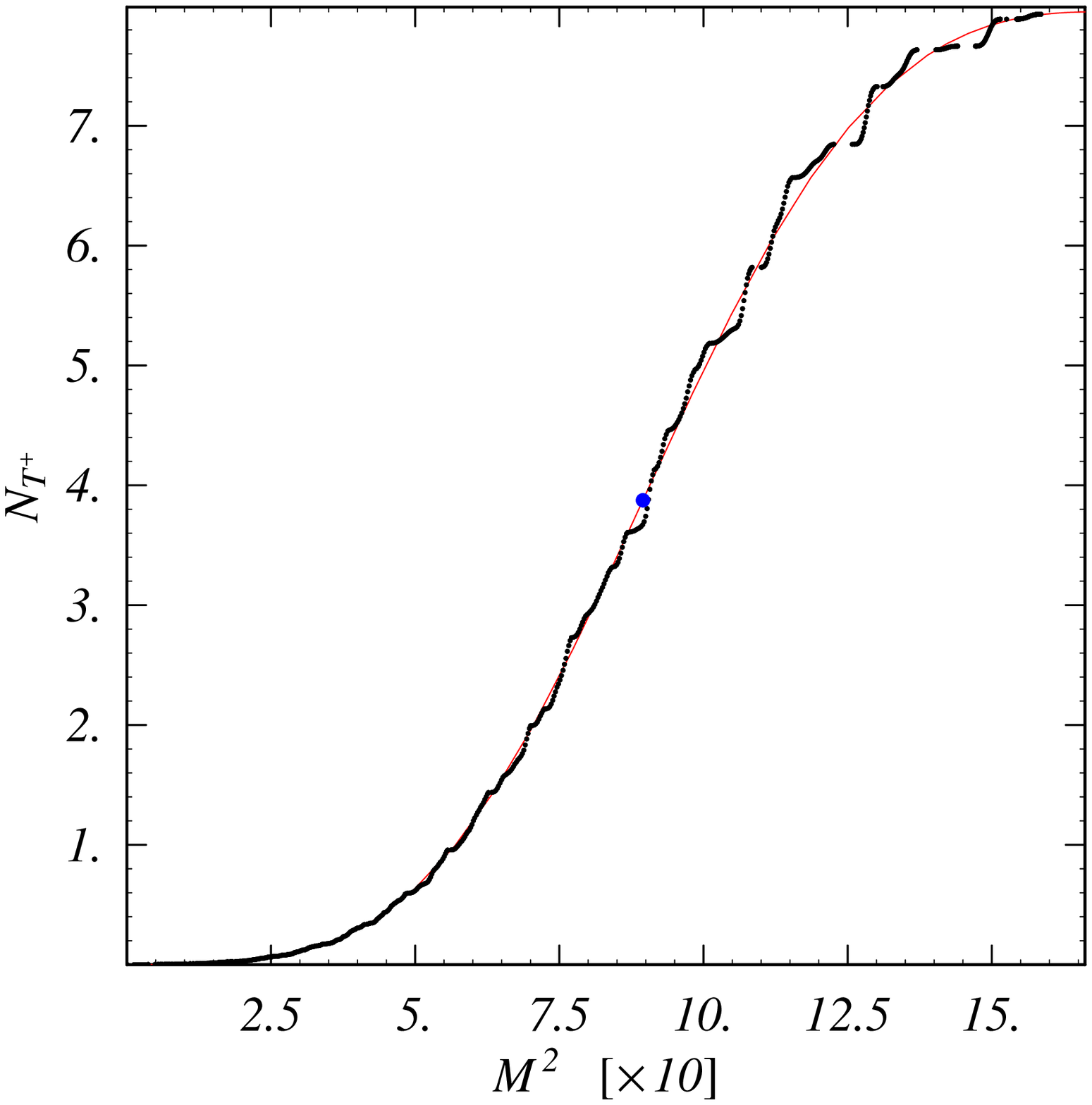}} &
      \resizebox{50mm}{!}{\includegraphics{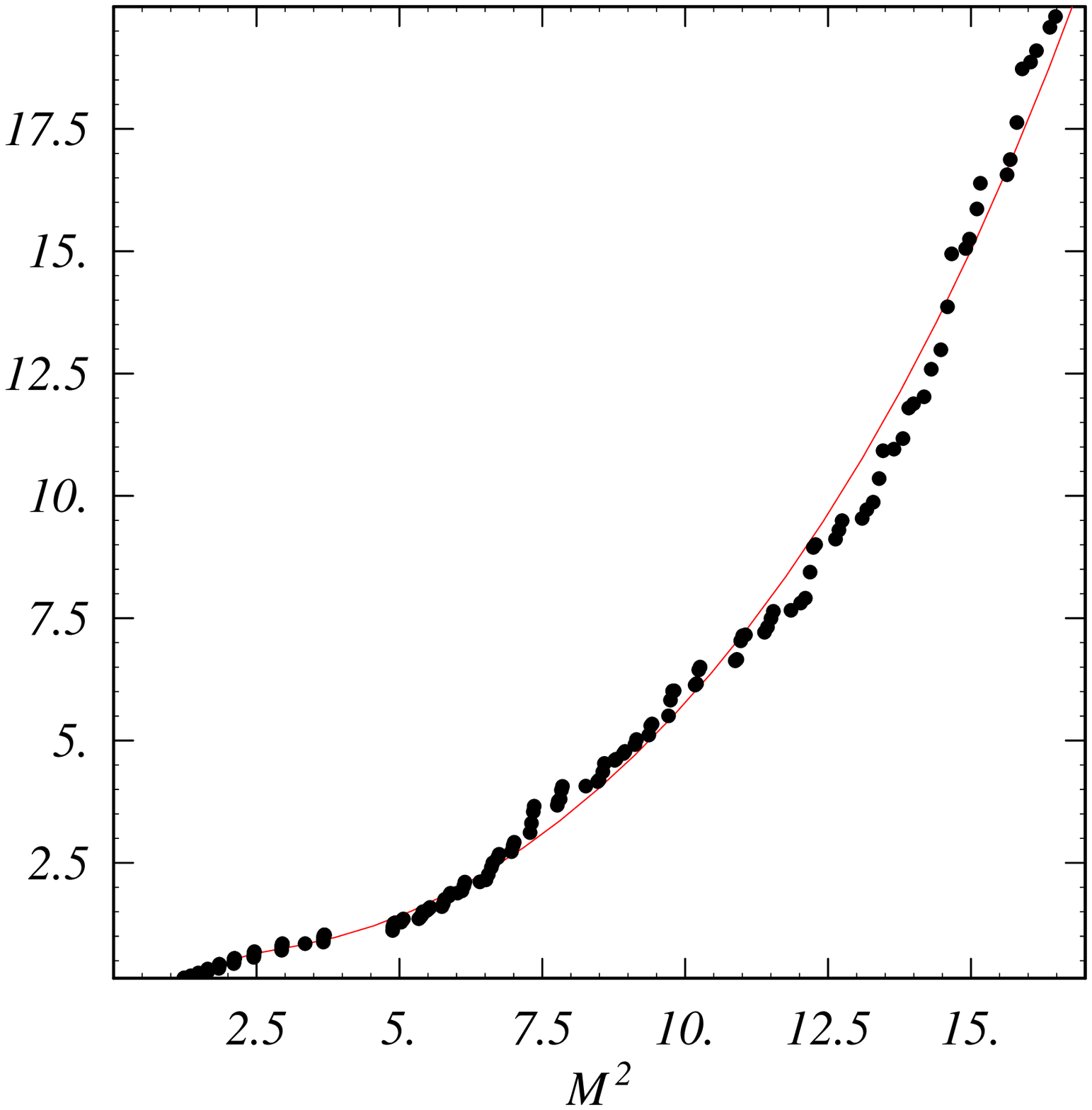}}  &
      \resizebox{48mm}{!}{\includegraphics{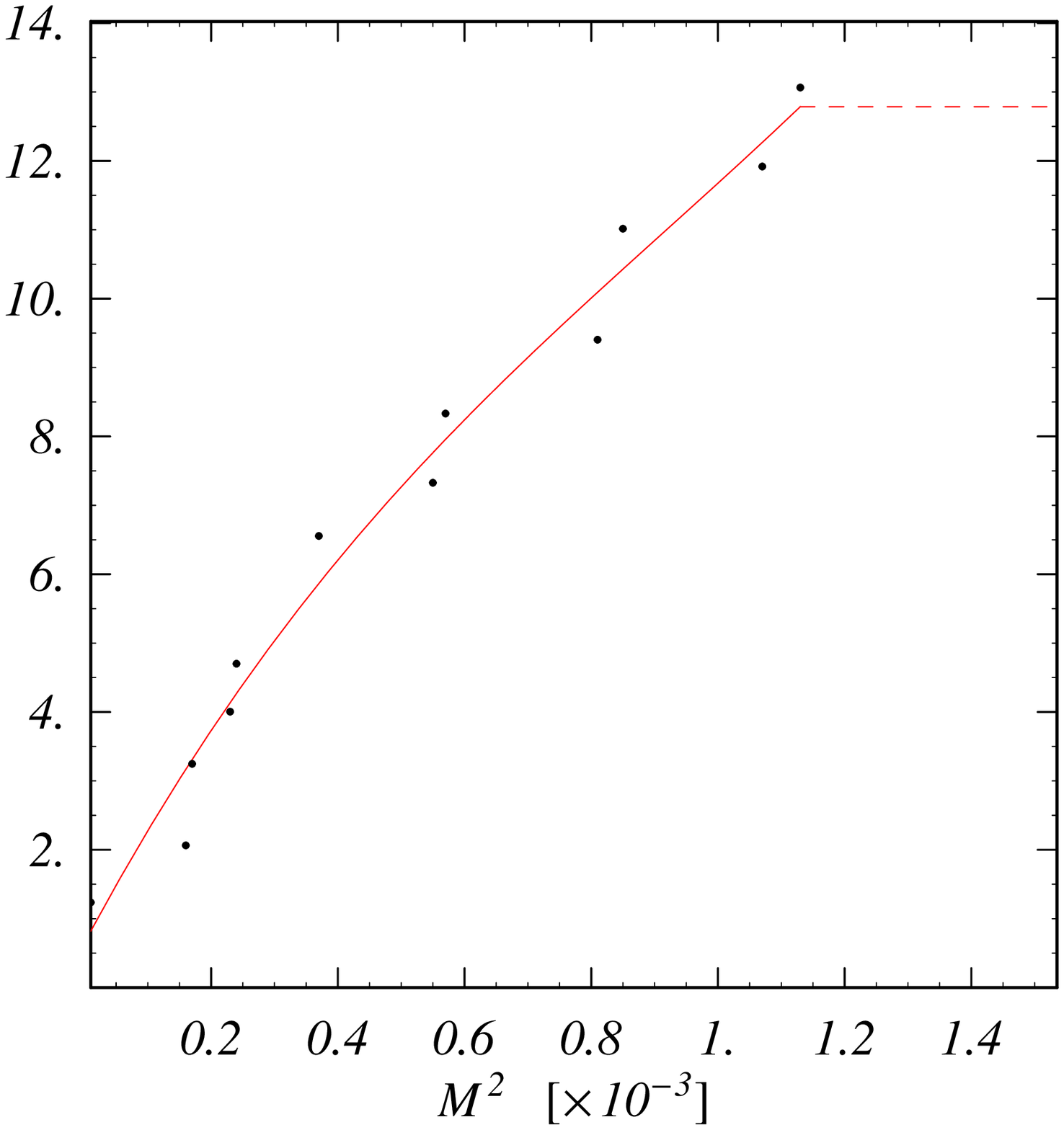}}\\
     \qquad (a)&\qquad (b) &\qquad (c)\\[3mm]
    %  \hline\hline
    \end{tabular}
\vspace{-6mm}
    \caption{CDF of the $\mathcal{T}$-even $(\mathcal{T}^{+})$ sector at $K=14$ and $g=0.1$. Shown
are data (dots) and a fit to the data: (a) all states in units of $10^5$
states; (b) range of masses just above the mass-gap in units of $10^2$
states; (c) states below the mass gap.}
\label{CDFandDoSA}
\end{center}
\end{figure}
\begin{figure}[H]
  \begin{center}
     \hspace*{-0.5cm}
     \begin{tabular}{ccc}
   % \hline\hline\\[3mm]
   \resizebox{50mm}{!}{\includegraphics{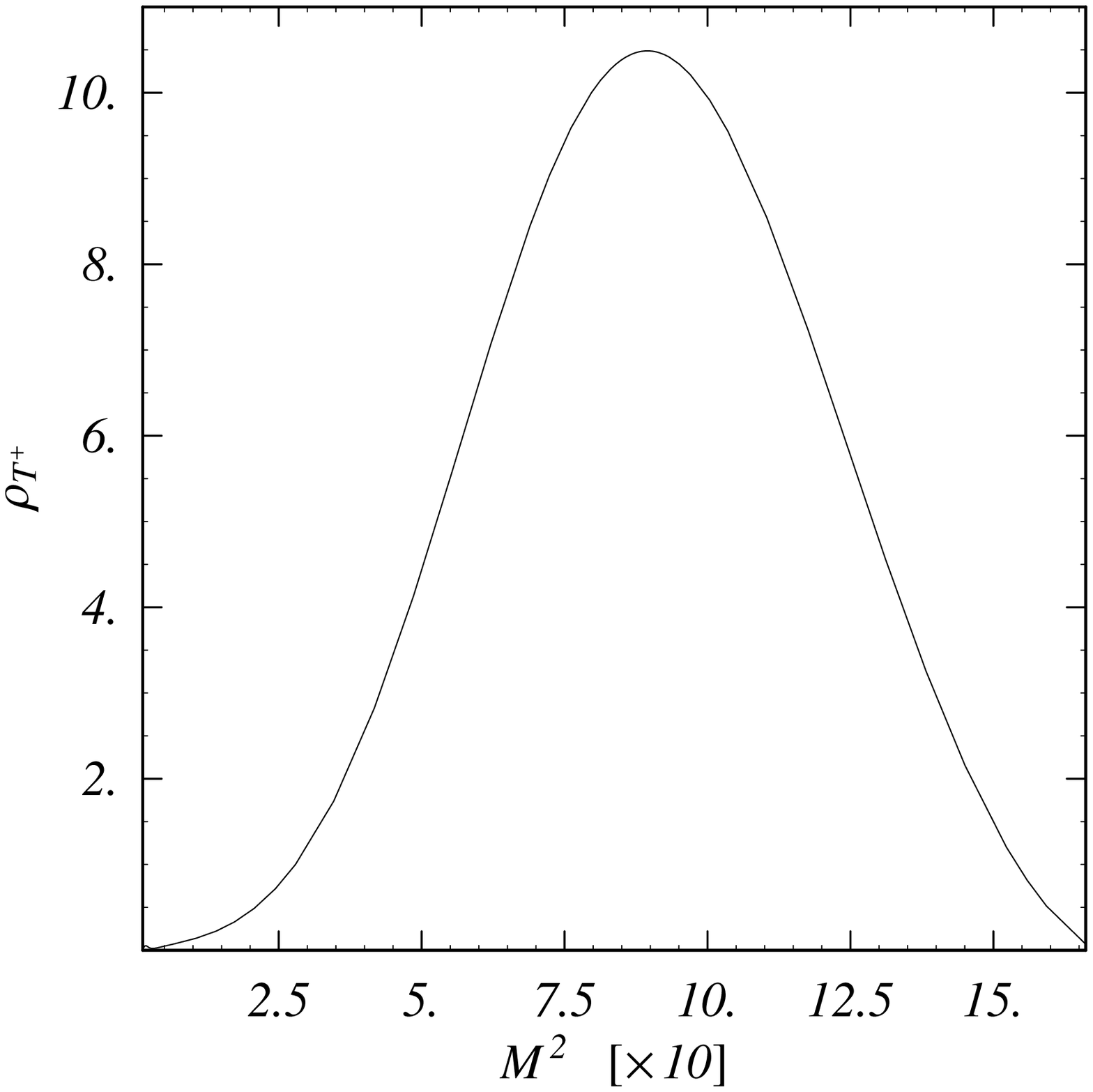}} &
      \resizebox{48mm}{!}{\includegraphics{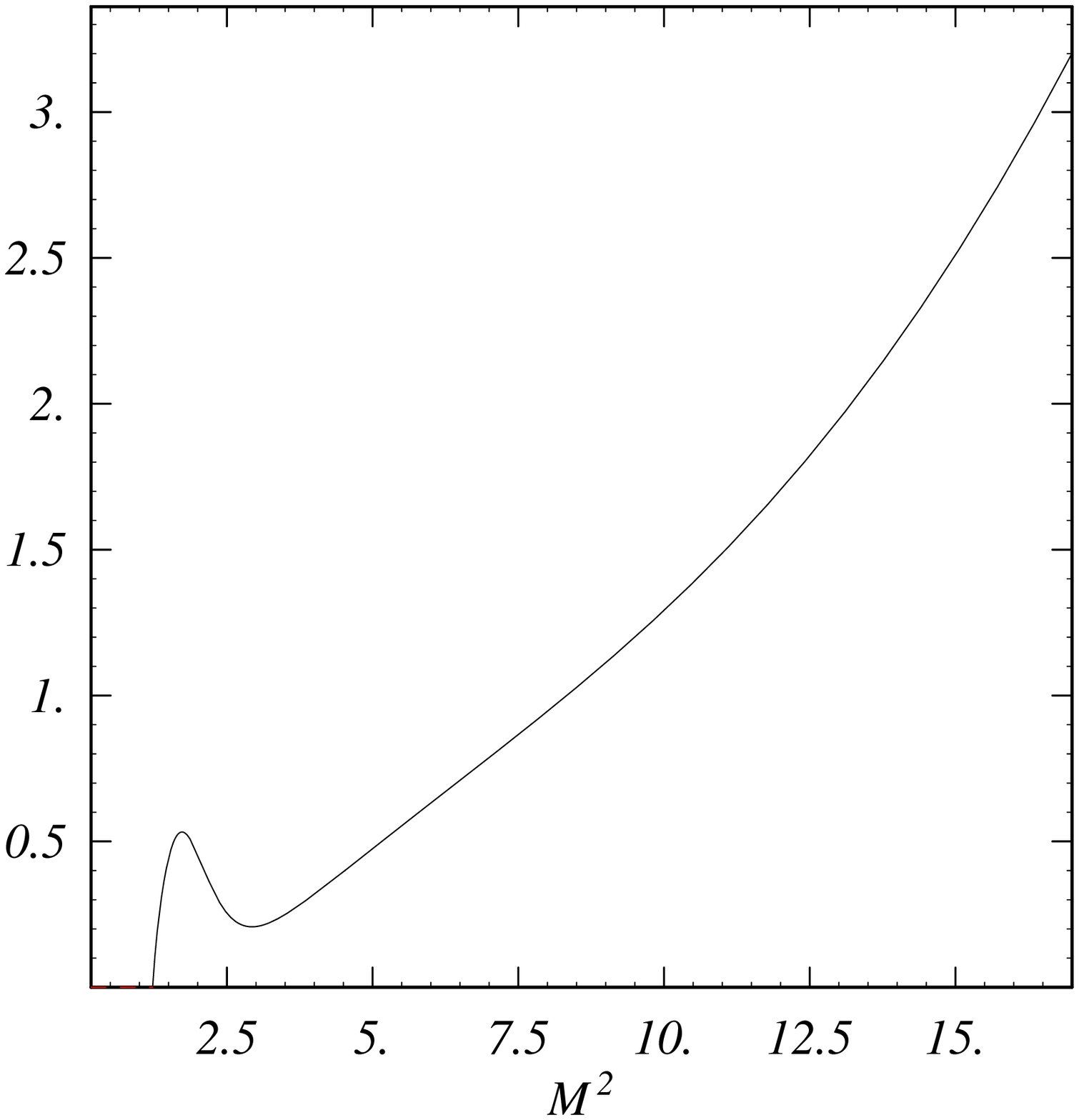}}  &
      \resizebox{48mm}{!}{\includegraphics{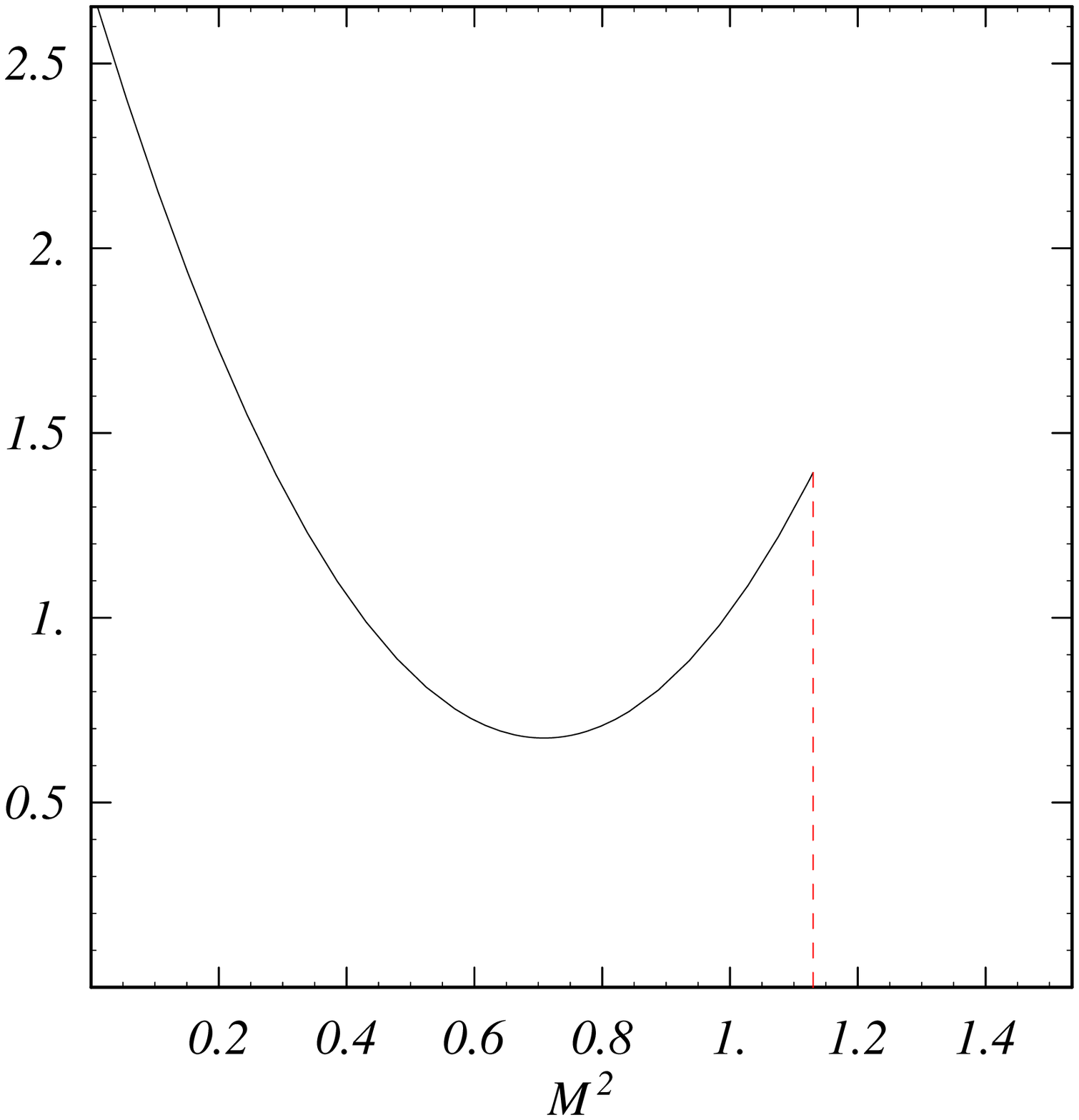}}\\
     \qquad (a)&\qquad (b) &\qquad (c)\\[3mm]
    %  \hline\hline
    \end{tabular}
\vspace{-6mm}
    \caption{Same as Fig.~\ref{CDFandDoSA} but for the density of
    states, in units of (a) $10^3$ states, (b) $10^2$ states, and (c) 1 state.
%Density of states (DoS) versus $M^2$: (a) entire DoS corresponding
% to Fig.~\ref{CDFandDoSA}(a) in units of $10^3$ states; (b) corresponds to
%Fig.~\ref{CDFandDoSA}(b) in units of $10^2$ states; (c) states below the
%mass gap associated with  Fig.~\ref{CDFandDoSA}(c).
}
\label{CDFandDoSB}
\end{center}
\end{figure}

\begin{figure}[H]
  \begin{center}
     \hspace*{-0.5cm}
     \begin{tabular}{ccc}
   % \hline\hline\\[3mm]
     \resizebox{50mm}{!}{\includegraphics{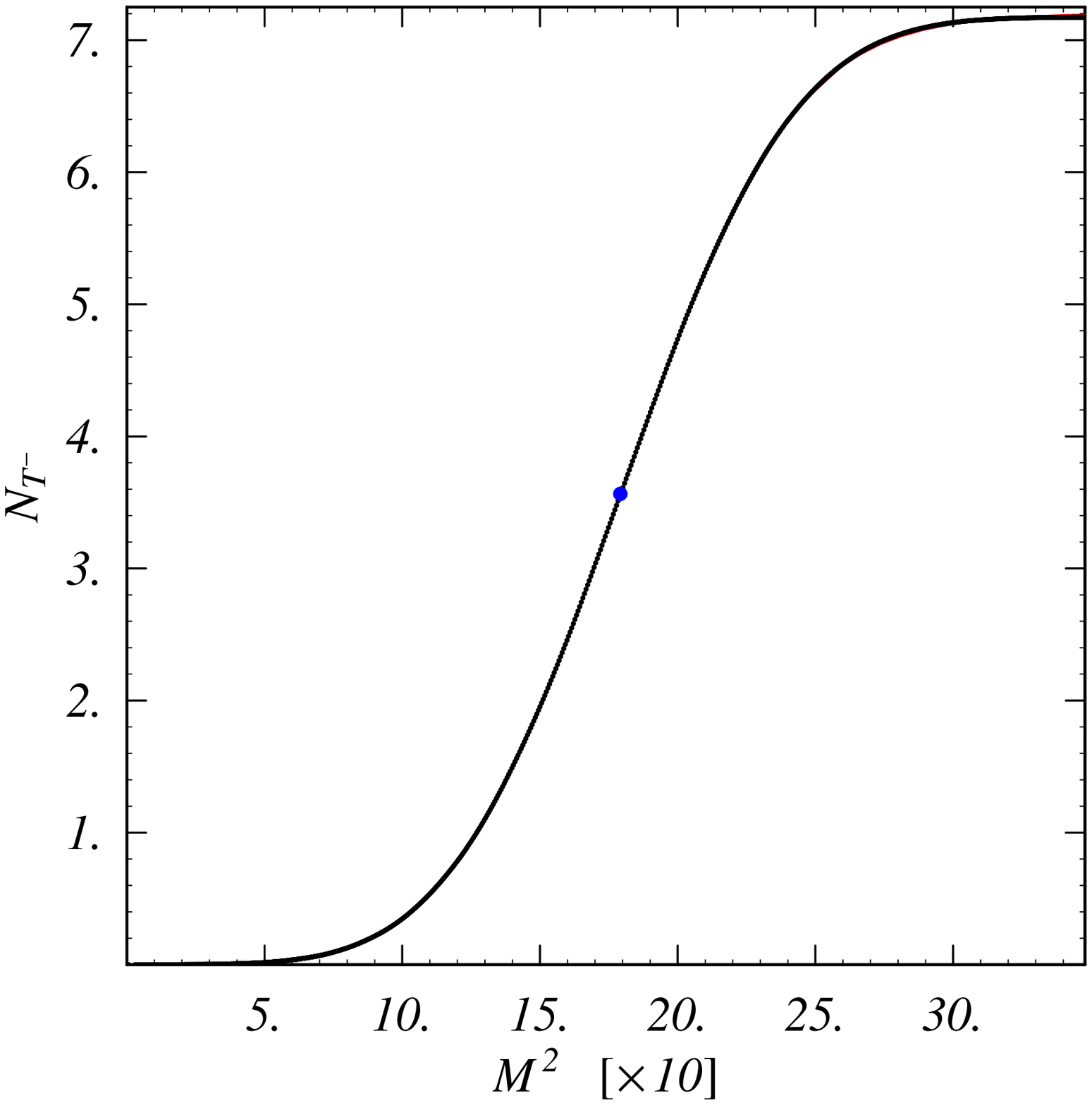}} &
      \resizebox{48mm}{!}{\includegraphics{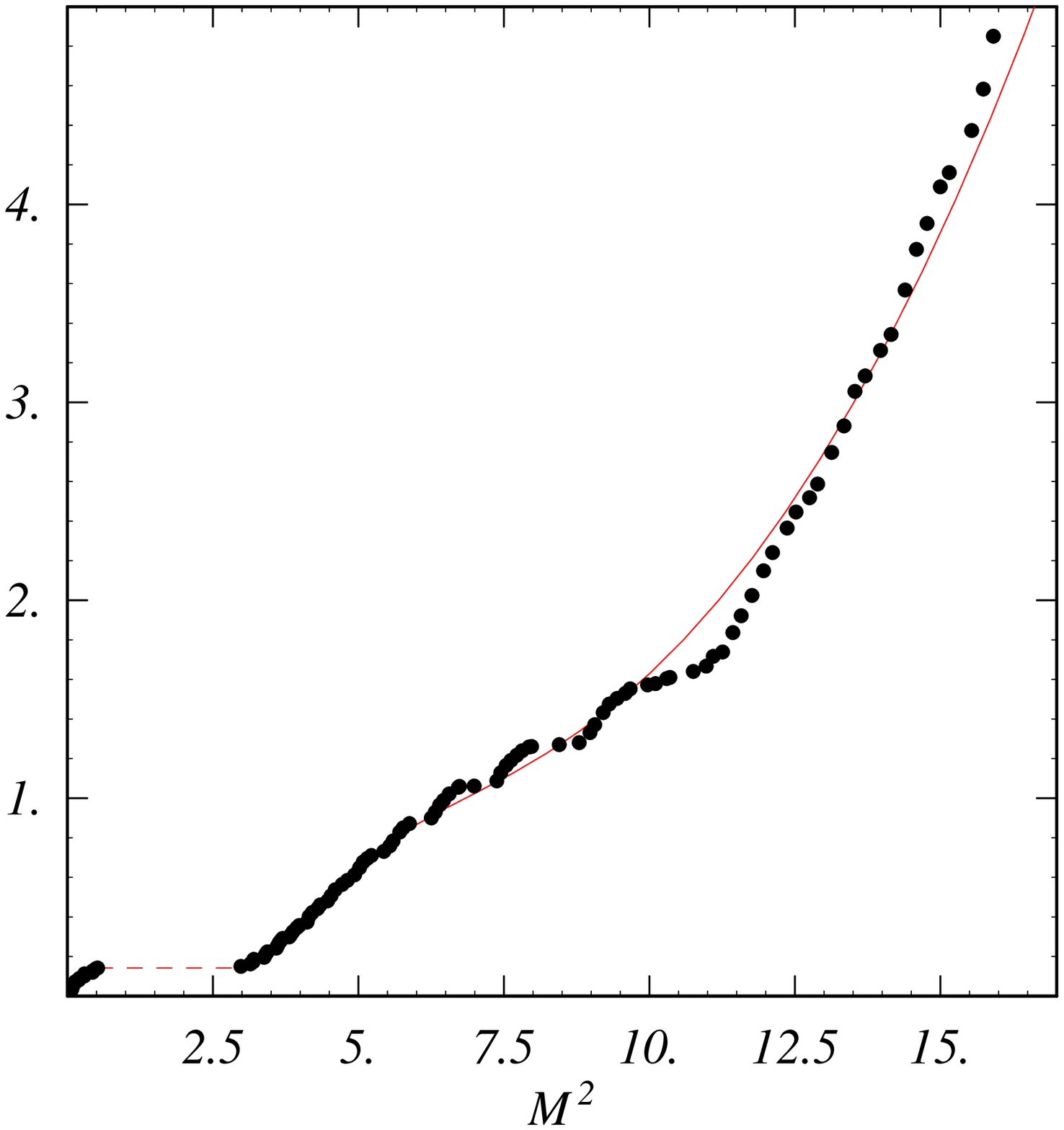}}  &
      \resizebox{49mm}{!}{\includegraphics{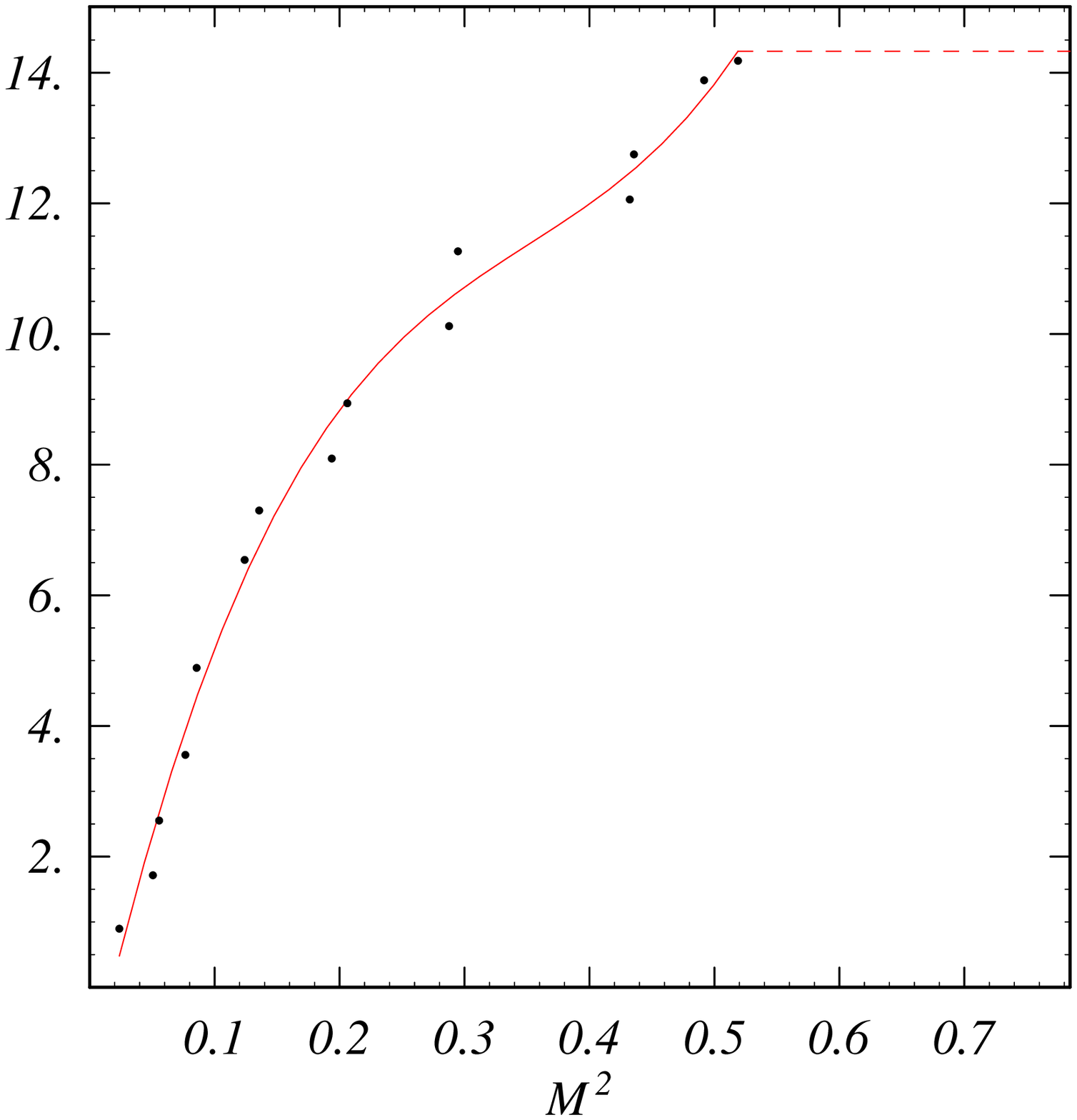}}\\
     \qquad (a)&\qquad (b) &\qquad (c)\\[3mm]
    %  \hline\hline
    \end{tabular}
\vspace{-6mm}
    \caption{
CDF of the $\cal T$-odd (${\cal T}^-$) sector at $K=16$ and $g=0.5$. Shown
are data (dots) and a fit to the data:
(a) all states in units of $10^6$
states with point of inflection;
(b) range of masses just above the mass-gap in units of $10^2$
states; (c) states below the mass gap.  The relatively poor fit
near $M^2=10$ in (b) does not have a significant effect on the results.}
\label{CDFandDoSC}
\end{center}
\end{figure}

\begin{figure}[H]
  \begin{center}
     \hspace*{-0.5cm}
     \begin{tabular}{ccc}
   % \hline\hline\\[3mm]
      \resizebox{50mm}{!}{\includegraphics{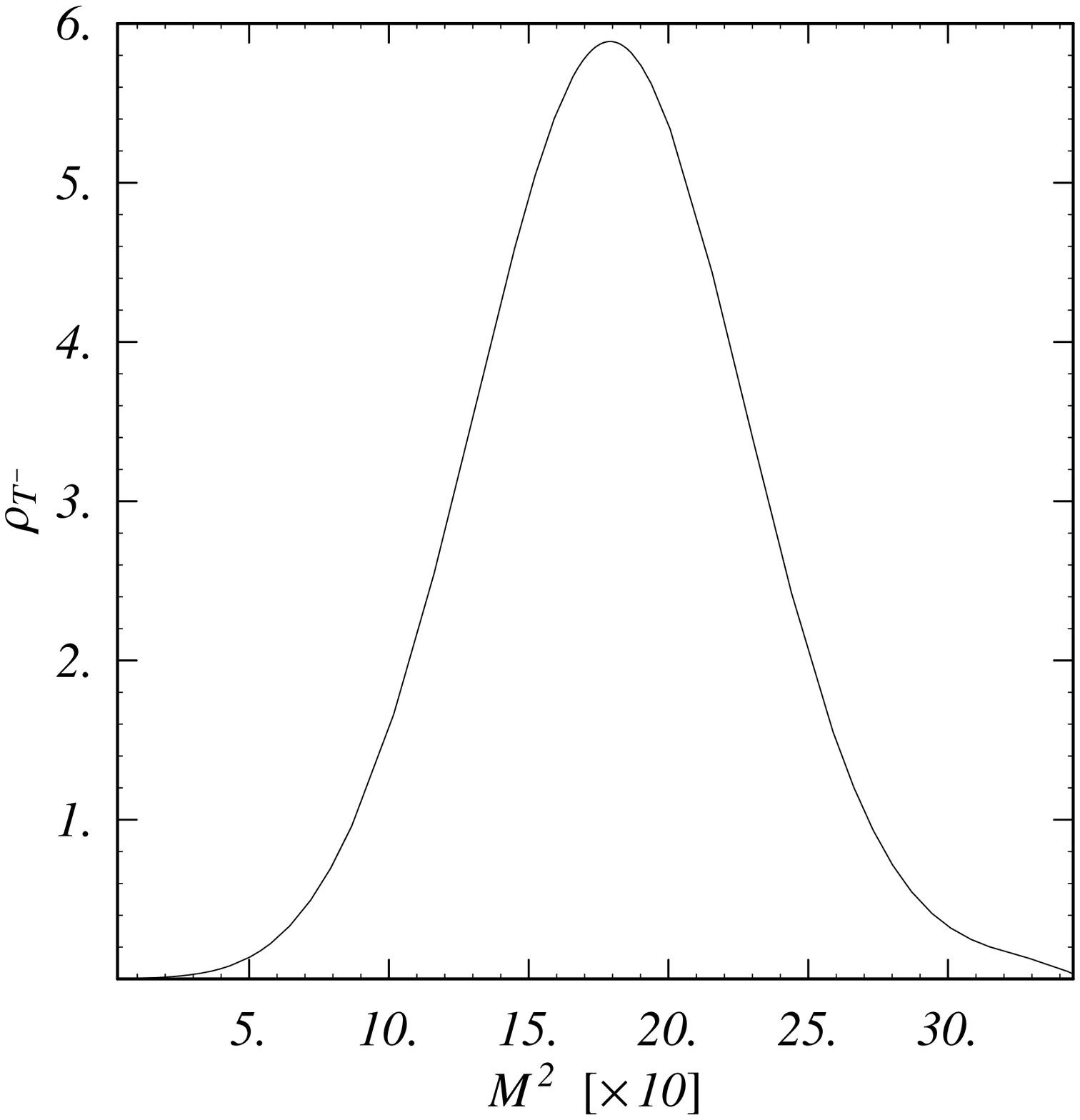}} &
      \resizebox{48mm}{!}{\includegraphics{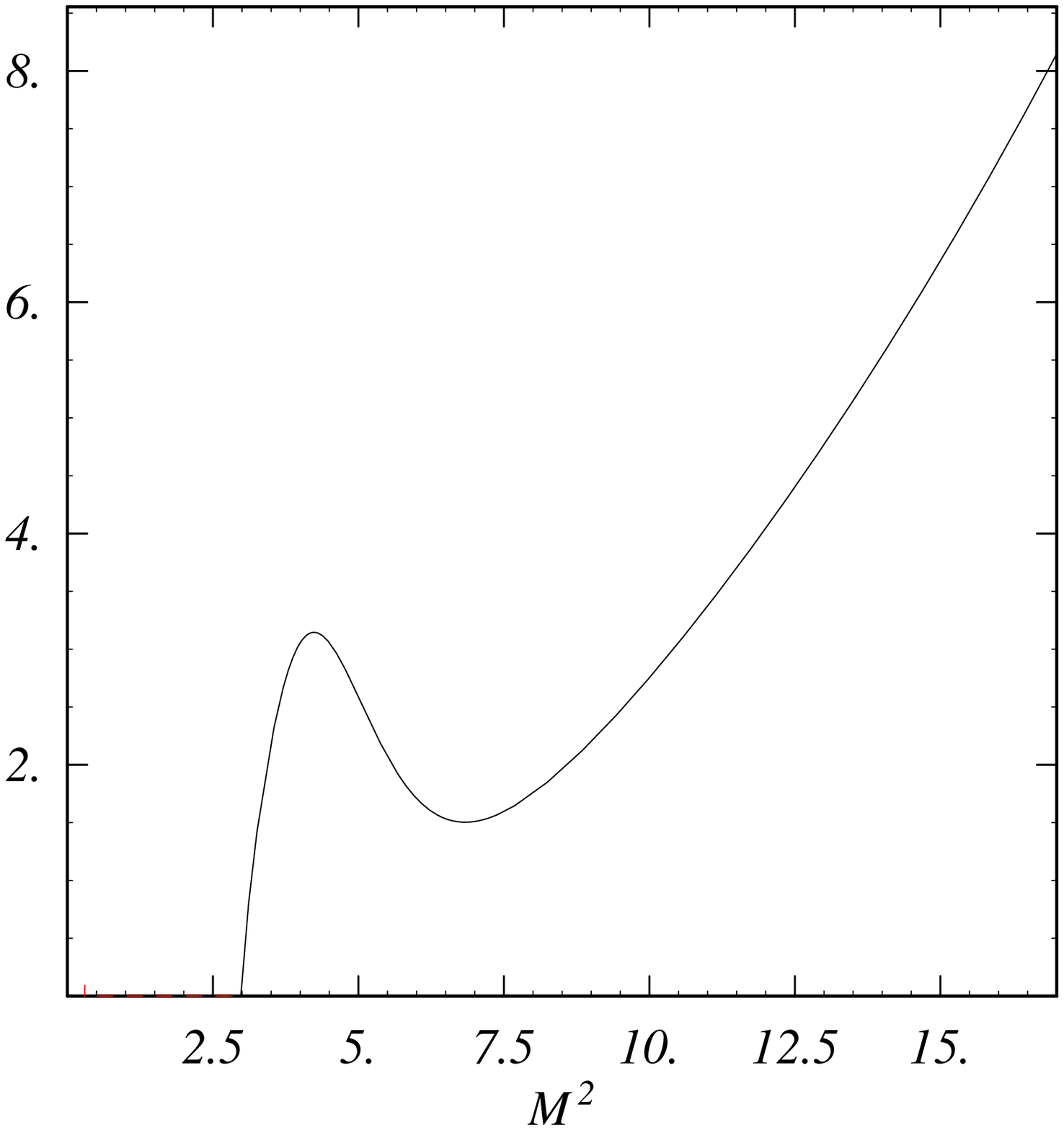}}  &
      \resizebox{48mm}{!}{\includegraphics{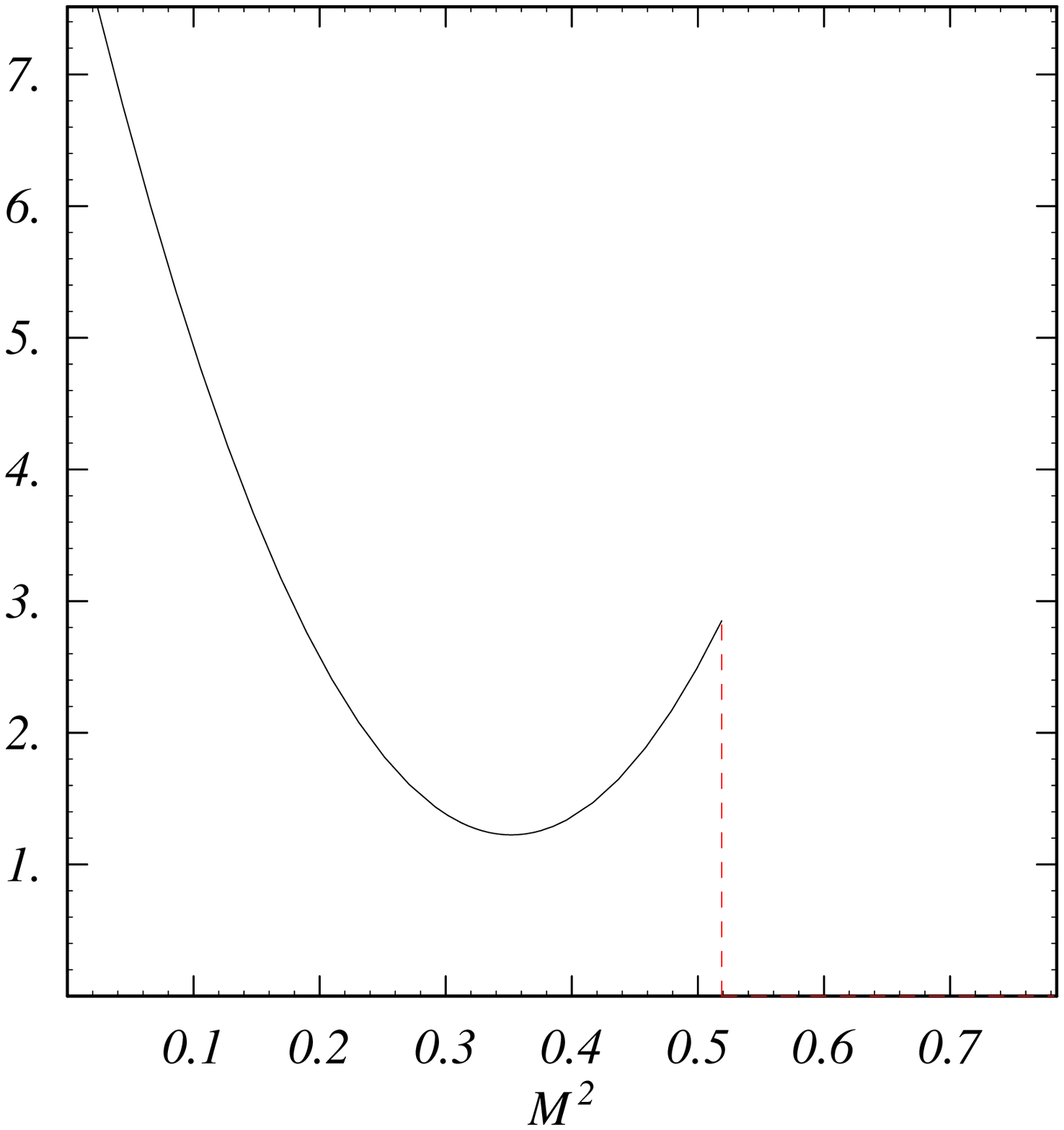}}\\
     \qquad (a)&\qquad (b) &\qquad (c)\\[3mm]
    %  \hline\hline
    \end{tabular}
\vspace{-6mm}
    \caption{Same as Fig.~\ref{CDFandDoSC}, but for the DoS, in units of
    (a) $10^4$ states, (b) and (c) 10 states.
    %DoS corresponding to Fig.~\ref{CDFandDoSC}.: (a) overall DoS
%in units of $10^4$ states; (b) DoS in units of $10$ states; (c) DoS for states below the mass gap %in units of $10$ states.
}
\label{CDFandDoSD}
\end{center}
\end{figure}

\begin{figure}[H]
  \begin{center}
     \hspace*{-0.5cm}
     \begin{tabular}{ccc}
   % \hline\hline\\[3mm]
     \resizebox{49mm}{!}{\includegraphics{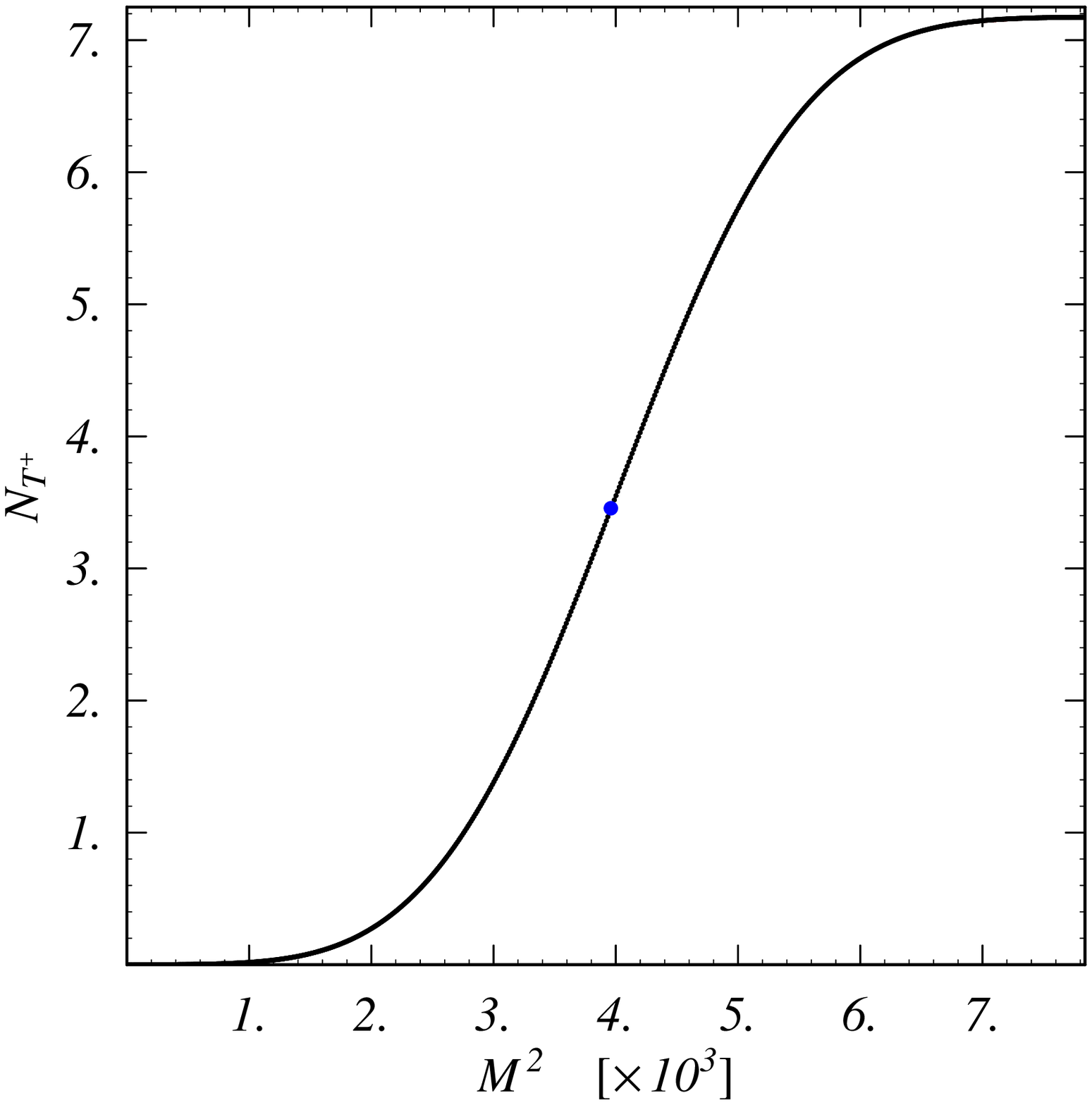}} &
      \resizebox{48mm}{!}{\includegraphics{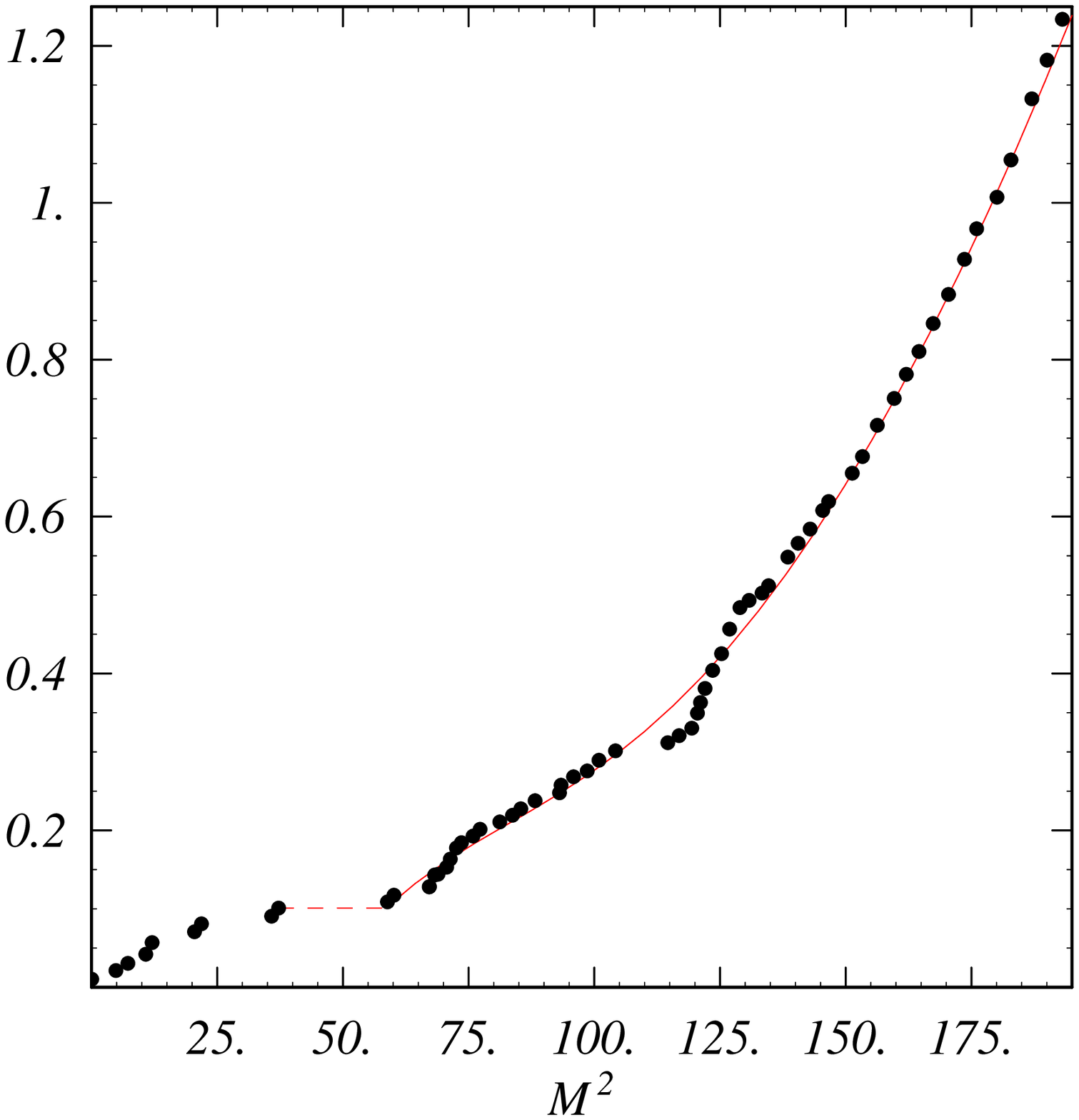}}  &
      \resizebox{48mm}{!}{\includegraphics{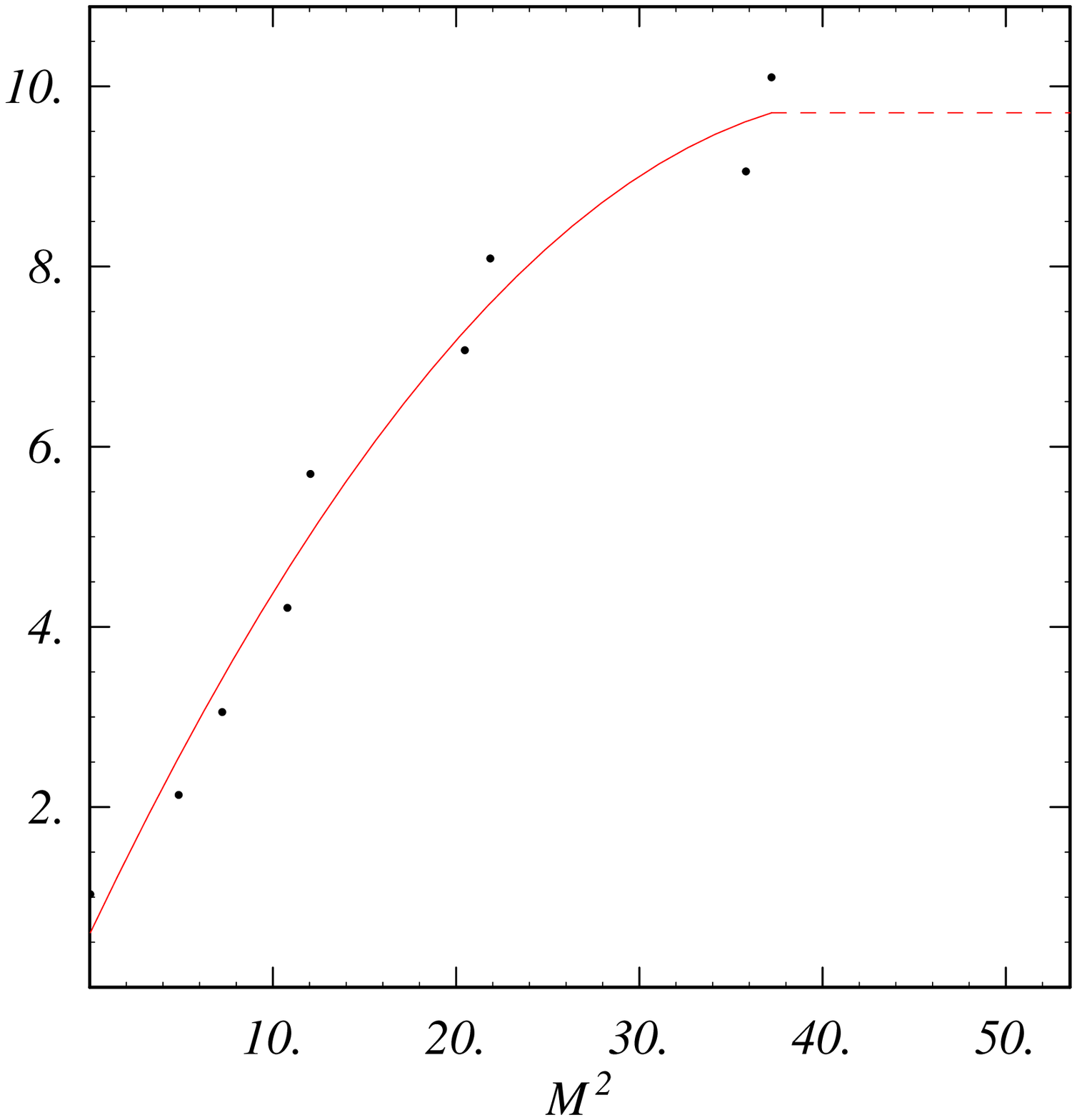}}\\
     \qquad (a)&\qquad (b) &\qquad (c)\\[3mm]
    %  \hline\hline
    \end{tabular}
\vspace{-6mm}
\caption{CDF of the $\cal T$-even sector at $K=16$ and $g=4.0$. Shown
are data (dots) and a fit to the data:
(a) all states in units of $10^6$
states; (b) range of masses just above the mass-gap in units of $10^2$
states; (c) states below the mass gap.}
\label{CDFandDoSE}
\end{center}
\end{figure}

\begin{figure}[H]
  \begin{center}
     \hspace*{-0.5cm}
     \begin{tabular}{ccc}
   % \hline\hline\\[3mm]
      \resizebox{50mm}{!}{\includegraphics{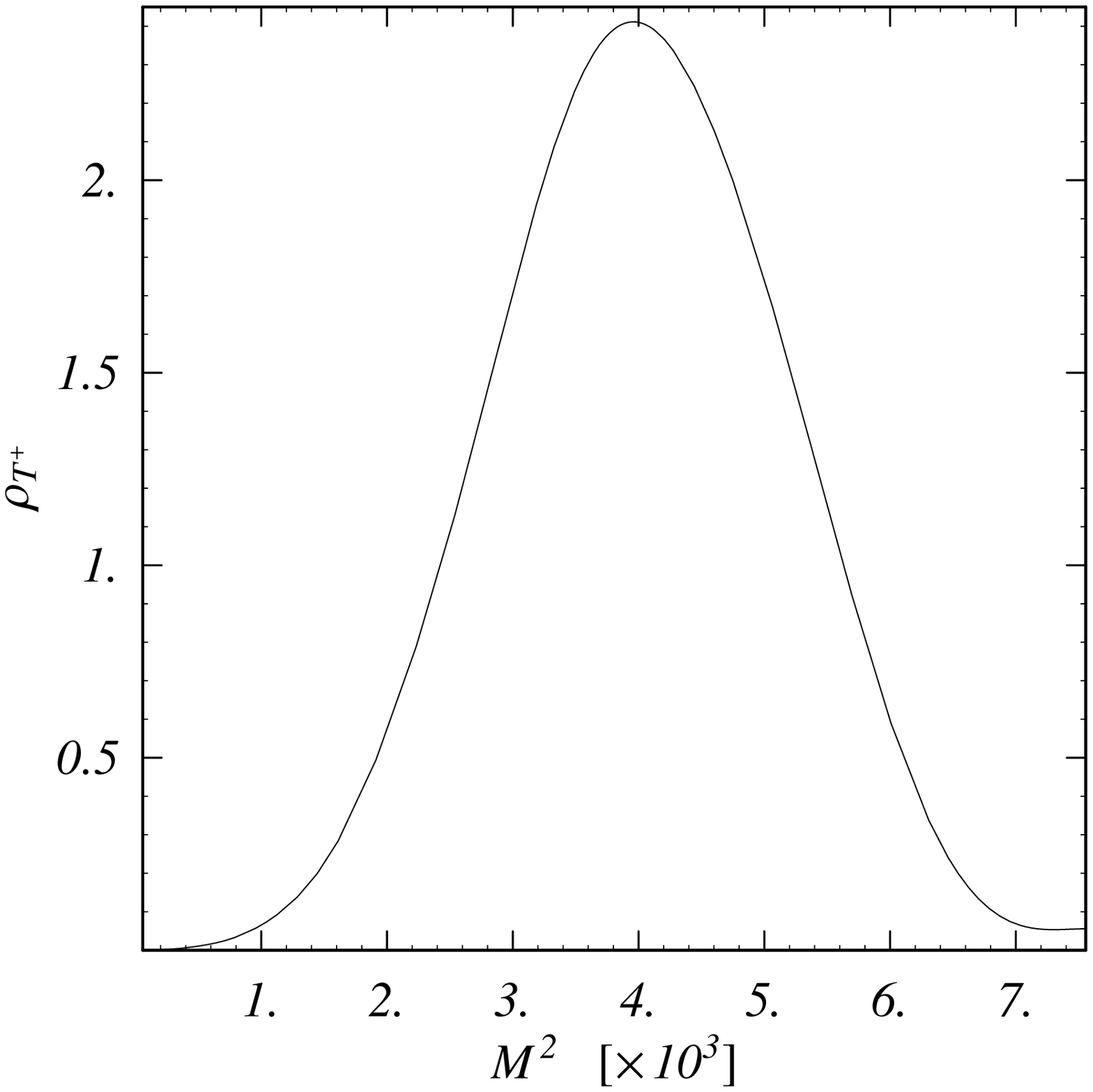}} &
      \resizebox{49mm}{!}{\includegraphics{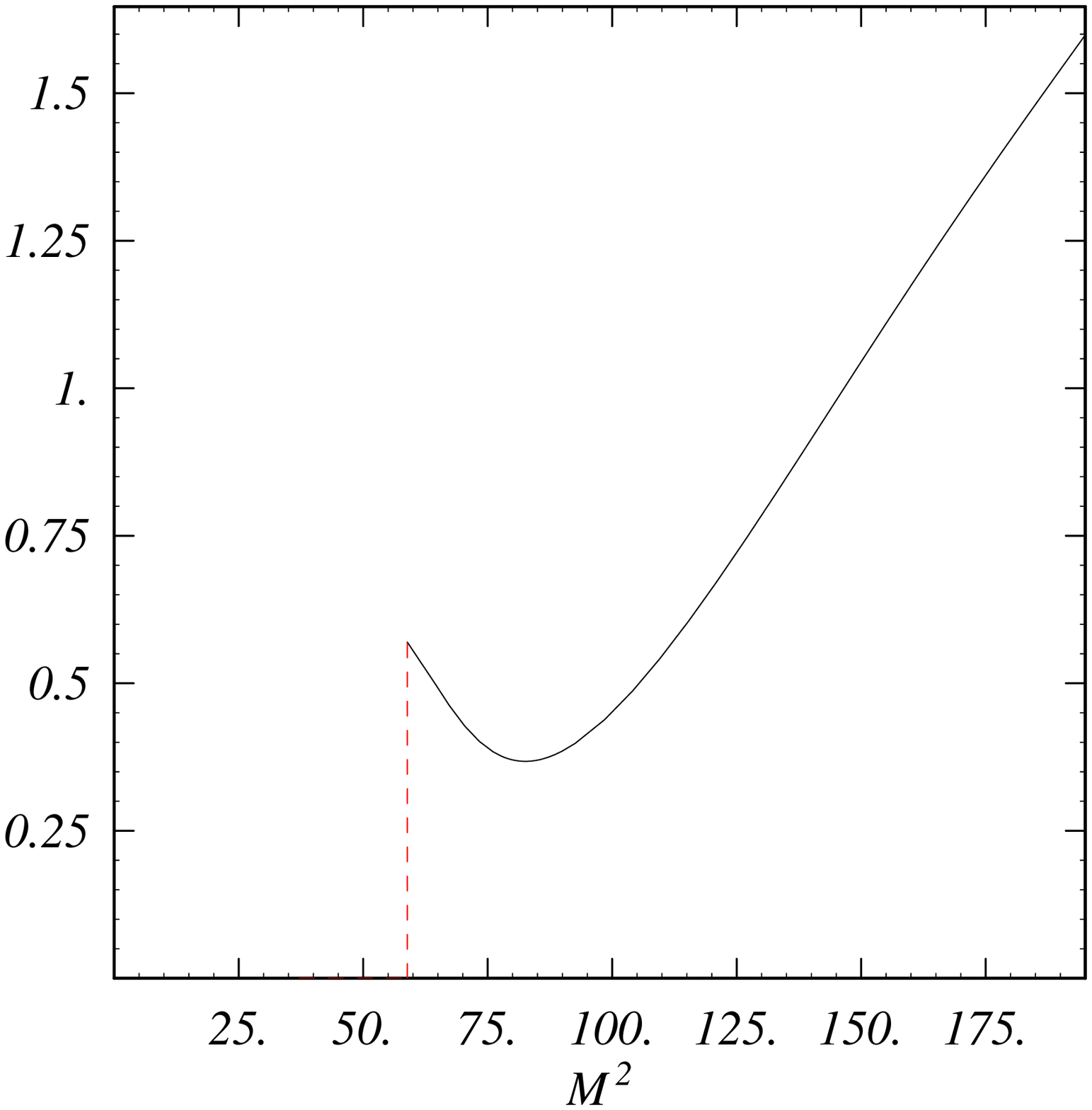}}  &
      \resizebox{48mm}{!}{\includegraphics{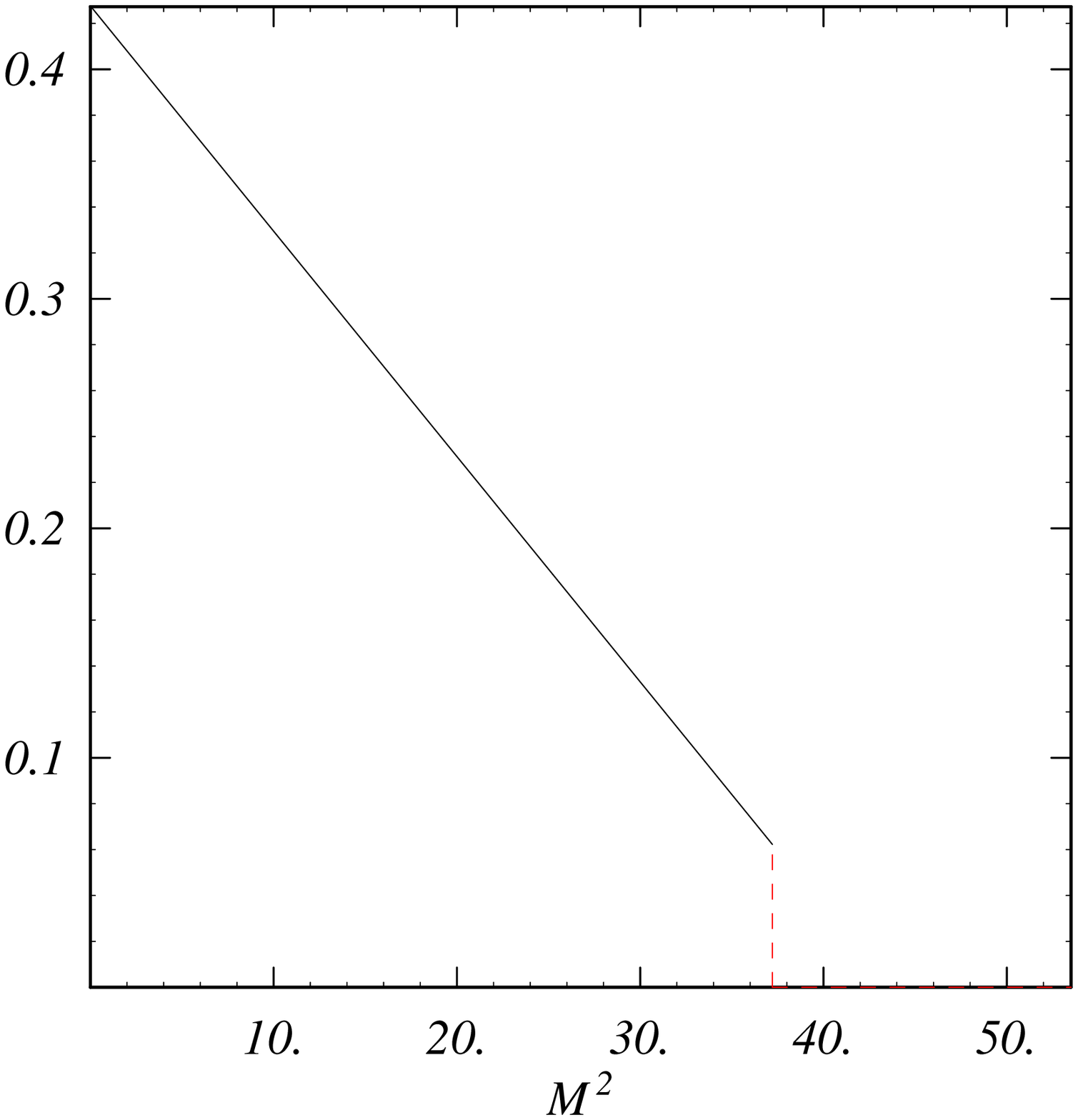}}\\
     \qquad (a)&\qquad (b) &\qquad (c)\\[3mm]
    %  \hline\hline
    \end{tabular}
\vspace{-6mm}
    \caption{Same as Fig.~\ref{CDFandDoSE}, but for the DoS, in units of
    (a) $10^3$ states, (b) and (c) 1 state.
%The above density of states plots correspond the CDF shown in
%Fig.~\ref{CDFandDoSE}.  Figure (a) above is given in units of $10^3$ states.
}
\label{CDFandDoSF}
\end{center}
\end{figure}

We fit the low-lying mass spectrum  of the CDF using the following function
\begin{align}
\label{eqn:fitfcnQ}
N_1(M^{2}_{l},K)&=\sum_{p=0}^{p(K)}\alpha_{p}\, M^{2p}_{l},
\end{align}
while the logarithm of the CDF for higher masses
is fit to the following function
\begin{align}
\label{eqn:fitfcnH}
\ln[N_2(M^2_{h},K)]&=(x_h+a_1)^{\gamma} \exp[-b_1 M_{h}^{2\delta}]\sum_{p=0}^{p(K)}\alpha_{p}\, M_{h}^{2p}
\end{align}
The other parameters in our fit functions are computed using standard non-linear fit algorithms.
Typical results are shown in Figs.~\ref{CDFandDoSA}-\ref{CDFandDoSF}.

%cdf&DoS;g=0.5

The spectrum exhibits some structure at relatively small $M^2$, as we
can see from
Figs.~\ref{CDFandDoSA}(b) and (c). These states
dominate the thermodynamics in the range of temperatures at which
we perform the free-energy calculations. Noteworthy is also the point of 
inflection in the CDF plot
and the peak in the DoS plot. The data beyond this point show the effect
of the
cutoff imposed by the resolution $K.$

>From the plots of the CDF and the DoS, and in particular the figures that depict the DoS for
$g\gtrsim 0.5,$ one may predict the result for the free energy $\cal F$. The
main contribution to $\cal F$ comes from the one nearly massless state  of
the spectrum. This state exists only in the $\cal T$-even sector, and
consequently
this sector dominates the thermodynamics at low temperatures. Therefore, we do
not expect the fit (e.g., see Fig.~\ref{CDFandDoSF}(c)) to give us an
accurate result for the free energy, because it essentially leaves out the
contribution from the nearly massless state. We will return to this point
when we discuss the numerical results for the free energy in
Sec.~\ref{subSec:NumResultsFE}.

%cdf&DoS;g=4.0

\subsection{Hagedorn temperature}
\label{subSec:Hag}
%%%%%%%%%%%%%%%%%%%%%%%%%%%%%%%%%%%%%%%%%%%%%%%%%%%%%%%%%%%%%%%%%%
We find that the physical spectrum in the theory  grows approximately
exponentially with the mass of the state
\begin{equation}
\label{eqn:rhoHagedorn}
\rho_{H}(M^2)\sim {\mathrm{exp}}(M/T_{\rm{H}}),
\end{equation}
and therefore has a Hagedorn
temperature $T_H$.
The partition function has the following general form
\begin{align}
\label{eqn:divpart}
\mathcal{Z} &\propto \int^{M} d\bar{M} \rho_{H}(\bar{M})
\exp\!{\bigg(\frac{-\bar{M}}{T}\bigg)}
%\sim
%\int^{M} d\bar{M}
%\exp\!{[\big(\frac{1}{T_{\mathrm{H}}}-\frac{1}{T}\big)\bar{M}]}\non\\
%&\propto
\propto
T_{\mathrm{H}}T \,\frac{
\exp\!{[M\big(\frac{1}{T_{\mathrm{H}}}-\frac{1}{T}\big)]}}
{T-T_{\mathrm{H}}}.
\end{align}
The partition function $\mathcal{Z}$ obviously
diverges as $T\to T_{\mathrm{H}}$, and
$T_H$ sets the region of validity for the calculation of thermodynamic
properties.  Thus $T_H$ serves as an
upper limit for the temperatures we can use to calculate the thermodynamic
functions.

We will calculate $T_H$ by fitting the CDF
with an exponential function, and then determining its exponent as a function
of the resolution and coupling. At fixed coupling we extrapolate
to infinite resolution and obtain the continuum Hagedorn temperature as a
function of the coupling. We fit the CDF in the region lying above the mass gap and
below the point of inflection. The number of states below the mass gap is
closely related to the number of free massless states and
therefore not a factor in the Hagedorn domain. The number of states above
the point of inflection are significantly reduced because of the cutoff
imposed
by the finite resolution and are therefore not useful in determining the
Hagedorn temperature.

\begin{figure}
  \begin{center}
     \hspace*{-0.5cm}
     \begin{tabular}{cc}
   % \hline\hline\\[3mm]
      \resizebox{76mm}{!}{\includegraphics{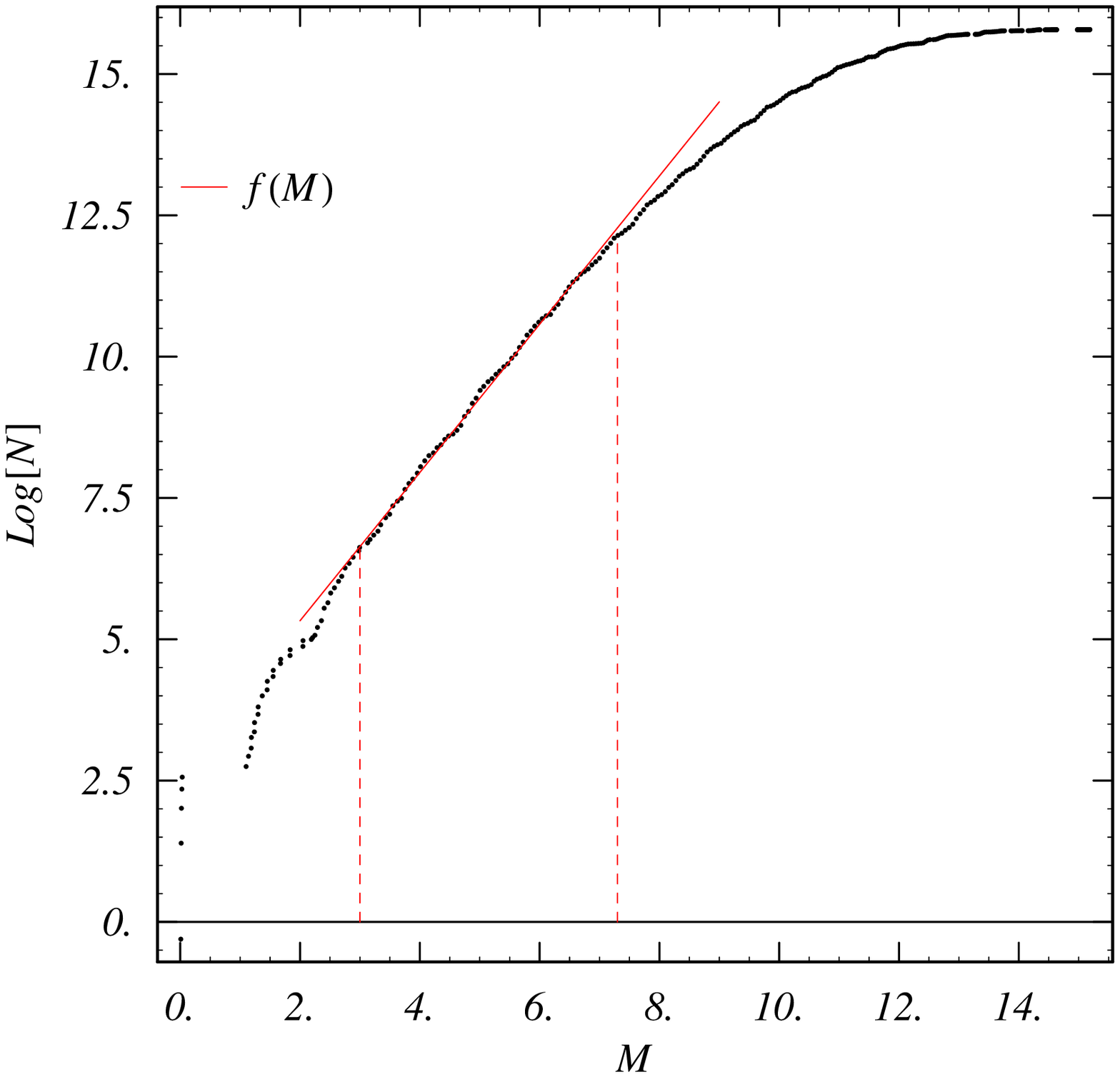}} &
      \resizebox{74mm}{!}{\includegraphics{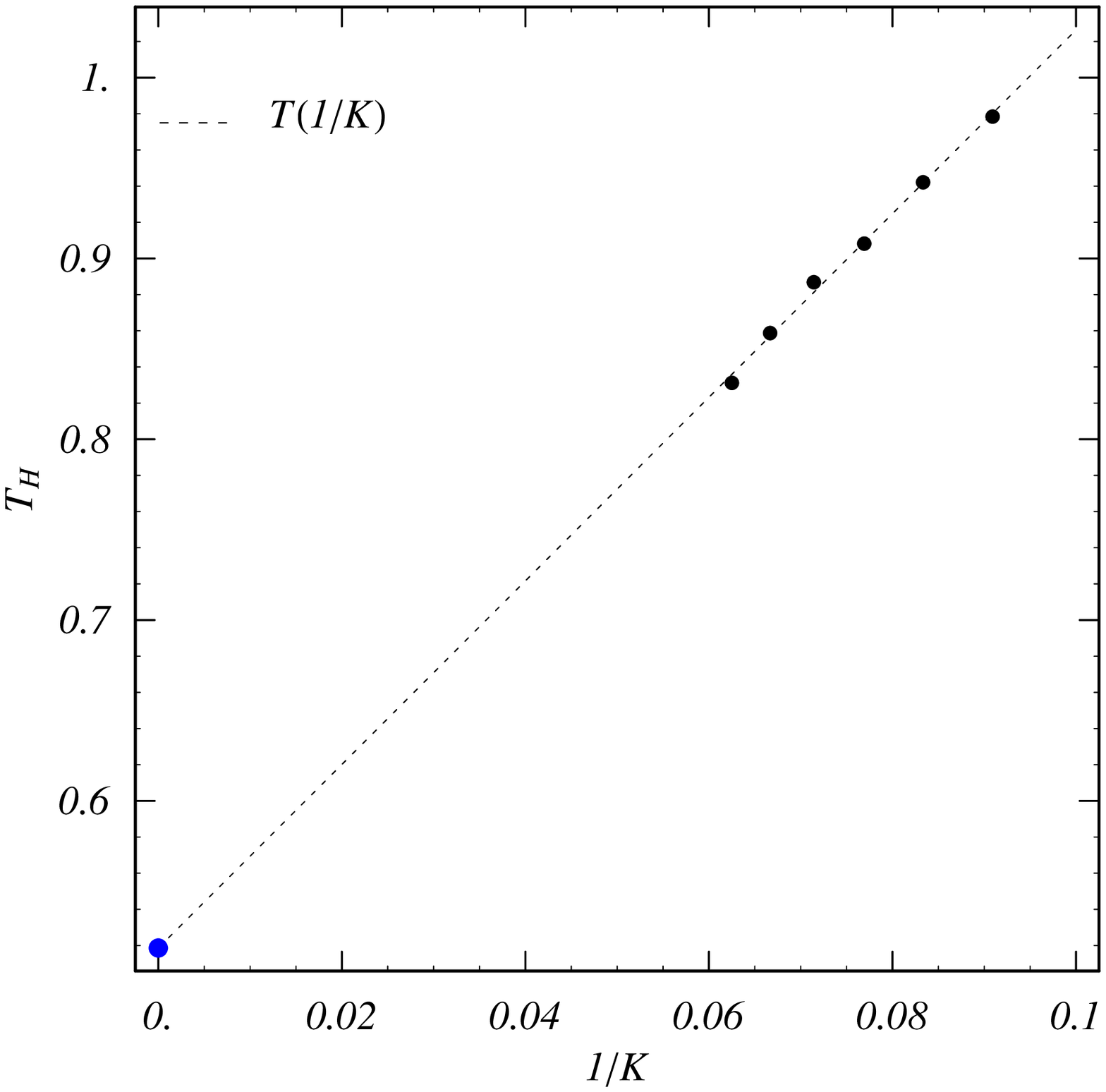}}\\
     \qquad (a)&\qquad (b)\\[3mm]
    %  \hline\hline
    \end{tabular}
\vspace*{-6mm}
    \caption{(a) Logarithm of the CDF versus $M$.
The approximately linear part of this
logarithmic CDF is fit to
    $f(M)=\a M+ \frac{1}{T_{H}}$. (b) Extrapolated Hagedorn
temperature for coupling $g=0$ with $T^{\infty}_{H}(0)\approx 0.52 \kappa.$ }
\label{LogFit}
\end{center}
\end{figure}

In Fig.~\ref{LogFit}(a) we show a typical fit to the CDF in this
region of $M^2$. The data is fit well with an exponential.
The particular figure deals with the
$\cal T$-odd sector of the spectrum at coupling $g=0.1$ and $K=13.$ A
similar behavior occurs for the $\cal T$-even sector and other values of $K$.
It is clear from our data that the spectral CDF and the DoS exhibit
a Hagedorn behavior. We plot the logarithm of the CDF
versus the bound-state
mass obtained from our numerical calculations. Then we estimate
%,
%\emph{visually},
the range of the mass values $M$ where the plot is approximately
linear, and fit this region to a linear fit of the form $\a M+
T^{-1}_{H}.$ The non-linear part of the distribution, for high values of $M,$  is
cut off because
of the finite resolution $K.$
The extrapolated result for the
$\cal T$-even sector is very similar to the result of the $\cal T$-odd sector,
so we take the average of the two values, for each $K.$

\begin{figure}
  \begin{center}
     \hspace*{-0.5cm}
     \begin{tabular}{cc}
   % \hline\hline\\[3mm]
      \resizebox{74mm}{!}{\includegraphics{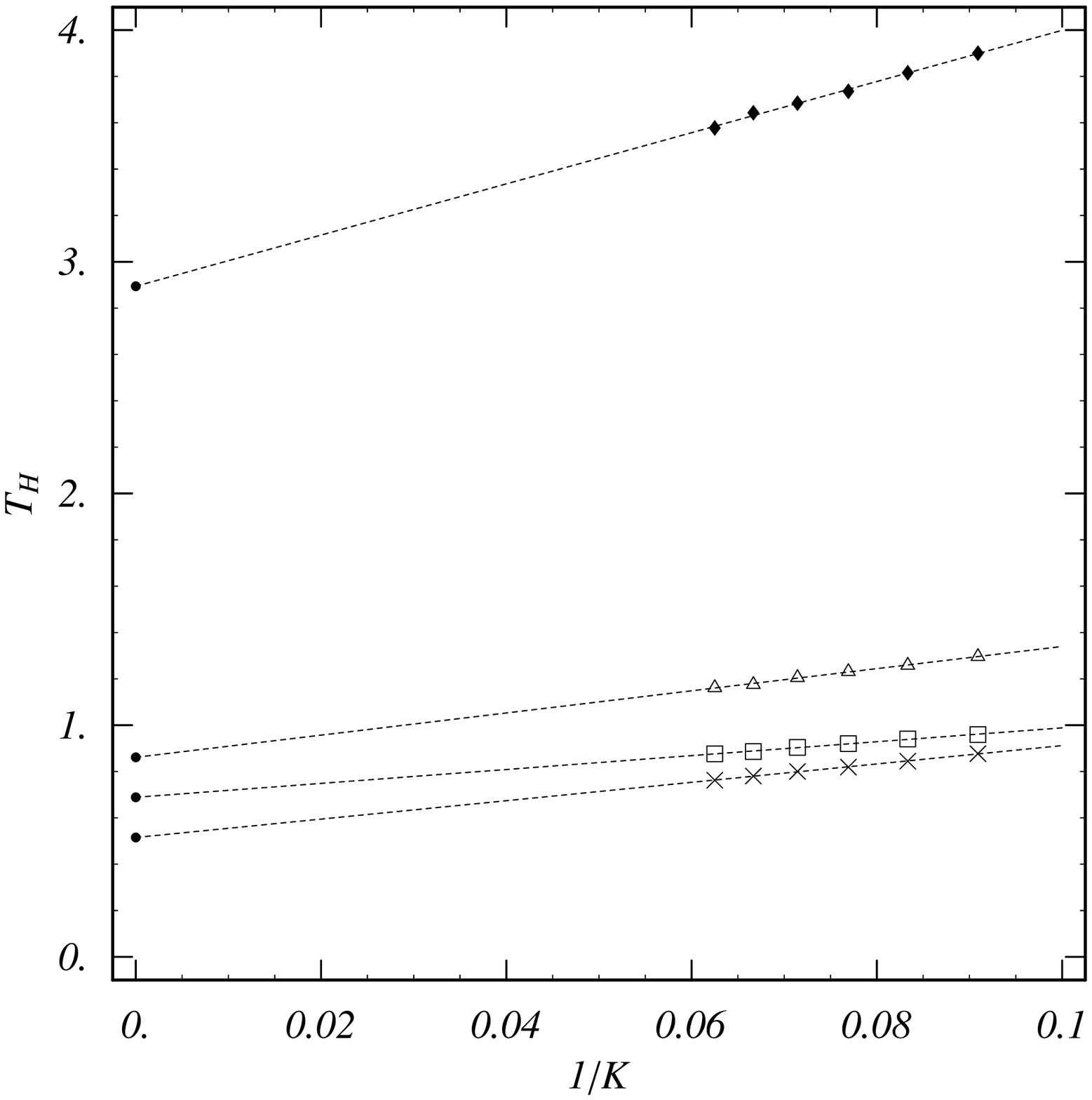}}&
      %\resizebox{74mm}{!}{\includegraphics{THagVSKinvallunsOUT.eps}} &
    %
      \resizebox{76.2mm}{!}{\includegraphics{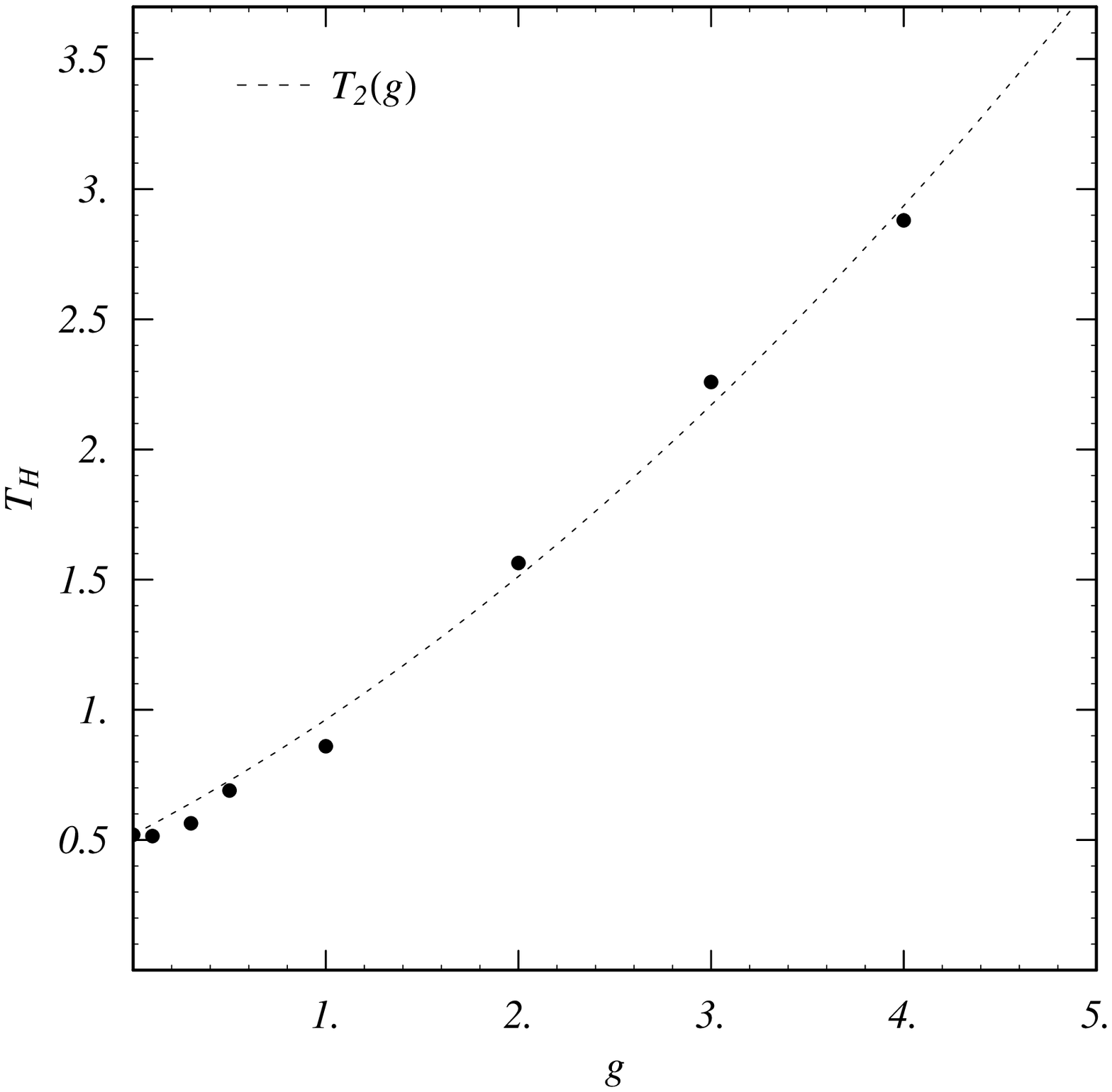}}\\
     \qquad (a)&\qquad (b)\\[3mm]
    %  \hline\hline
    \end{tabular}
\vspace{-6mm}
    \caption{Hagedorn temperature (a) plotted versus $1/K$
at couplings $g=0.1$ (crosses), 0.5(boxes), 1.0(triangles), 4.0(diamonds) in units where
$\kappa=1,$ and
(b) extrapolated in $K$ as function of $g$ with a
fit to $T_{2}(g)=0.52+ 0.39 g + 0.054 g^2$ (dashed line).
In (a), the dots at $1/K=0$ are the continuum values. 
For clarity we have included only four representative values of $g$ in (a).}
\label{ThagfitAll}
\end{center}
\end{figure}

In  Fig.~\ref{LogFit}(b) we show the Hagedorn temperature as a function of
the inverse resolution for the massive, free theory  $(g=0, \kappa=1)$.
The data appear
to be following a straight line and have been extrapolated to the continuum, where
 we found $T_{H}(g=0)\approx 0.52$ in units where $\kappa=1.$  We show
the Hagedorn temperature in Fig.~\ref{ThagfitAll} for several cases where
$(g\neq 0, \kappa=1)$ and for resolution $K \in [11,16].$
As a check we note that for the cases  $g=0$ and $g=0.1$ one expects the
corresponding
extrapolated Hagedorn temperatures to be comparable.

For values of $g\in\{0,4.0\}$ that are considered here, the upper
bound
for temperatures is set by the Hagedorn temperature of the free theory,
$T^{\infty}_{H}(0)\approx 0.52 \kappa$ and thus we calculate the
thermodynamic properties of the theory
below
this limit. From Fig.~\ref{ThagfitAll} we glean that
$T^{\infty}_{H}(g)$ grows with the coupling. For larger values of
$g$ we may therefore access a significantly larger region
in $T.$ However, we will leave the discussion of these cases for future work.
%

%%%%%%%%%%%%%%%%%%%%%%%%%%%%%%%%%%%%%%%%%%%%%%%%%%%%%%%%%%%%%%%%%%
\section{Finite  temperature results in 1+1 dimensions}
\label{sec:Finite temp}

%%%%%%%%%%%%%%%%%%%%%%%%%%%%%%%%%%%%%%%%%%%%%%%%%%%%%%%%%%%%%%%%%%
\subsection{The free energy}
\label{subSect:BasicFormul}

%%%%%%%%%%%%%%%%%%%%%%%%%%%%%%%%%%%%%%%%%%%%%%%%%%%%%%%%%%%%%%%%%%

We now introduce the basic formulation necessary for our finite
temperature calculations. Note that our approach here deviates slightly
from our earlier work~\cite{Hiller:2004ft}, mainly in the way the free energy
and the mass squared are normalized. We
consider a system with constant volume which is in contact with a heat
bath of constant temperature. The free energy,\footnote{Although our focus is
the free energy, $\mathcal{F}$, it is straightforward to
use our numerical techniques to calculate other thermodynamic
functions, such as the internal
energy $\mathcal{E}(T,V)=T^2\left(\frac{\partial{\ln \mathcal{Z}}}
{\partial{T}}\right)_\mathrm{V}$,the entropy
$S=(\mathcal{E}-\mathcal{F})/T$, and the heat capacity
$\mathcal{C}_{\mathrm{V}}(T,V)=\left(\frac{\partial{{\mathcal{
E}}}}{\partial{T}}\right)_{\mathrm{V}},$ cf.~\cite{Hiller:2004ft}.} in units 
where $k_{B}=1$, is given by
\begin{equation}
\label{eqn:FEdef}
\mathcal{F}(T,V)=-T\ln\mathcal{Z}.
\end{equation}
For the large-$N_c$ system at hand, the thermodynamics is
described by a canonical ensemble of
non-interacting glueball and meson-like states. The bound states of the theory
constitute a supersymmetric two-dimensional free gas.
The canonical free energy for such a gas in $D$ space-time dimensions is
given by
\begin{align}
\label{eqn:canfree}
\mathcal{F}_{b}&= T\sum_{n=1}^{\infty} \int \frac{d^{D-1}
\mathbf{p}}{(2\pi )^{D-1}} \ln \bigl(
1-\e^{-\frac{1}{T}\sqrt{\mathbf{p}^{2}+M_{n}^{2}}} \bigr) \\
\mathcal{F}_{f}&= -T\sum_{n=1}^{\infty}\int \frac{d^{D-1}
\mathbf{p}}{(2\pi )^{D-1}} \ln \bigl(
1+\e^{-\frac{1}{T}\sqrt{\mathbf{p}^{2}+M_{n}^{2}}} \bigr)\label{ferm}
\end{align}
for bosons and fermions, respectively,
where $M_{n}^{2}$ in the expression for $\mathcal{F}_{b}$ ($\mathcal{F}_{b}$) 
is the invariant bosonic (fermionic) mass spectrum. 
The integral is performed by expanding  the logarithm and 
using the integral representation of the
modified Bessel function of the second kind $\mathbf{K}_{\nu}(x)$, to find
\begin{align}
\label{eqn:FEbos}
\mathcal{F}_{b,M}&=-2V_{D-1}\sum_{n=1}^{\infty}\sum_{q=1}^{\infty}
\biggl(\frac{M_{n}T}{2
\pi q}\biggr)^{D/2}\mathbf{K}_{D/2}\bigl(\frac{qM_n}{T}\bigr),\\
\label{eqn:FEfer}
\mathcal{F}_{f,M}&=-2V_{D-1}\sum_{n=1}^{\infty}
\sum_{q=1}^{\infty}(-1)^{q+1}
\biggl(\frac{M_{n}T}{2
\pi q}\biggr)^{D/2}\mathbf{K}_{D/2}\bigl(\frac{qM_n}{T}\bigr).
\end{align}
Here $V_{D-1}$ is the volume in $D-1$ space dimensions.
The total free energy is obtained by adding these two expressions.
Because the spectrum is supersymmetric, the sums over masses
$M_n$ traverse the same spectrum, and
the total free energy takes the form
\begin{align}
\label{eqn:FEtot}
\mathcal{F}_{tot,M}&=-4V_{D-1}\sum_{n=1}^{\infty}\sum_{q=0}^{\infty}
\biggl(\frac{M_{n}T}{2\pi
(2q+1)}\biggr)^{D/2}\mathbf{K}_{D/2}\biggl(\frac{(2q+1)M_n}{T}\biggr).
\end{align}
For our calculations we use a rescaled form of Eq.~$(\ref{eqn:FEtot})$,
with $D=2$
\begin{align}
\label{eqn:FEtotD2}
\tilde{\mathcal{F}}\equiv
-\frac{\mathcal{F}_{tot,M}}{{4(K-1)L}}&=\frac{1}{2(K-1)}\sum_{n=1}^{\infty}
\sum_{q=0}^{\infty}\frac{M_{n}T}{
(2q+1)\pi}\mathbf{K}_{1}\biggl(\frac{(2q+1)M_n}{T}\biggr)\non\\
&=\frac{1}{2(K-1)}\sum_{k=1}^{\infty}
\sum_{q=0}^{\infty}\frac{d_{k}M_{k}T}{
(2q+1)\pi}\mathbf{K}_{1}\biggl(\frac{(2q+1)M_k}{T}\biggr).
\end{align}
In the last line we introduced the factor $d_{k}$ which counts degeneracies of
mass eigenvalues. This equation is most efficient in the present calculation,
because it expresses the free energy solely as a function of the numerically
evaluated bound-state masses $M_k$.
We have chosen to normalize by $(-{4(K-1)L})^{-1}$,
since $2 n_{0}\equiv 4(K-1)$ is the total number of massless
states of the free, massive theory (with $g=0$ and $\kappa=1$).
In practice, we can truncate the sum over Bessel functions at $q=10$ due to
fast convergence. Obviously, the sum over states is finite at any finite $K$.

The contribution of the massless states to the free energy
in $D=2$ dimensions can be calculated analytically to be
\begin{align}
\label{eqn:nomassFE}
\mathcal{F}_{b,0}&= -\frac{n_{b}L}{\pi}\int_{0}^{\infty}dp_{o} 
           \frac{p_{o}}{\e^{\, p_{o}/T}-1}=-n_{0}L T^{2}\frac{\pi}{6} \\
\mathcal{F}_{f,0}&=- \frac{n_{f}L}{\pi}\int_{0}^{\infty}dp_{o}
\frac{p_{o}}{\e^{\, p_{o}/T}+1}=-n_{0}L T^{2}\frac{\pi}{12} \\
\tilde{\mathcal{F}}_{0}&=n_{0} T^{2}\frac{\pi}{16(K-1)}.
\end{align}
for bosons, fermions, and the contribution to the rescaled total,
respectively.
Thus one may separate this contribution from the rest of
Eq.~$(\ref{eqn:FEtotD2}).$

Finally, the sum over the states is replaced by an integral over the density of
states, and Eq.~$(\ref{eqn:FEtotD2})$ becomes
\begin{align}
\label{eqn:FEusedforCalculation}
\tilde{\mathcal{F}}&=\frac{1}{2(K-1)}\int^{M^{2}}
 d\bar{M}^{2}\rho(\bar{M}^{2})\biggl\{\sum_{q=0}^{\infty}\frac{\bar{M} T}{(2q+1)
\pi }\mathbf{K}_{1}\biggl(\frac{(2q+1)\bar{M}}{T}\biggr)\biggr\},
\end{align}
where $\rho(M^2)=\rho_{b}(M^2)=\rho_{f}(M^2)$ for supersymmetric systems.
The $\cal T$ symmetry splits the bosonic and fermionic sectors into halves.
Thus when calculating the free energy, or other thermodynamic properties,
we can write
\begin{align}
\label{eqn:FEtrick}
\tilde{\mathcal{F}}&=
(\tilde{\mathcal{F}}_{b}+\tilde{\mathcal{F}}_{f})_{\mathcal{ T}^{+}}
+(\tilde{\mathcal{F}}_{b}+\tilde{\mathcal{F}}_{f})_{\mathcal{ T}^{-}}
=\tilde{\mathcal{F}}_{tot\,\mathcal{T}^{+}}+
\tilde{\mathcal{F}}_{tot\,\mathcal{T}^{-}}.
\end{align}
%

%%%%%%%%%%%%%%%%%%%%%%%%%%%%%%%%%%%%%%%%%%%%%%%%%%%%%%%%%%%%%%%%%%
\subsection{An analytic result: The free energy for the free theory}
\label{subSec:analytical}
%%%%%%%%%%%%%%%%%%%%%%%%%%%%%%%%%%%%%%%%%%%%%%%%%%%%%%%%%%%%%%%%%%

Let us start by exploring the free, massive theory ($g=0$, $\kappa=1,$)
which can be solved analytically.
We will compare the contributions of the meson and glueball sectors
to the free energy. In particular, using the
free-meson sector we can check the validity of our approach
to replace the sums over discrete spectra with a density of states function
and check how good our numerical results are compared to an
analytic calculation.

\begin{figure}
  \begin{center}
     \hspace*{-0.5cm}
     \begin{tabular}{cc}
   % \hline\hline\\[3mm]
      \resizebox{75mm}{!}{\includegraphics{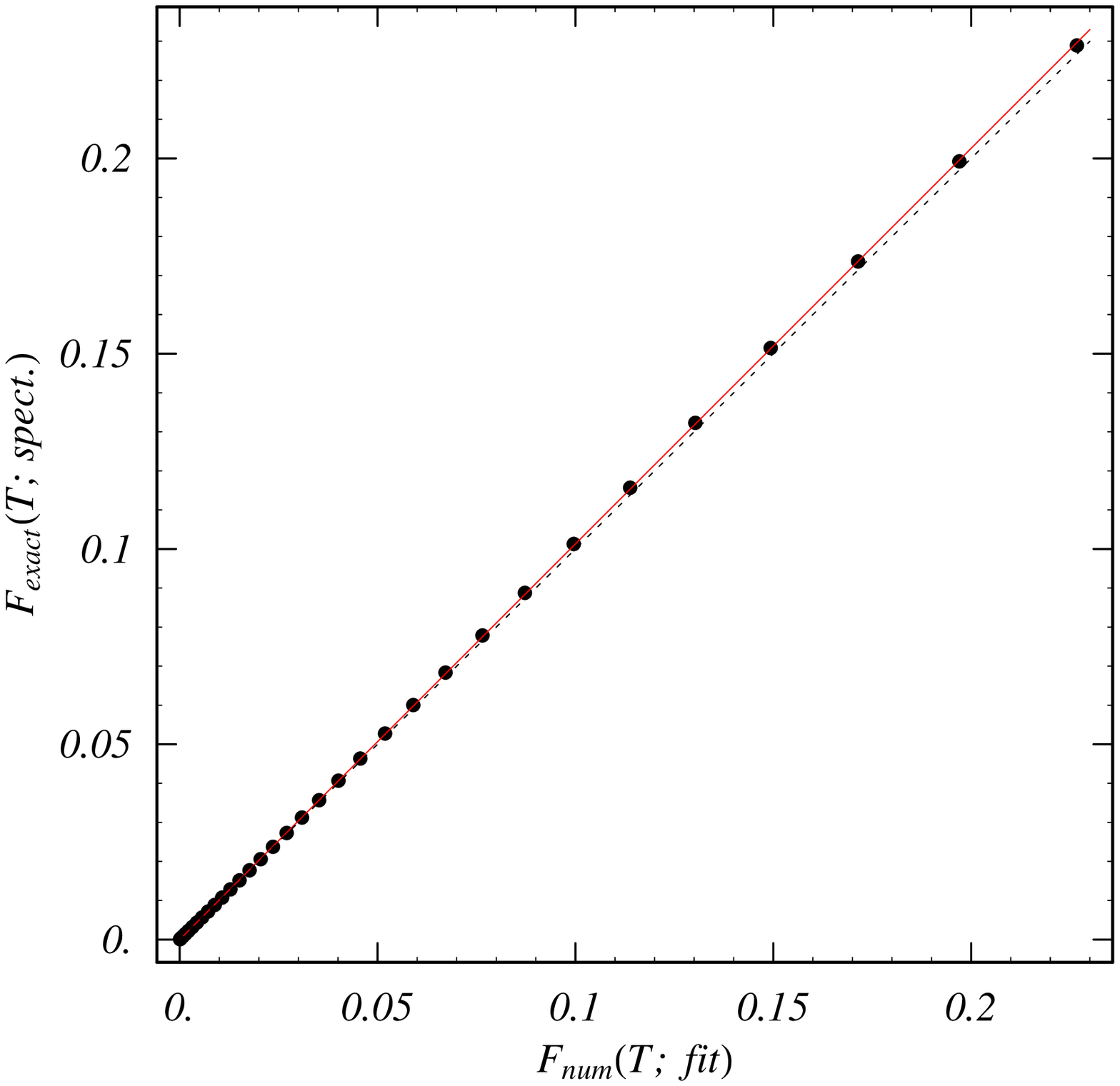}} &
    %
      %\resizebox{75mm}{!}{\includegraphics{MlessK16bOUT.eps}}
      %\resizebox{75mm}{!}{\includegraphics{g00FEvsKinvBOUT.eps}}
      \resizebox{75mm}{!}{\includegraphics{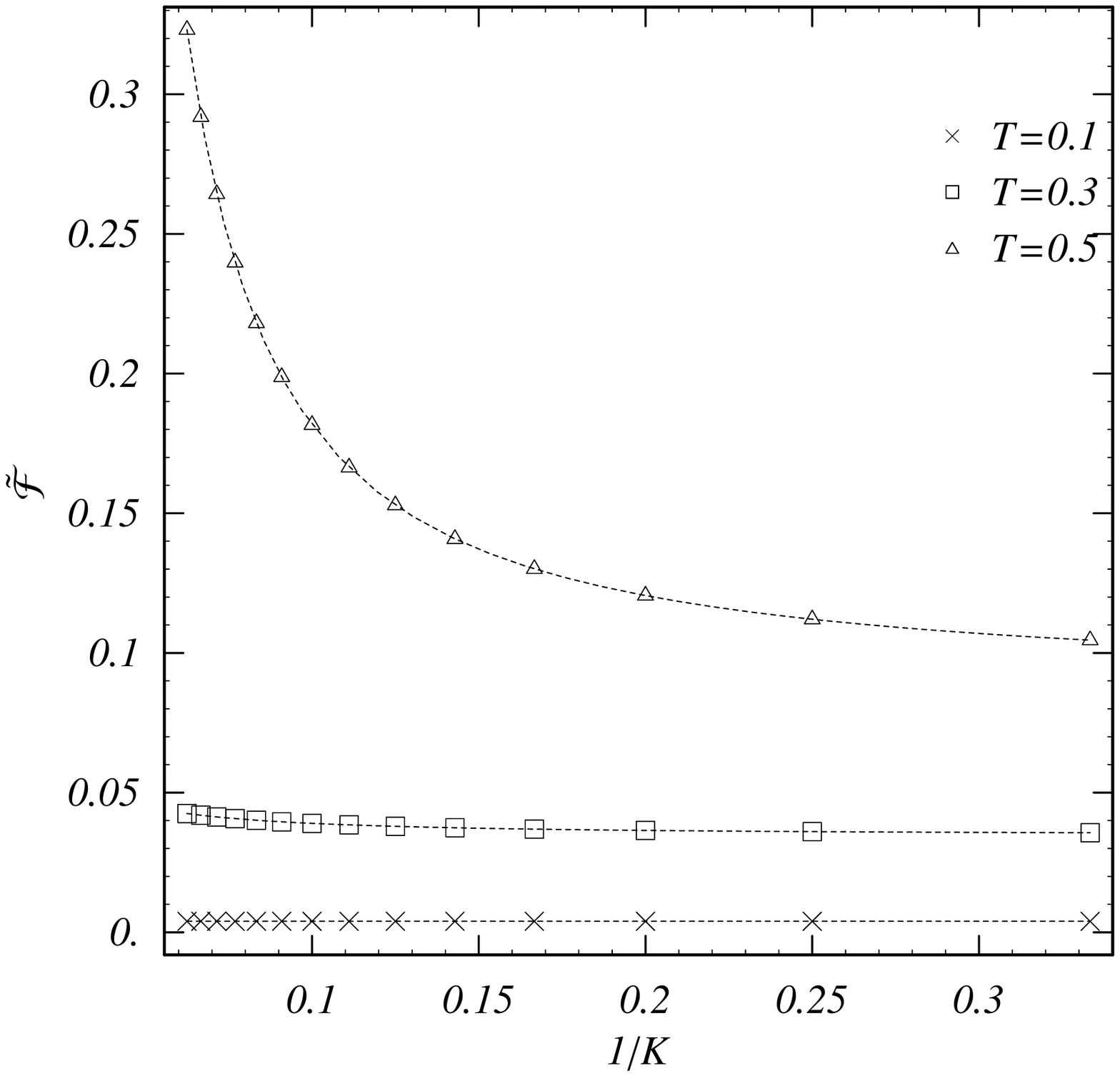}}
      \\
     \qquad (a)&\qquad (b)\\[3mm]
    %  \hline\hline
    \end{tabular}
\vspace{-8mm}
    \caption{The free energy $\tilde{\mathcal{F}}$ for the
    free theory ($g=0$, $\kappa=1$) as
    (a) compared between the analytic and numerical results
    and (b) a function of $1/K$.
    In (a) the temperature ranges from $0.015$ to $0.5$,
in units where $\kappa=1$, by steps of $\Delta T=0.015\kappa$.
The solid line represents an exact match, the dotted line the actual
relation between $\tilde{\mathcal{F}}_{spect.}$ and $\tilde{\mathcal{F}}_{fit}$.
In (b) the temperatures are $T=0.1\kappa$ (crosses), $0.3\kappa$ (boxes), 
$0.5\kappa$ (triangles). }
\label{g0Comparison}
\end{center}
\end{figure}

First, we compare the free energy 
obtained by the means of the analytic method outlined above, $\tilde{\mathcal{F}}_{spect.}$,
to the free energy  extracted from the numerical
approach,  $\tilde{\mathcal{F}}_{fit}$, using the DoS. The graph presented in
Fig.~\ref{g0Comparison}(a) compares analytic and numerical results at
$K=16$ for temperatures $0.015\le T\le 0.5,$ in units where $\kappa=1.$
We deduce from the plot that the agreement
is within $1\%$, which is a typical result. Cutoff dependence is very mild:
$\tilde{\mathcal{F}}_{spect.}/\tilde{\mathcal{F}}_{fit}(K=13) = 1.012$, while
$\tilde{\mathcal{F}}_{spect.}/\tilde{\mathcal{F}}_{fit}(K=16) = 1.015$.
%%
%\begin{figure}[H]
% \begin{center}
%  \resizebox{100mm}{!}{\includegraphics{g00FEvsKinvBOUT.eps}}
%      \\[-2mm]
%    \caption{The $g=0$ free energy $\tilde{\mathcal{F}}$ versus
%     $1/K$ at temperatures $T=0.1\kappa,0.3\kappa, 0.5\kappa.$ }
%\label{fig:g00vsKinv}
%  \end{center}
%\end{figure}
%
In  Fig.~\ref{g0Comparison}(b) we show the free energy of the free, massive
theory as a function
of the inverse resolution at different temperatures.
It seems that the free energy $\tilde{\mathcal{F}}$ converges for low and
intermediate temperatures, while it diverges for temperatures close to the
Hagedorn transition in the continuum limit, as expected.

We can extract the contributions to the free energy of different parts
of the spectrum by using Eq.~$(\ref{eqn:FEtotD2})$. It is interesting
to compare the contributions of the two non-interacting sectors of the
theory. At
temperature $T=0.5 \kappa$ and $K=5$, the three-parton meson states
contribute
$1.9\times 10^{-2}\kappa^2$ to the free energy, while the
corresponding $K=5,$ three-parton glueball
state contributes only $3.9\times 10^{-4}\kappa^2.$
Results at different temperatures are listed in
Table~\ref{tab:tableFreeMsGl}. The free energy
$\tilde{\mathcal{F}}_{mesons}$ associated with the free meson sector
dominates the corresponding  $\tilde{\mathcal{F}}_{glueball}$.
This is a consequence of the fact that the mesonic spectrum  has
$4(K-1)$ massless states that contribute $\frac{\pi}{8}T^2$
to $\tilde{\mathcal{F}}$ and that the meson sector possesses
many more states than
the glueball sector, especially low-mass states which are important
in the present calculation.

\begin{table}[h]
\begin{center}
\begin{tabular}{ccccccccc}
\hline\hline \\[-4.0mm]
$T/\kappa$& 0.20 & 0.25 & 0.30 & 0.35 & 0.40 & 0.45 & 0.50 & 0.55\\
\hline\\[-4mm]
$\tilde{\mathcal{F}}_{meson}/\kappa^2$ & 0.009 & 0.016 & 0.026 & 0.040 & 0.059 & 0.089 & 0.133 & 0.199 \\
$\tilde{\mathcal{F}}_{glueball}/\kappa^2$ & 0.000 & 0.000 & 0.001 & 0.004 & 0.011 & 0.026 & 0.053 & 0.101 \\[0.5mm]
\hline \hline
\end{tabular}
\caption{The free energy as a function of the temperature in the meson and the
glueball sectors in the free theory.}
\label{tab:tableFreeMsGl}
\end{center}
\end{table}
%

%%%%%%%%%%%%%%%%%%%%%%%%%%%%%%%%%%%%%%%%%%%%%%%%%%%%%%%%%%%%%%%%%%%%%%%%%
\subsection{Numerical results for nonzero coupling}
\label{subSec:NumResultsFE}
%%%%%%%%%%%%%%%%%%%%%%%%%%%%%%%%%%%%%%%%%%%%%%%%%%%%%%%%%%%%%%%%%%%%%%%%%
We now discuss numerical results for the free energy at finite values of
$g.$ These results were obtained by applying the DoS method, i.e. replacing
the sum over states by an integral over the bound-state masses times
a density of states computed from a fit to the numerically obtained CDF.
First, we discuss the temperature dependence
and then the coupling dependence of the free energy
$\tilde{\mathcal{F}},$ for temperatures that lie below the
zero-coupling Hagedorn temperature $T_{H}(g=0)\approx 0.52\kappa.$

%%%%%%%%%%%%%%%%%%%%%%%%%%%%%%%%%%%%%%%%%%%%%%%%%%%%%%%%%%%%%%%%%%%%%%%%%
\subsubsection{Temperature dependence of the free energy}
\label{subsubSec:TempDep}
%%%%%%%%%%%%%%%%%%%%%%%%%%%%%%%%%%%%%%%%%%%%%%%%%%%%%%%%%%%%%%%%%%%%%%%%%

The DoS method works also for the interacting theory ($g>0$).
Using either the discrete spectrum approach (sum over the states;
$(\ref{eqn:FEtotD2})$), or the DoS fit to the spectrum yields good agreement,
at least for weak to medium couplings. The disagreement
between the two approaches  is typically below 1\%,
as seen in Table~\ref{tab:tableFreeEnK14g01}.
This table also presents contributions
to the free energy from about a thousand states up to $M^2=9.10283\kappa^2,$
which belong to the $\cal T$-even sector at resolution $K=14$ and weak
coupling $g=0.1.$
We also consider, in Table~\ref{tab:tableFreeEnK16g40}, the free energy
at resolution $K=16$ for large coupling $g=4.0.$

\begin{figure}
  \begin{center}
     \hspace*{-0.5cm}
     \begin{tabular}{cc}
   % \hline\hline\\[3mm]
   \resizebox{75.5mm}{!}{\includegraphics{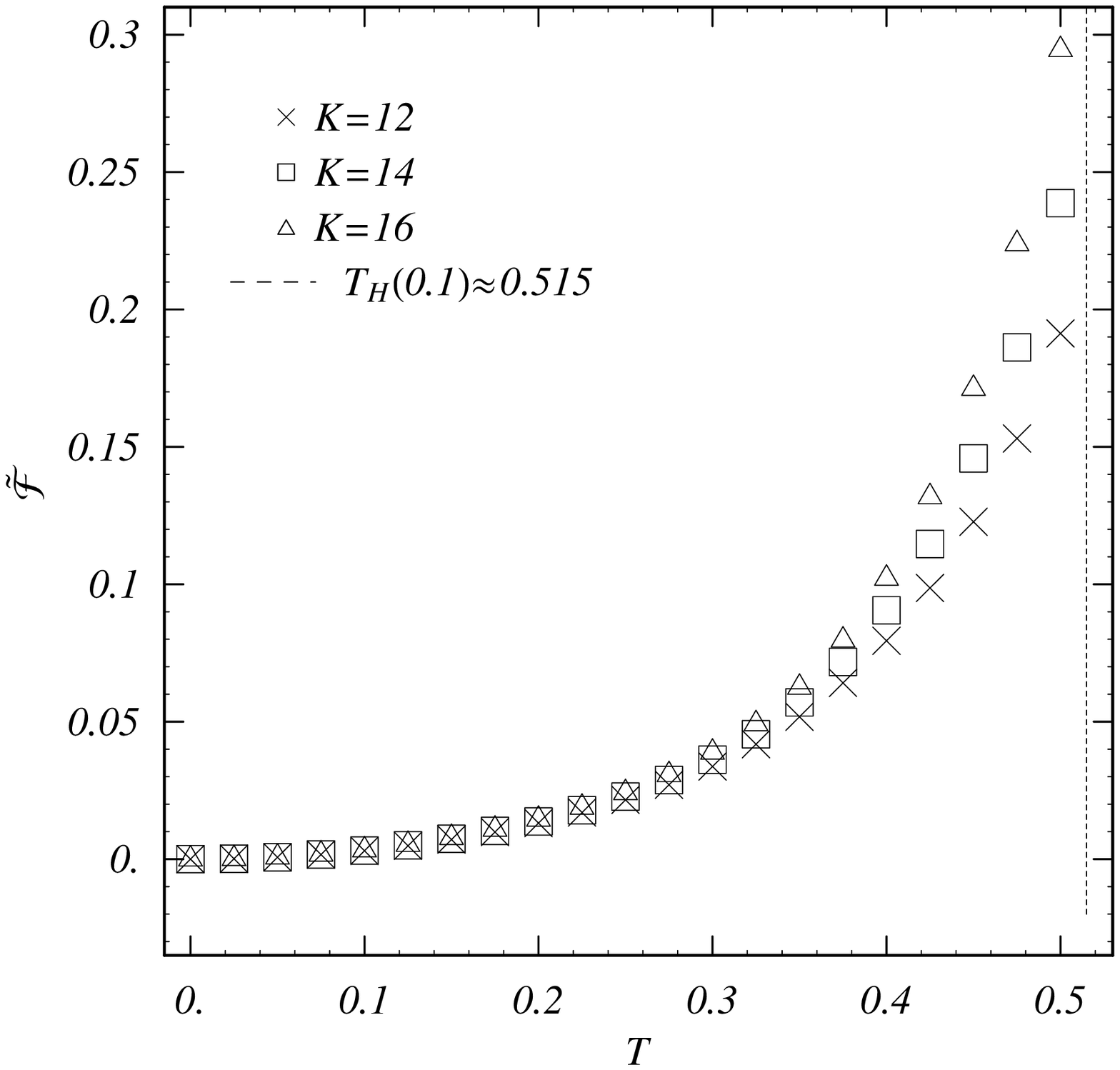}} &
     % \resizebox{75.5mm}{!}{\includegraphics{FEvsTaug01unscaledOUT.eps}} &
    %
      \resizebox{76mm}{!}{\includegraphics{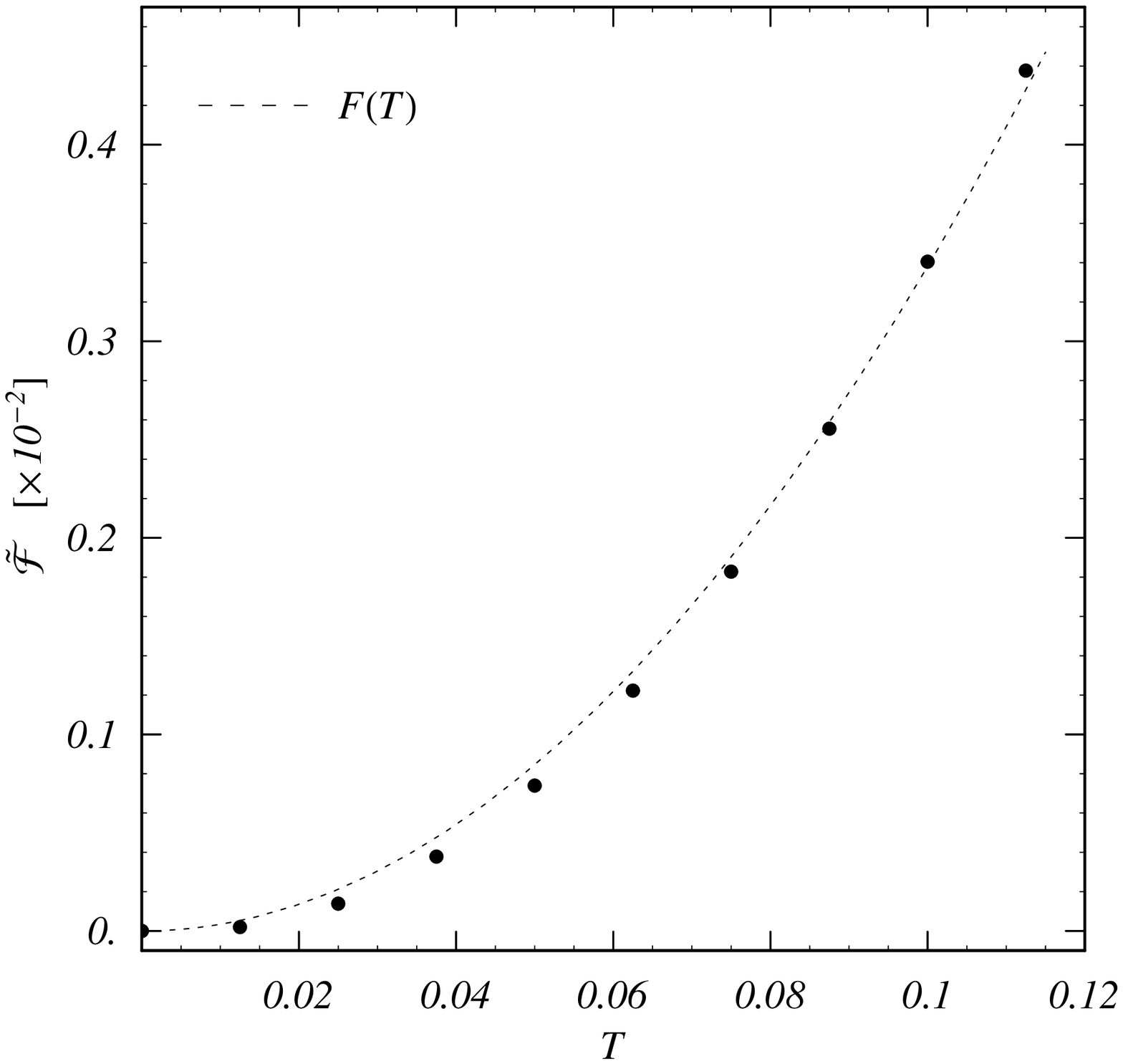}}\\
     \qquad (a)&\qquad (b)\\[3mm]
    %  \hline\hline
    \end{tabular}
\vspace{-6mm}
    \caption{Free energy $\tilde{\mathcal{F}}$ at weak coupling ($g=0.1$)  as a function of $T$ at $K=12$ (crosses), 14 (boxes), 16 (triangles) 
for (a) all temperatures $T<T_H$ (dashed vertical line) and (b) low
temperatures with a quadratic fit $F(T)=\alpha_{w} T^2$.}
\label{FEvsTg4}
\end{center}
\end{figure}

\begin{figure}
  \begin{center}
     \hspace*{-0.5cm}
     \begin{tabular}{cc}
   % \hline\hline\\[3mm]
   \resizebox{75.5mm}{!}{\includegraphics{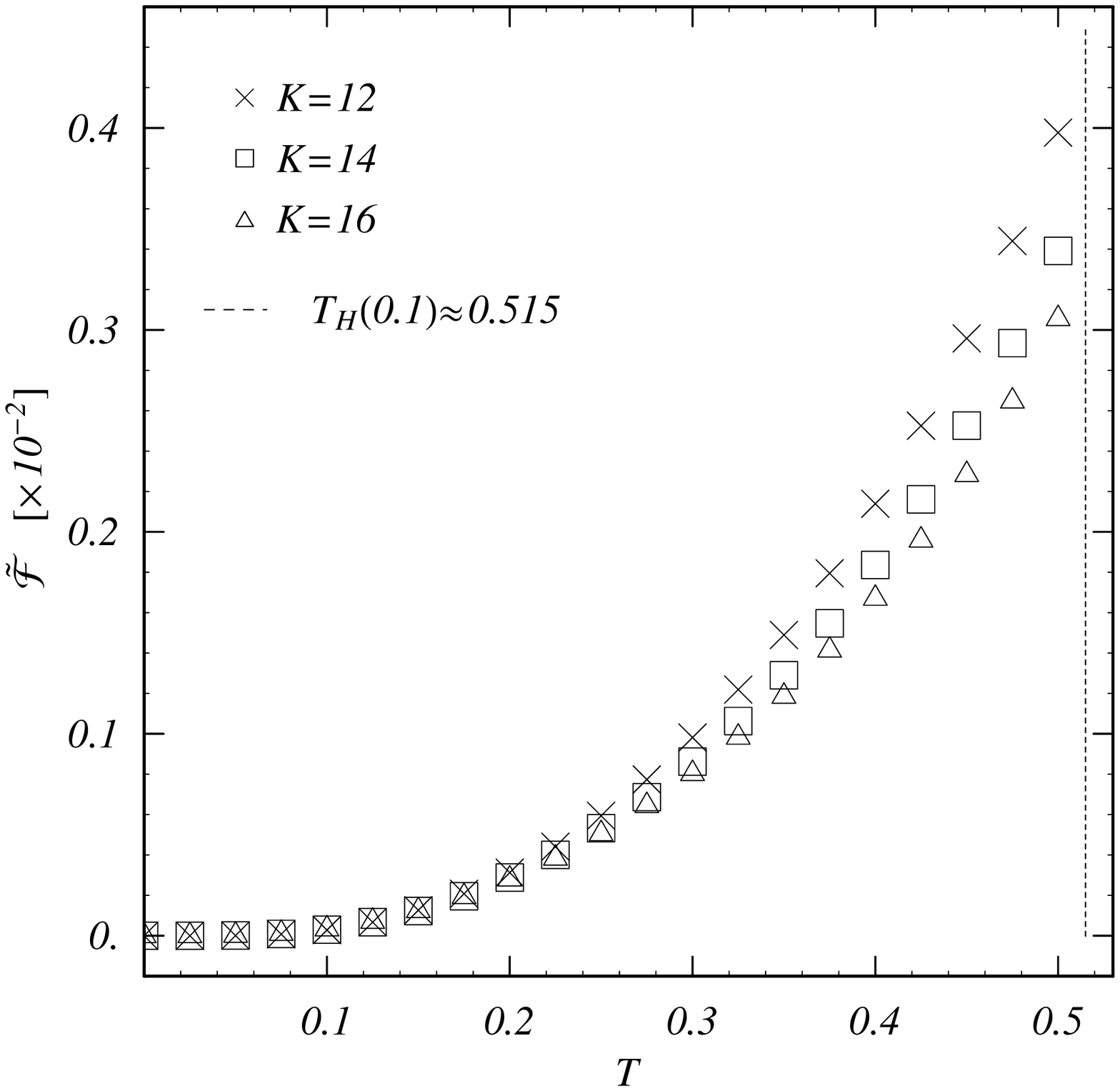}} &
   %   \resizebox{75.5mm}{!}{\includegraphics{FEvsTaug40unscaledOUT.eps}} &
    %
      \resizebox{78mm}{!}{\includegraphics{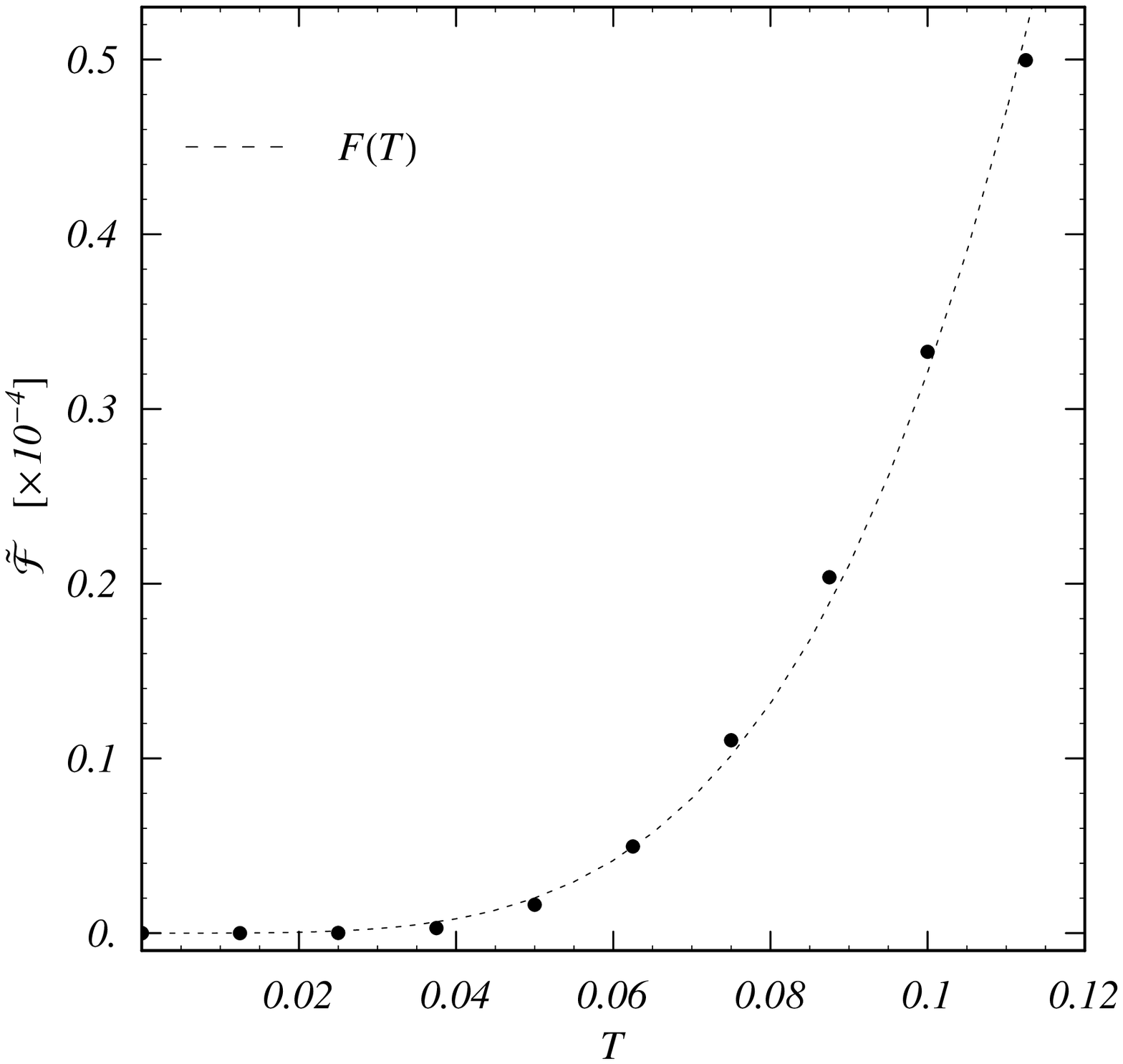}}\\
     \qquad (a)&\qquad (b)\\[3mm]
    %  \hline\hline
    \end{tabular}
\vspace{-6mm}
   \caption{Same as Fig.~\ref{FEvsTg4}, but for strong coupling ($g=4.0$). In 
(b) the low temperature behavior is described by a quartic fit $F(T)=\alpha_{s} T^4$.
%Free energy $\tilde{\mathcal{F}}$ at strong
%coupling ($g=4.0$)  as a
% function of $T$ at $K=16$ for (a) all temperatures $T<T_H$ and (b) low
%temperatures with a quadratic fit $F(T)=\alpha_{w} T^2$.
}
\label{FEvsg4fits}
\end{center}
\end{figure}

It is verified  from these
tables that, at low temperatures, the major contribution to the free
energy comes from the low-lying states, i.e. the states below the
mass gap. In the case of Table~\ref{tab:tableFreeEnK16g40},
we have ten such states. As we increase the temperature,
more states from above the mass gap will contribute significantly.
The results are shown in Fig.~\ref{FEvsTg4} for small coupling, $g=0.1$,
and in Fig.~\ref{FEvsg4fits} for large coupling, $g=4.0.$ For
$K=16$ we find that, at a coupling value of $g=0.1$, the
free energy is quadratic in $T.$ This is expected since for low
values of temperature the quasi-massless modes dominate and their
contribution should be similar to the massless states of the
free theory. The fourth column of
Table~\ref{tab:tableFreeEnK14g01}, which considers contributions
from states below the mass gap (e.g., $M^2 \in [0.0001,0.00113]\kappa^2$),
shows clearly that these states dominate.

At large coupling ($g=4.0$) we obtain free energies
that are about a hundred times smaller than the available weak coupling
free energies. This is less than the free
energy  $\tilde{\mathcal{F}}_{0}$ that would be contributed from
a single massless state!
However, it can be justified from the spectrum at these temperatures.
Recall Figs.~\ref{CDFandDoSE} and \ref{CDFandDoSF}, especially
those plots
depicting the  spectrum below the mass gap. For the example shown
in Table~\ref{tab:tableFreeEnK16g40}, the dominant symmetry sector is
$\cal T$-even, the sector which includes the lightest state. 
We denote the contribution of this lightest state by $\mathcal{F}_1$.
This single state accounts for
almost all the contribution to the free energy for $0\leq T\leq 0.5\kappa$.
The latter result is close to the
contribution $\tilde{\mathcal{F}}_{0}$, that would be made by
a single, exactly massless
mode, suggesting that this very light state may be approximated with a
massless state. Finally, at this coupling and $K=16$, we
show in Fig.~\ref{FEvsg4fits}(b) the behavior of the free energy at
low $T,$ which appears to be quartic.
%

%1total
\begin{table}[t]
\begin{center}
\begin{tabular}{ccccc}
\hline \hline\\[-2.0mm]
$T$ & $\tilde{\mathcal{F}}_{spect.}$ & $\tilde{\mathcal{F}}_{fit}$
&$\tilde{\mathcal{F}}^{<}_{fit}$&$\tilde{\mathcal{F}}_{0}$
\\[1.0mm]
$[\kappa]$& $[10^{-2} \kappa^2]$&$[10^{-2} \kappa^2]$&$[10^{-2} \kappa^2]$& $[10^{-2} \kappa^2]$\\[2.0mm]
\hline\\[-0.5mm]
0 & 0 & 0 & 0 &0\\ 
0.025 & 0.00658 & 0.00654 & 0.00654& 0.0009\\ 
0.05 & 0.03459 & 0.03448 & 0.03448 &0.0038\\ 
0.075 & 0.08522 & 0.08499 & 0.08499 &0.0085\\ 
0.1 & 0.15852 & 0.15810 & 0.15809 &0.0151\\ 
$\vdots$ & $\vdots$ & $\vdots$ & $\vdots$ & $\vdots$ \\ 
0.225 & 0.89782 & 0.89487 & 0.86245 &0.0765\\ 
0.25 & 1.14656 & 1.14269 & 1.07109 &0.0944\\ 
0.275 & 1.44823 & 1.44337 & 1.30232 &0.1142\\ 
0.3 & 1.81661 & 1.81078 & 1.55613 &0.1359\\ 
$\vdots$ & $\vdots$ & $\vdots$ & $\vdots$ & $\vdots$ \\ 
0.425 & 5.33793 & 5.33113 & 3.16396 &0.2728\\ 
0.45 & 6.54702 & 6.54176 & 3.55327 &0.3059\\ 
0.475 & 7.98870 & 7.98597 & 3.96515 &0.3408\\ 
0.5 & 9.69247 & 9.69351 & 4.39962 &0.3776\\ 
\hline \hline
\end{tabular}
\caption{Free energy as a function of temperature $T$ at $K=14$ in the
$\cal T$-even sector for weak coupling, $g=0.1$:
$\tilde{\mathcal{F}}_{spect.}$ is obtained by
summing over the eigenvalues in the interval
$M^2 \,\in\, [0.00001,9.10283]\kappa^2$;  $\tilde{\mathcal{F}}_{fit}$
is obtained by the DoS method described in Sec.~\ref{EstimatingDoS}.
$\tilde{\mathcal{F}}^<_{fit}$ and
 $\tilde{\mathcal{F}}_{0}$ are the contributions to the latter of the
states below the mass gap  (i.e., $M^2<0.00113\kappa^2$)
and of a single supersymmetric massless state, respectively.}
\label{tab:tableFreeEnK14g01}
\end{center}
\end{table}

%
%2quasi
\begin{table}[t]
\begin{center}
\begin{tabular}{ccccc}
\hline \hline\\[-2.0mm]
$T$ & $\tilde{\mathcal{F}}$ & $\tilde{\mathcal{F}}_{\mathcal{T}^+}$
&$\tilde{\mathcal{F}}_{1}$& $\tilde{\mathcal{F}}_{0}$
\\[1.0mm]
$[\kappa]$& $[10^{-2} \kappa^2]$&$[10^{-2} \kappa^2]$&$[10^{-2} \kappa^2]$&$[10^{-2} \kappa^2]$\\[2.0mm]
\hline\\[-0.5mm]
         0 &     0 &     0 &     0 &     0 \\
     0.025 &     0 &     0 &     0 &     0.0008 \\
      0.05 &     0.0002 &     0.0002 &     0.0002 &     0.0033 \\
     0.075 &     0.0011 &     0.0011 &     0.0011 &     0.0074 \\
       0.1 &     0.0033 &     0.0033 &     0.0033 &     0.0131 \\
     $\vdots$ &     $\vdots$ &     $\vdots$ &     $\vdots$ &     $\vdots$ \\
     0.225 &     0.0384 &     0.0383 &     0.0383 &     0.0663 \\
      0.25 &     0.0505 &     0.0503 &     0.0503 &     0.0818 \\
     0.275 &     0.0644 &     0.0641 &     0.0639 &     0.0990 \\
       0.3 &     0.0803 &     0.0796 &     0.0793 &     0.1178 \\
     $\vdots$ &     $\vdots$ &     $\vdots$ &     $\vdots$ &     $\vdots$ \\
     0.425 &     0.1959 &     0.1867 &     0.1813 &     0.2364 \\
      0.45 &     0.2283 &     0.2149 &     0.2068 &     0.2651 \\
     0.475 &     0.2647 &     0.2458 &     0.2340 &     0.2953 \\
       0.5 &     0.3056 &     0.2795 &     0.2628 &     0.3272 \\
\hline \hline
\end{tabular}
\caption{Results for free energy as a function of temperature $T$ at $K=16$ and strong coupling, $g=4.0.$ $\tilde{\mathcal{F}}$ corresponds to the overall free energy including both symmetry sectors. The third column shows the overall contribution of the $\cal T$-even sector, the fourth the contribution from the single nearly massless state, $M^2\approx 0.0362\kappa^2,$ and the last column is the contribution that a supersymmetric massless state
would make, if it were present.}
\label{tab:tableFreeEnK16g40}
\end{center}
\end{table}
%
%a
%
%
%b
%
\subsubsection{Coupling dependence of the free energy}
\label{subsubSec:CouplDep}
\begin{figure}
  \begin{center}
     \hspace*{-0.5cm}
     \begin{tabular}{cc}
   % \hline\hline\\[3mm]
 \resizebox{75mm}{!}{\includegraphics{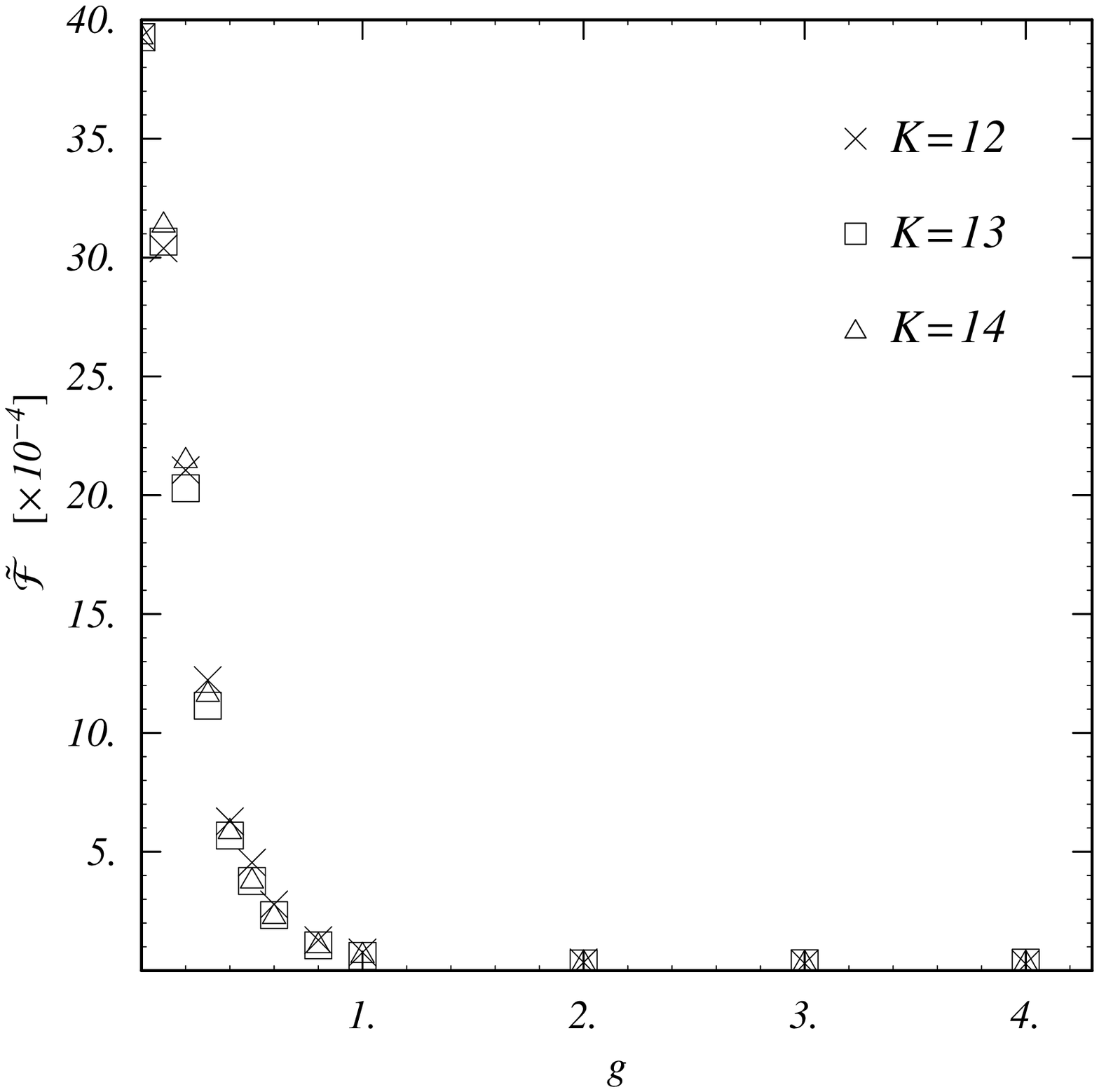}} &
% \resizebox{75mm}{!}{\includegraphics{gFEatfixedK1213tau01OUT.eps}} &
    %
\resizebox{75mm}{!}{\includegraphics{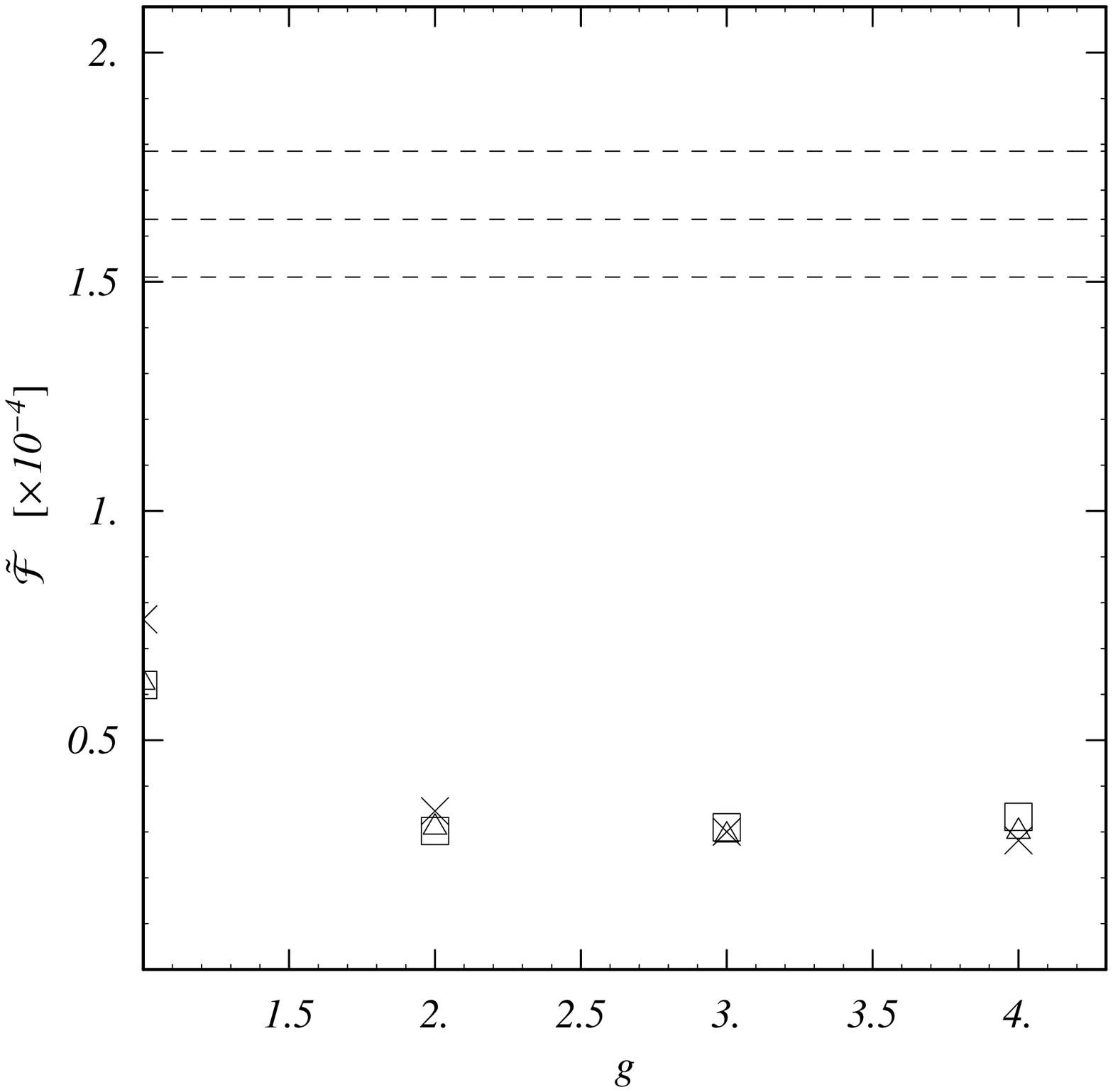}}\\
% \resizebox{75mm}{!}{\includegraphics{gFEatfixedK1213tau01modOUT.eps}}\\
     \qquad (a)&\qquad (b)\\[3mm]
    %  \hline\hline
    \end{tabular}
\vspace{-6mm}
    \caption{The free energy as a function of the coupling $g$ at temperature
 $T=0.1\kappa$ (with $\kappa=1$) and for resolutions $K=12$ (crosses), 13 (boxes), 14 (triangles) with two different
 vertical scales. In (b) we see along with the data the contribution to the free
energy that would be made by a pair of exactly massless superpartners. For fairly large
values of $g$ and at this temperature, the overall free energy is small 
compared to  the contribution of a single pair of massless states. This is expected at this coupling region
because the masses are very large  and are suppressed by the modified Bessel function, $\mathbf{K}_{1}(x)$.}
\label{FEvsgA}
  \end{center}
\end{figure}

\begin{figure}
  \begin{center}
     \hspace*{-0.5cm}
     \begin{tabular}{cc}
   % \hline\hline\\[3mm]
\resizebox{75mm}{!}{\includegraphics{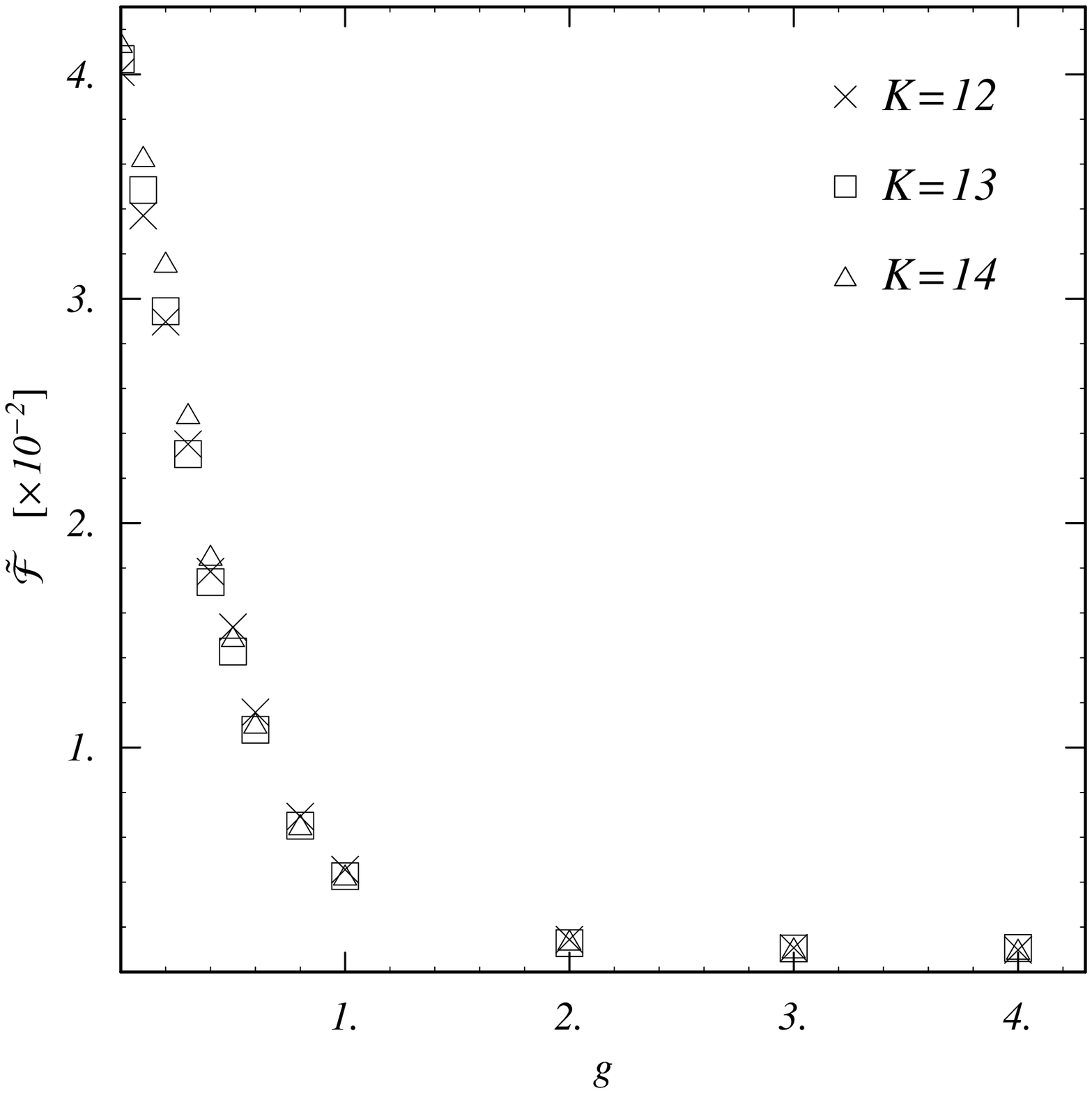}} &
% \resizebox{75.5mm}{!}{\includegraphics{gFEatfixedK1213tau03OUT.eps}} &
    %
\resizebox{77mm}{!}{\includegraphics{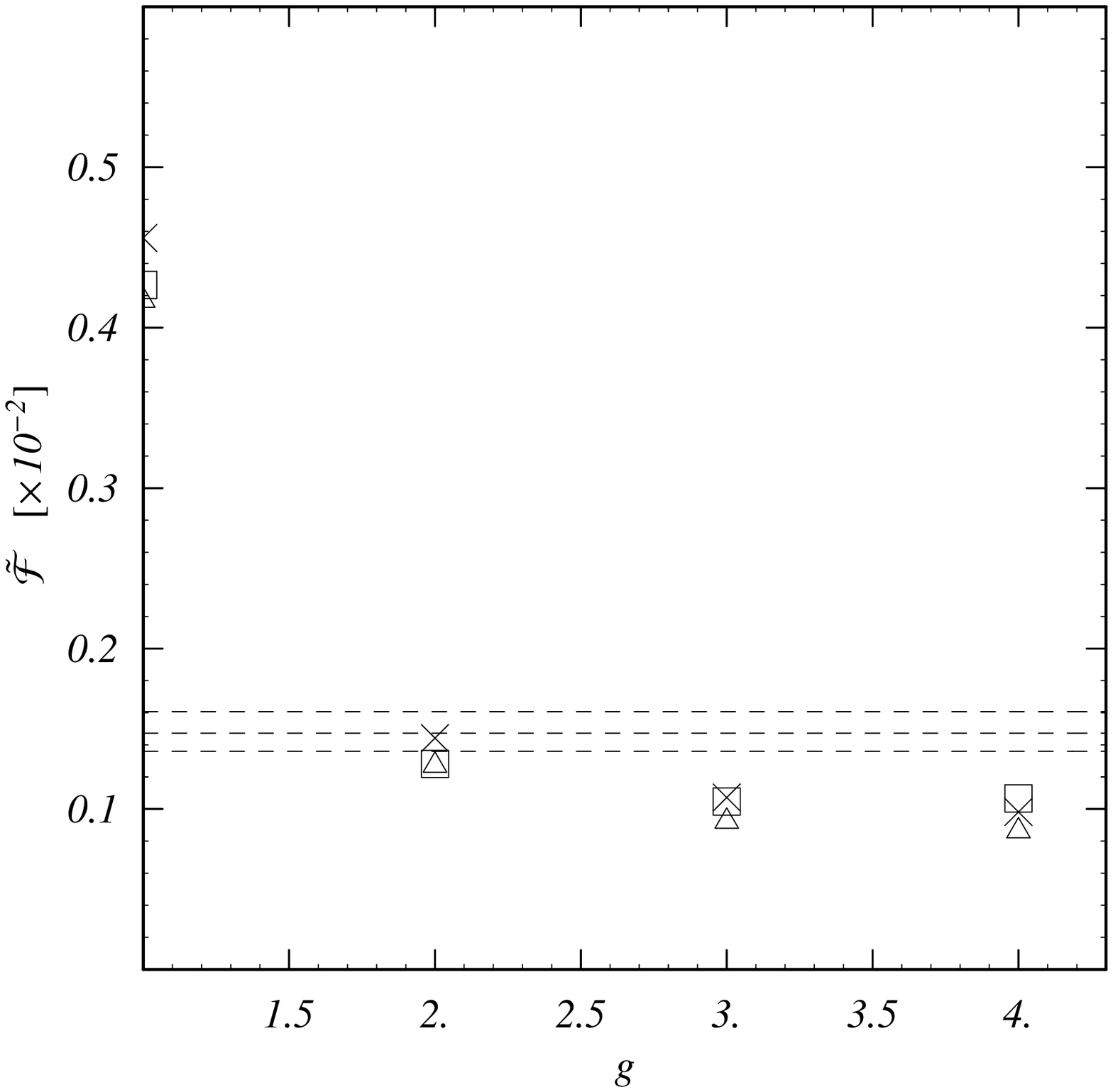}}\\
% \resizebox{77mm}{!}{\includegraphics{gFEatfixedK1213tau03modOUT.eps}}\\
     \qquad (a)&\qquad (b)\\[3mm]
    %  \hline\hline
    \end{tabular}
\vspace{-6mm}
    \caption{Same as Fig.~\ref{FEvsgA}, but for $T=0.3$.
    From (b) it is clear that for values of $g>1.0$ and at this temperature,
    the contribution to the free energy resembles the one from the nearly
    massless state in the $\cal T$-even sector.}
\label{FEvsgB}
  \end{center}
\end{figure}

\begin{figure}
  \begin{center}
     \hspace*{-0.5cm}
     \begin{tabular}{cc}
   % \hline\hline\\[3mm]
\resizebox{75mm}{!}{\includegraphics{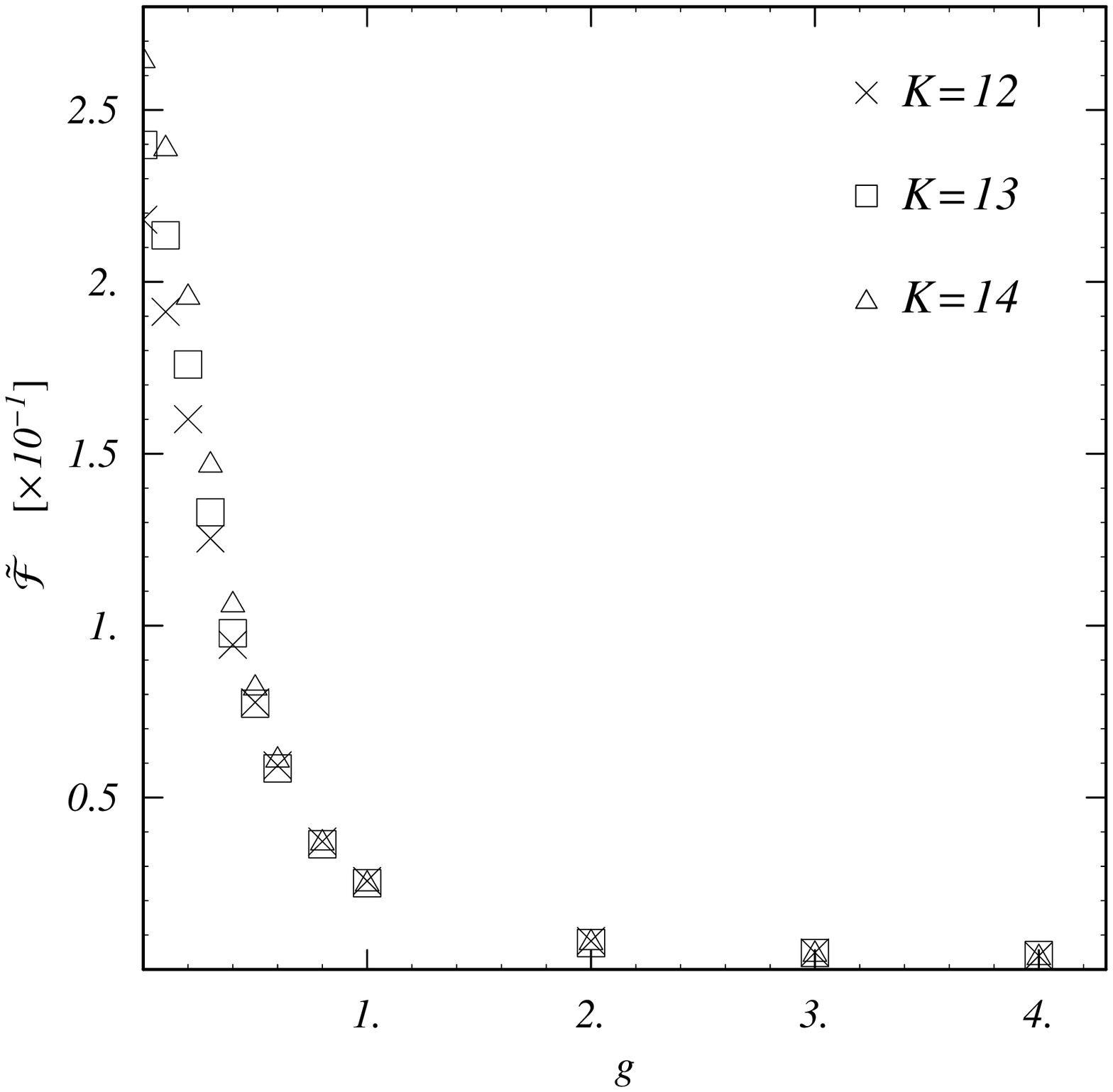}} &
% \resizebox{75.5mm}{!}{\includegraphics{gFEatfixedK1213tau05OUT.eps}} &
    %
 \resizebox{77mm}{!}{\includegraphics{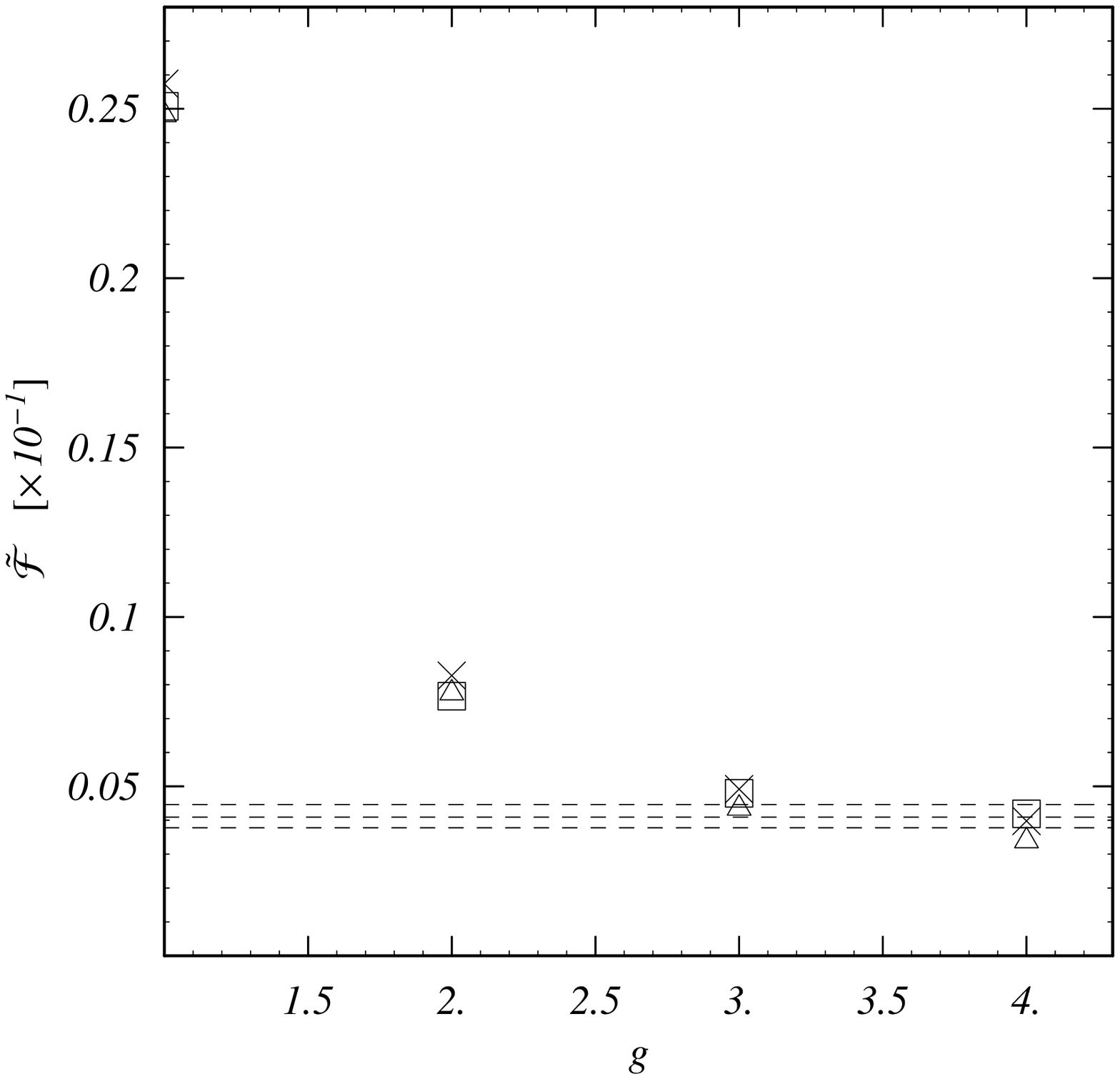}}\\
% \resizebox{77mm}{!}{\includegraphics{gFEatfixedK1213tau05modOUT.eps}}\\
     \qquad (a)&\qquad (b)\\[3mm]
    %  \hline\hline
    \end{tabular}
\vspace{-6mm}
    \caption{Same as Fig.~\ref{FEvsgA}, but for $T=0.5\kappa$.
    From (b) it is clear that more states contribute to the free energy 
    at higher temperature, and  at relatively large values of $g$.}
\label{FEvsgC}
  \end{center}
\end{figure}

The behavior of the free energy as a function of the coupling
is summarized in Figs.~\ref{FEvsgA}-\ref{FEvsgC}.
For relatively  low temperatures ($T\approx 0.1$)
and for values of $g$ on the order of one and
above (see Fig.~\ref{FEvsgA}),
the DoS fit misses the most important contribution, which is
expected from the single lightest state in the $\cal T$-even sector.
For instance, for the
resolution $K=16$ at coupling $g=4.0,$ the lightest bound state has
$M^2=0.0362 \kappa^2$ and the next available state is at $M^2=4.86\kappa^2$.
Although the fit in Fig.~\ref{CDFandDoSE}(c)
seems to capture quite well the behavior of the states
below the mass gap, states which are expected to dominate the
thermodynamics at low temperatures, it yields 
$\tilde{\mathcal{F}}(T=0.1)\approx 10^{-6} \kappa^2.$ This
is not what we expect from the CDF data. The free energy should be close
to the contribution of a pair of massless supersymmetric partners,
$\frac{\pi}{16(K-1)}T^2.$
A way to improve the calculation of
the free energy is to use the discrete spectrum
and sum over the states instead of approximating this part of the spectrum
with a fit function.
By extracting the states' degeneracies from the CDF data and by
utilizing Eq.~$(\ref{eqn:FEtotD2})$, we obtain
$\tilde{\mathcal{F}}(T=0.1)\approx 3.33\times 10^{-5}
\kappa^2,$ which matches the expectations much better.
In fact, the value of $\tilde {\cal F}$ is the contribution 
of one supersymmetric
$\cal T$-even state. The $\cal T$-odd sector
does not contribute significantly to $\tilde{\cal F}$, since its lightest
state ($M^2=3.651\kappa^2$) is heavily suppressed due to the Bessel factor
$\mathbf{K}_{1}(M/T)$ at $T=0.1$.

Although failing here, generally (at relatively weak couplings and small temperatures)
the fit does a good job, mainly because the states
below the mass gap
are very light compared to those for large $g$, and therefore not suppressed 
by the Bessel function,  $\mathbf{K}_{1}(M/T),$ of Eqs.~$(\ref{eqn:FEtotD2})$ 
and
 $(\ref{eqn:FEusedforCalculation}).$  Therefore, for  large values of the
coupling, we see  that  as the temperature is  gradually being increased,
the contribution to the free energy becomes similar to the one coming from
an exact massless mode; this contribution is included
as dotted lines in
Figs.~\ref{FEvsgA}---\ref{FEvsgC}. These results are also in accord
with the results presented in
Tables~\ref{tab:tableFreeEnK14g01} and \ref{tab:tableFreeEnK16g40}.
%1
%
%
%2
%
%
%3
%

As a check, we have compared results for the massive, strongly coupled theory
($g$ large, $\kappa=1$) and the massless theory ($g=1$, $\kappa=0$),
where $g$ is the only scale factor.  We expect the strongly coupled theory
to have masses $M$ related to the masses $M_*$ of the massless theory
by $M^2=g^2 M_*^2$.  At $g=4.0$, we find $M^{2}\approx 4.05^{2} M^{2}_{*}$.
The free energies are related by
\[\tilde{\mathcal{F}}(T,M^2,K)=g^2 \tilde{\mathcal{F}}_{*}(T/g,M^2_{*},K).\]
  We also calculated the free
energy with the DoS method described earlier, and we found that it matches
quite well the free energy of the theory with  $g=4.0$ and $\kappa=1.$
%
%4
%
\begin{figure}
  \begin{center}
     \hspace*{-0.5cm}
     \begin{tabular}{cc}
   % \hline\hline\\[3mm]
 \resizebox{73mm}{!}{\includegraphics{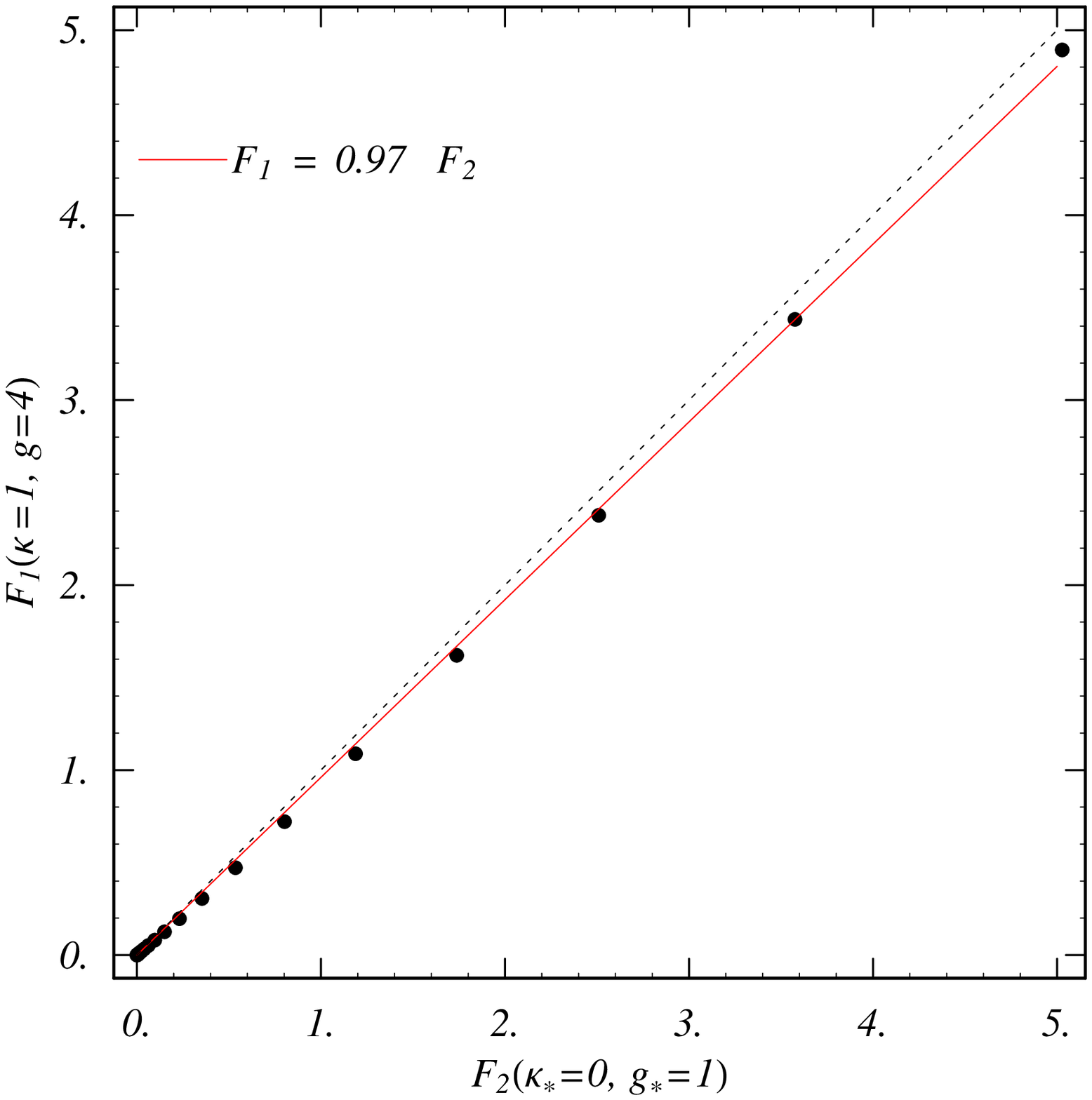}} &
 \resizebox{77mm}{!}{\includegraphics{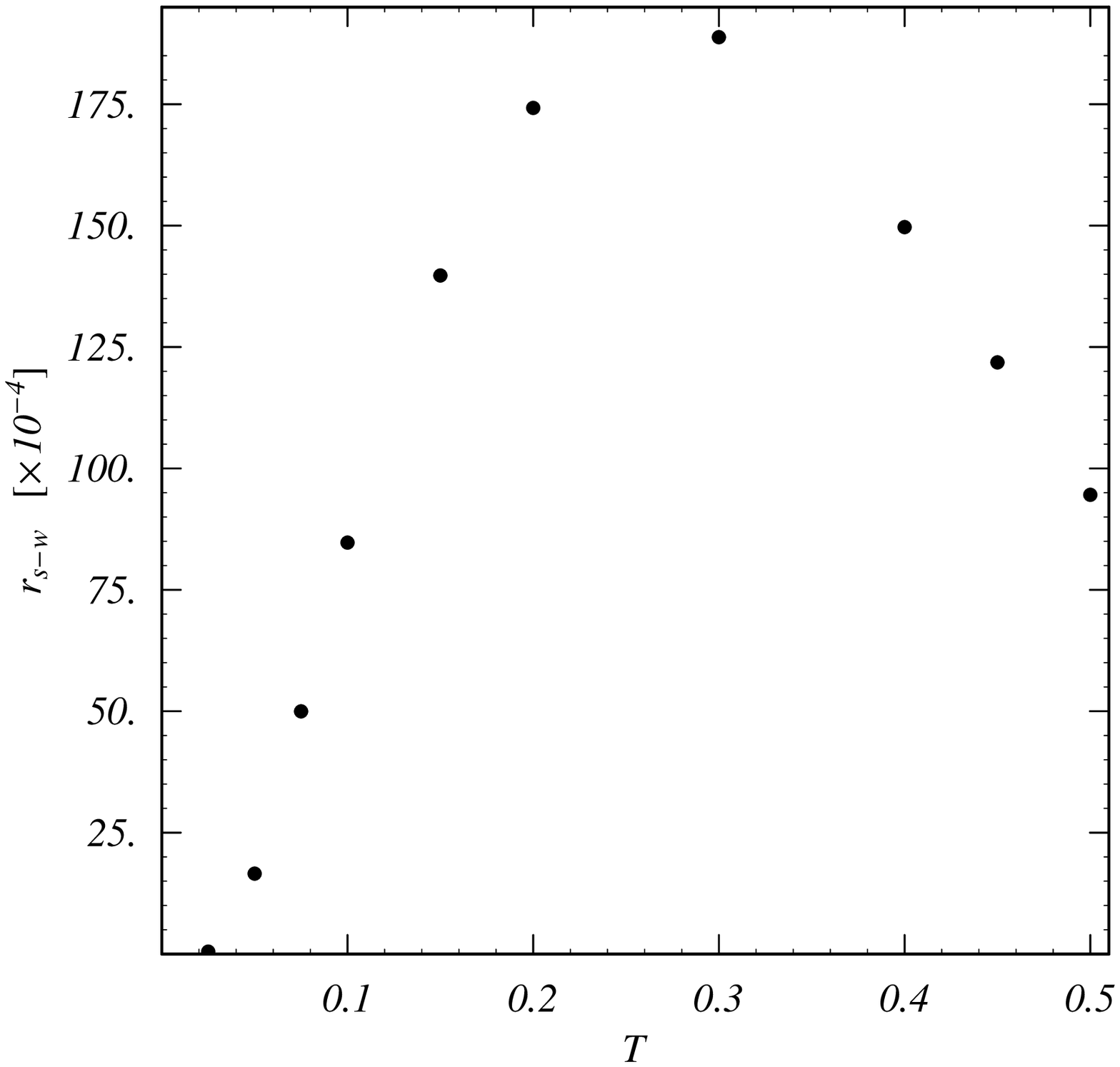}}\\
     \qquad (a)&\qquad (b)\\[3mm]
    %  \hline\hline
    \end{tabular}
\vspace{-6mm}
    \caption{(a) Comparison of the free energies $\tilde{\mathcal{F}}(g=4, \kappa=1)$ and
$\tilde{\mathcal{F}}_{*}(g_{*}=1, \kappa_{*}=0)$ at various temperatures between $0.015\kappa$ and $0.5\kappa$ with steps of  $\Delta T=0.015\kappa,$ for $\kappa=1.$ The dashed line represents a perfect match, the
solid line the best linear fit $\tilde{\mathcal{F}}(g=4, \kappa=1)=0.97 \tilde{\mathcal{F}}_{*}(g_{*}=1, \kappa_{*}=0)$.
(b) Strong/weak coupling ratio $r_{s-w}$ as
a function of the temperature at $K=16$.}
\label{fig:DUALspectra}
  \end{center}
\end{figure}
This is shown in Fig.~\ref{fig:DUALspectra}(a).
Therefore, by solving a numerically less challenging problem,
i.e. the model with no CS term, we were able to determine the strong
coupling behavior of the theory with a CS term.

Having established
that $g=4.0$ is a relatively strong  coupling, and by knowing the exact,
weak-coupling ($g=0$) free energy, let us calculate the
strong/weak coupling free-energy ratio $r_{s-w}$ at $K=16,$
the highest available resolution in our calculations.
At low
temperature, $T=0.1\kappa$, we get
$r_{s-w}(K=16)\approx 8.47 \times 10^{-3},$ and at $T=0.5\kappa$ we
obtain $r_{s-w}(K=16)\approx 9.46 \times 10^{-3}$.
Results for several temperatures are summarized in
Fig.~\ref{fig:DUALspectra}(b).
A quartic fit to the strong-coupling data
reveals that, at resolution $K=16$, the ratio is
\[
r_{s-w}=\frac{\tilde{\mathcal{F}}_s}{\tilde{\mathcal{F}}_w}\approx 0.85 T^2.
\]
This is consistent with the fact that at low temperatures and weak coupling, $g,$ the massless states dominate and make a contribution
proportional to $T^2$ to $\tilde{\mathcal{F}}_{w}.$
Our CDF data suggest that in the continuum limit, however, the
lightest (nearly massless) state of the strongly coupled theory will become exactly  massless, and also
yield a contribution  proportional to $T^2$ at low
temperatures. On the other hand, we know that in the weak coupling ($g\sim 0$) theory for finite $K$ 
there are exactly $2(K-1)$ massless pairs of bound states and the ratio $r_{s-w},$ for large $K,$ becomes
$
r_{s-w}(K)=({2(K-1)})^{-1}.
$
Thus, we conclude that in the continuum limit
\[
r_{s-w}\overset{K\to\infty}{\longrightarrow}0,
\]
and the discrepancy between the strongly and weakly coupled theories
becomes maximal. However, we cannot exclude the possibility that we may have more than one pair of massless states in the strongly coupled sector of the theory for large values of $K$, namely a number of massless states proportional to $K.$
Although from Fig.~$\ref{CS0g1}(a),$ which refers to the strongly coupled system (dual theory), one may try to argue in favor of the latter statement that the mass of states at small K seem to follow a trend towards the massless limit for relatively large $K$. However, this is not a definitive result, at least from our data, because the highest resolution we have is only up to $K=16.$ Therefore, as far as thermodynamics is concerned in this paper we will just assume that we only have one massless pair in the continuum limit of the strongly interacting sector.
Further results for $r_{s-w}$ for several values of $T$ and $K$ are
presented in Fig.~\ref{fig:DUALspectra}(b) and Table~\ref{tab:K15rws}.
It seems that $r_{s-w}$ decreases with $K$ starting at medium temperatures.
%K=15:ratio
\begin{table}[t]
\begin{center}
\begin{tabular}{ccccc}
\hline \hline\\[-4.0mm]
$T$ & $r_{s-w}(K=13)$ & $r_{s-w}(K=14)$ & $r_{s-w}(K=15)$ & $r_{s-w}(K=16)$ \\[1mm]
$[\kappa]$& $[10^{-4}]$& $[10^{-4}]$& $[10^{-4}]$& $[10^{-4}]$\\[2mm]
\hline\\[-2.5mm]
      0.025 &       0.12 &       0.19 &       0.29 & 0.50\\
      0.05 &      10.44 &      11.32 &      12.89 &   16.56\\
     0.075 &      42.95 &      41.24 &      42.71 &    49.97\\
       0.1 &      84.74 &      76.62 &      75.77 &    84.72\\
      0.15 &     161.82 &     138.05 &     130.57 &    139.74\\
       0.2 &     217.32 &     180.08 &     166.48 &    174.25\\
       0.3 &     262.02 &     208.95 &     186.77 &    188.81\\
       0.4 &     235.75 &     180.94 &     154.97 &    149.70\\
      0.45 &     207.06 &     155.59 &     129.71 &    121.84\\
       0.5 &     174.62 &     128.31 &     103.76 &    94.57\\
\hline \hline
\end{tabular}
\caption{Data for strong to weak coupling ratio $r_{s-w}$ for $K=13,14,15,16$
at various temperatures. We show 
$r_{s-w}(K=16)$ as a function of $T$ in Fig.~\ref{fig:DUALspectra}(b).}
\label{tab:K15rws}
\end{center}
\end{table}
%

%%%%%%%%%%%%%%%%%%%%%%%%%%%%%%%%%%%%%%%%%%%%%%%%%%%%%%%%%%%%%%%%%%
\section{Discussion}
\label{sec:discussion}
%%%%%%%%%%%%%%%%%%%%%%%%%%%%%%%%%%%%%%%%%%%%%%%%%%%%%%%%%%%%%%%%%%%%%%%%%%%%
We have studied the thermodynamics of   ${\cal N}=(1,1)$ super
Yang--Mills theory in 1+1 dimensions with fundamentals and a Chern--Simons term
that gives mass to the adjoint partons. We used SDLCQ to solve the theory in
the large-$N_c$ approximation. The theory has
two classes of bound states: glueballs, which form a closed string in
color space, and meson-like states, which form open strings.
In the large-$N_c$  approximation, these two sectors do not interact with each
other, and make independent contributions to the
thermodynamics. We previously calculated the contribution of the
glueball sector but without a CS term. We found that the meson-like
sector dominates the glueball sector for combinatorial reasons,
and, therefore, the results presented here represent the full
thermodynamics of the theory. Adding a CS term to the theory
introduces an additional parameter, and thus allows us
to inquire about the coupling dependence of the theory.

We have been able to take the calculation up to resolution $K=16$,
which effectively means that we are diagonalizing matrices that are of
order $7 \times 10^6$ by $7 \times 10^6$ in our approximation of the
continuum field theory. We introduced a new Lanczos method, which is
particularly valuable in our calculation of the Hagedorn
temperature.

It is interesting that the spectrum for this theory has
a mass gap, which we have discussed extensively. The states
below the mass gap dominate the low temperature behavior of the
theory while the states above the mass gap and below the point of inflection of the CDF determine the Hagedorn
temperature. In fact, the very low temperature behavior is dominated
by a few massless or nearly massless states in the theory.

The determination of the Hagedorn temperature from the states beyond the
mass gap requires a detailed understanding of the SDLCQ spectrum and
careful fitting techniques. We have checked our numerical methods by
comparing the solutions of the free, massive
theory obtained numerically to those extracted analytically.

We extrapolated the Hagedorn temperature at fixed coupling to
the continuum limit. The process is repeated at various
values of the coupling to determine the coupling dependence of the
Hagedorn temperature. We find that it increases with the coupling from a
value of about $T_H=\frac{1}{2}\kappa$ at $g=0$ to a value
of nearly $3.0\kappa$ at a coupling of $g=4.0$.

We calculate the free energy of the theory as a
function of both the temperature and the
coupling. As the coupling vanishes, the bound
state spectrum can be obtained analytically; the analytic results
agree with our SDLCQ calculations. The theory has
$4 (K-1)$ massless fermionic bound states and an equal number of
bosonic bound states. At low temperature and near-zero coupling,
the free energy is simply given by the contribution of these massless
states, which can be calculated analytically. As the temperature
increases, the free energy grows quadratically and
starts to diverge as the temperature
approaches the Hagedorn temperature.  As we discussed above, this
point of divergence increases with the coupling.

At strong coupling
and very low temperature, the nearly
massless bound states dominate the free energy. We find one
such fermionic and one such bosonic state. These states have
very small masses at the highest resolution. Their masses appear to
decrease with increasing resolution, suggesting that they will become
massless in the continuum limit. 
%Here the number of the states is independent of the resolution. 
Since the free energy at low temperatures is proportional
to the number of nearly massless states, 
the free energy at strong coupling is independent of the 
resolution and therefore has this fixed value in the continuum limit. 
On the other hand, at weak
coupling the number of light states grows with the resolution and
diverges in the continuum limit, as does the free energy. We therefore 
find that at low temperatures the ratio of the free energies
at strong and weak coupling goes to zero as we approach the continuum.

A number of interesting extensions of SYM theory with fundamentals and CS
term exist,
for which an SDLCQ mass spectrum can be computed. For instance, one can
increase the number of dimensions or increase the
number of supersymmetries. It would be interesting and 
straightforward to extract
the thermodynamic properties of these extended theories.

%%%%%%%%%%%%%%%%%%%%%%%%%%%%%%%%%%%%%%%%%%%%%%%%%%%%%%%%%%%%%%%%%%
\section*{Acknowledgments}
This work was supported in part by the U.S. Department of Energy and the
Minnesota Supercomputing Institute. One of the authors (U.T.) would like
to thank the Research Corporation for supporting his work.
%%%%%%%%%%%%%%%%%%%%%%%%%%%%%%%%%%%%%%%%%%%%%%%%%%%%%%%%%%%%%%%%%%

%----------------------------------------------------------------------%
%                     BIBLIOGRAPHY
%----------------------------------------------------------------------%


\begin{thebibliography}{99}
%\cite{Gubser:1996de}
\bibitem{Gubser:1996de}
S.~S.~Gubser, I.~R.~Klebanov, and A.~W.~Peet,
%``Entropy and Temperature of Black 3-Branes,''
Phys.\ Rev.\ D {\bf 54}, 3915 (1996) [arXiv:hep-th/9602135]. %%
%\cite{Li:1998kd}
\bibitem{Li:1998kd}
M.~Li,
%``Evidence for large N phase transition in N = 4 super
%Yang--Mills theory  at finite temperature,''
JHEP {\bf 9903}, 004 (1999) [arXiv:hep-th/9807196]; A.A.~Tseytlin
and S.~Yankielowicz,
%``Free energy of N=4 super Yang--Mills in Higgs phase and non-extremal
%D3-brane
%interactions,''
Nucl.\ Phys.\ B {\bf 541}, 145-162 (1999)
[arXiv:hep-th/9809032]; A.~Fotopoulos and T.R.~Taylor,
%``Remarks on two-loop free energy in N=4 supersymmetric Yang-Mills theory
%at finite temperature,''
Phys.\ Rev.\ D {\bf 59}, 061701 (1999) [arXiv:hep-th/9811224].
%
%\cite{Das:rx}
\bibitem{Das:rx}
A.~Das and M.~Kaku,
%``Supersymmetry At High Temperatures,''
Phys.\ Rev.\ D {\bf 18}, 4540 (1978).
%
\bibitem{sakai}
Y.~Matsumura, N.~Sakai, and T.~Sakai,
%``Mass spectra of supersymmetric Yang-Mills theories in (1+1)-dimensions,''
Phys.\ Rev.\ D {\bf 52}, 2446 (1995) [arXiv:hep-th/9504150].
%
\bibitem{Lunin:1999ib}
O.~Lunin and S.~Pinsky,
%``SDLCQ: Supersymmetric discrete light cone quantization,''
AIP Conf.\ Proc.\  {\bf 494}, 140 (1999) [arXiv:hep-th/9910222].
%
\bibitem{lattice}
A.~G.~Cohen, D.~B.~Kaplan, E.~Katz, and M.~Unsal,
%``Supersymmetry on a Euclidean spacetime lattice. I:
%A target theory with  four supercharges,''
[arXiv:hep-lat/0302017]
%
%\cite{Hiller:2004ft}
\bibitem{Hiller:2004ft}
  J.~R.~Hiller, S.~S.~Pinsky, Y.~Proestos and N.~Salwen,
  %``${\cal N}=(1,1)$ super Yang--Mills theory in 1+1 dimensions at finite temperature,''
  Phys.\ Rev.\ D {\bf 70}, 065012 (2004)
  [arXiv:hep-th/0407076].
  %%CITATION = HEP-TH 0407076;%%
%
%\cite{Brodsky:2001bx}
\bibitem{Brodsky:2001bx}
S.~J.~Brodsky,
%``Physics at the light-front,''
Nucl.\ Phys.\ Proc.\ Suppl.\  {\bf 108}, 327 (2002)
[arXiv:hep-ph/0112309].
%
%\cite{Alves:2002tx}
\bibitem{Alves:2002tx}
V.~S.~Alves, A.~Das, and S.~Perez,
%``Light-front field theories at finite temperature,''
Phys.\ Rev.\ D {\bf 66}, 125008 (2002) [arXiv:hep-th/0209036];
%\cite{Das:2003mf}
%\bibitem{Das:2003mf}
A.~Das and X.~X.~Zhou,
%``Light-front Schwinger model at finite temperature,''
Phys.\ Rev.\ D {\bf 68}, 065017 (2003) [arXiv:hep-th/0305097];
%%CITATION = HEP-TH 0305097;%
%
%\cite{Das:2003sx}
%\bibitem{Das:2003sx}
A.~Das,
%``Finite temperature field theories on the light-front,''
arXiv:hep-th/0310247.
%
%\cite{Weldon:aq}
\bibitem{Weldon:aq}
H.~A.~Weldon,
%``Covariant Calculations At Finite Temperature: The Relativistic Plasma,''
Phys.\ Rev.\ D {\bf 26}, 1394 (1982);
%
%
%\cite{Beyer:2003qb}
\bibitem{Beyer:2003qb}
M.~Beyer, S.~Mattiello, T.~Frederico, and H.~J.~Weber,
%``Light front thermal field theory at finite temperature and density,''
[arXiv:hep-ph/0310222]. %%
%\cite{Elser:1996tq}
\bibitem{Elser:1996tq}
S.~Elser and A.~C.~Kalloniatis,
%``QED(1+1) at Finite Temperature -- a Study with Light-Cone Quantisation,''
Phys.\ Lett.\ B {\bf 375}, 285 (1996) [arXiv:hep-th/9601045].
%
%\cite{pb85}
\bibitem{pb85}
H.-C.~Pauli and S.J.~Brodsky, Phys.\ Rev.\ D {\bf 32}, 1993
(1985); {\bf 32}, 2001 (1985);
%
%\cite{bpp98}
\bibitem{bpp98}
S.J.~Brodsky, H.-C.~Pauli, and S.S.~Pinsky,
%``Quantum Chromodynamics and Other Field Theories on the Light Cone,''
Phys.\ Rep.\ {\bf 301}, 299 (1998) [arXiv:hep-ph/9705477].
%
%\cite{Antonuccio:1998kz}
\bibitem{Antonuccio:1998kz}
F.~Antonuccio, O.~Lunin, and S.~S.~Pinsky,
%``Bound states of dimensionally reduced SYM(2+1) at finite N,''
Phys.\ Lett.\ B {\bf 429}, 327 (1998) [arXiv:hep-th/9803027].
%
%\cite{Antonuccio:1999zu}
\bibitem{Antonuccio:1999zu}
F.~Antonuccio, O.~Lunin, and S.~Pinsky,
%``Super Yang--Mills at weak, intermediate and strong coupling,''
Phys.\ Rev.\ D {\bf 59}, 085001 (1999) [arXiv:hep-th/9811083].
%
%\cite{Haney:2000tk}
\bibitem{Haney:2000tk}
P.~Haney, J.~R.~Hiller, O.~Lunin, S.~Pinsky, and U.~Trittmann,
%``The mass spectrum of N = 1 SYM(2+1) at strong coupling,''
Phys.\ Rev.\ D {\bf 62}, 075002 (2000) [arXiv:hep-th/9911243].
%
%\cite{hpt2001}
\bibitem{hpt2001}
J.~R.~Hiller, S.~Pinsky, and U.~Trittmann,
%``Wave functions and properties of massive
%states in three-dimensional  supersymmetric Yang--Mills theory,''
Phys.\ Rev.\ D {\bf 64}, 105027 (2001) [arXiv:hep-th/0106193].
%
%\cite{Hagedorn_NC_65_68}
\bibitem{Hagedorn_NC_65_68}
R. Hagedorn, Nuovo Cimento Suppl. {\bf 3}, 147 (1965); R.
Hagedorn, Nuovo Cimento {\bf 56A}, 1027 (1968).
%
%\cite{Kutasov:1993gq}
\bibitem{Kutasov:1993gq}
  D.~Kutasov,
  %``Two-dimensional QCD coupled to adjoint matter and string theory,''
  Nucl.\ Phys.\ B {\bf 414}, 33 (1994)
  [arXiv:hep-th/9306013].
  %%CITATION = HEP-TH 9306013;%%
%
%\cite{Gross:1997mx}
\bibitem{Gross:1997mx}
  D.~J.~Gross, A.~Hashimoto and I.~R.~Klebanov,
  %``The spectrum of a large N gauge theory near transition from confinement  to
  %screening,''
  Phys.\ Rev.\ D {\bf 57}, 6420 (1998)
  [arXiv:hep-th/9710240].
  %%CITATION = HEP-TH 9710240;
%
%\cite{anp98}
\bibitem{Antonuccio:1998uz}
  F.~Antonuccio and S.~Pinsky,
  %``On the transition from confinement to screening in QCD(1+1) coupled to
  %adjoint fermions at finite N,''
  Phys.\ Lett.\ B {\bf 439}, 142 (1998)
  [arXiv:hep-th/9805188].
  %%CITATION = HEP-TH 9805188;%%
%
%\cite{Hiller:2002cu}
\bibitem{Hiller:2002cu}
  J.~R.~Hiller, S.~S.~Pinsky and U.~Trittmann,
  %``Approximate BPS states,''
  Phys.\ Rev.\ Lett.\  {\bf 89}, 181602 (2002)
  [arXiv:hep-th/0203162].
  %%CITATION = HEP-TH 0203162;%%
%
%\cite{Hiller:2004rb}
\bibitem{Hiller:2004rb}
  J.~R.~Hiller, M.~Harada, S.~S.~Pinsky, N.~Salwen and U.~Trittmann,
  %``Two-dimensional super Yang-Mills theory investigated with improved
  %resolution,''
  Phys.\ Rev.\ D {\bf 71}, 085008 (2005)
  [arXiv:hep-th/0411220].
  %%CITATION = HEP-TH 0411220;%%
%\cite{Hiller:2003jd}
\bibitem{Hiller:2003jd}
  J.~R.~Hiller, S.~S.~Pinsky and U.~Trittmann,
  %``Spectrum of N = 1 massive super Yang-Mills theory with fundamental  matter
  %in 1+1 dimensions,''
  Phys.\ Rev.\ D {\bf 67} (2003) 115005
  [arXiv:hep-ph/0304147].
  %%CITATION = HEP-PH 0304147;%%
%\cite{Hiller:2003qe}
\bibitem{Hiller:2003qe}
  J.~R.~Hiller, S.~S.~Pinsky and U.~Trittmann,
  %``Anomalously light mesons in a (1+1)-dimensional supersymmetric theory  with
  %fundamental matter,''
  Nucl.\ Phys.\ B {\bf 661} (2003) 99
  [arXiv:hep-ph/0302119].
  %%CITATION = HEP-PH 0302119;%%
%
% dosalg.tex Lanczos Alg
%
\bibitem{AlbenHams} R. Alben, M. Blume, H. Krakauer, and L. Schwartz,
Phys.\ Rev.\ B {\bf 12}, 4090 (1975);
A. Hams and H. De Raedt,
Phys.\ Rev.\ E {\bf 62}, 4365 (2000).

\bibitem{Iitaka} T. Iitaka and T. Ebisuzaki,
Phys.\ Rev.\ E {\bf 69}, 057701 (2004).

\bibitem{JaklicAichhorn} J. Jakli\v{c} and P. Prelov\v{s}ek,
Phys.\ Rev.\ B {\bf 49}, 5065(R) (1994);
M. Aichhorn, M. Daghofer, H.G. Evertz, and W. von der Linden,
Phys.\ Rev.\ B {\bf 67}, 161103(R) (2003).
%
%
\end{thebibliography}
\end{document}